\pdfoutput=1
\documentclass[11pt]{article}
\usepackage[margin=1in]{geometry}
\usepackage[utf8]{inputenc}			
\usepackage{bbm}
\usepackage[labelfont=sc]{caption}
\usepackage[singlespacing]{setspace}
\usepackage[font=small,labelfont=normalfont,subrefformat=parens]{subcaption}
\usepackage[usenames,svgnames]{xcolor}
\usepackage[bookmarks=true,colorlinks=true,allcolors=MidnightBlue]{hyperref}

\usepackage{multicol}

\usepackage{tikz}
\usetikzlibrary{arrows.meta, positioning, calc, fit, backgrounds}

\usepackage{lscape} 							
\usepackage{lipsum} 							
\usepackage{booktabs} 						
\usepackage{amsmath,amsfonts,amssymb} 	
\usepackage{array}							
\usepackage{textcomp}						
\usepackage{exscale} 							
\usepackage{tabularx} 						
\usepackage{natbib}							
\usepackage{ragged2e} 						
\usepackage{multirow}           					
\usepackage{eurosym} 						
\usepackage{verbatim}						
\usepackage{float}							
\usepackage{longtable}
\usepackage[toc,page]{appendix}
\usepackage{pdfpages}
\usepackage{rotating}
\usepackage{caption}
\usepackage{adjustbox}
\usepackage{longtable,lscape}
\usepackage{smartdiagram}
\usepackage[flushleft]{threeparttable}
\usepackage{numprint} 
\usepackage{siunitx}
\usepackage{tabularx}
\usepackage{numprint}
\usepackage{adjustbox}
\usepackage{placeins}
\usepackage{enumitem}
\usepackage[edges]{forest}
\definecolor{illured}{RGB}{192,0,0}     
\definecolor{illugold}{RGB}{191,144,0}  

\tikzset{
	title/.style={draw=illured, line width=1.1pt, rounded corners=5pt, fill=white,
		minimum width=3.0cm, minimum height=1.0cm, font=\Large\bfseries\boldmath},
	doc/.style={draw=illured, line width=1.1pt, rounded corners=3pt, fill=white,
		minimum size=0.72cm, inner sep=0pt},
	pool/.style={draw=illured, line width=1.1pt, rounded corners=7pt, fill=white, inner sep=6pt},
	sim/.style={-{Stealth[length=7pt, width=5pt]}, draw=illured, line width=0.9pt},
	sel/.style={-{Stealth[length=9pt, width=7pt]}, draw=illugold, line width=1.8pt},
	fieldline/.style={line width=2pt},
	lab/.style={font=\Large\bfseries\boldmath},          
	win/.style={font=\large\bfseries\boldmath},          
	legend/.style={font=\Large\bfseries\boldmath},       
	alab/.style={font=\large\bfseries\boldmath, inner sep=1.5pt},   
}

\newcommand{\legendline}[3]{
	\node[legend] (#1) at (#2) {#3};
	\draw[illured, line width=1.1pt] ($(#1.south west)+(0,-3pt)$) -- ($(#1.south east)+(0,-3pt)$);
}

\usepackage{booktabs, caption, makecell}

\usepackage{threeparttable}

\usepackage{longtable}

\title{{\Large \textbf{Did a feedback mechanism between propositional and prescriptive knowledge create modern growth?\thanks{I am indebted to Joel Mokyr for kindling my interest in the historical knowledge economy through his academic writings and his personal encouragement and invaluable input that has made this project possible. I would further like to express my gratitude to my PhD supervisors Max-Stephan Schulze and Jeremiah Dittmar at the LSE for their guidance and support. Moreover, I am grateful to everyone at Northwestern University who supported me during my stay in the fall of 2022. Further thanks go to seminar participants at the LSE graduate economic history seminar, the World Economic History Congress in Lund, the Annual Meeting of the Economic History Society, London, and the workshops on ``Growth, History and Development'' at the University of Southern Denmark,   ``Technology and Long-Term Development'' at the University of Valencia, ``Economic History in the Age of AI'' at Stellenbosch University, ``History LLMs'' in Zurich, the ``Workshop in Economic History'' in Uppsala, ``Methodological Advances in Quantitative Economic History'' at the University of Warwick, the 41st Congress of the European Economic Association in Dublin, the Economic and Social History Seminar Series at the University of Oxford, and the departmental seminar at St. Andrews. Additionally, I want to thank Bruno Caprettini, Lukas Rosenberger, Erik Hornung, Caterina Chiopris, Christian Vedel, Alexandra de Pleijt, and David de la Croix for their detailed and most helpful feedback. Finally, I am obliged to Alessandro Nuvolari and Valentina Tartari for kindly sharing their data on English patents.}
	}}\\
}

\author{Julius Koschnick (University of Southern Denmark)\footnote{University of Southern Denmark, Economics Department, Campusvej 55, Odense, Denmark}}

\begin{document}
	\onehalfspacing
	
	\maketitle
	\thispagestyle{empty}
	
	\vspace{4ex}

	\vspace{6mm}
	
	\begin{abstract}		
		Joel Mokyr has argued that self-sustained modern growth originated from a feedback loop between propositional (theoretical) and prescriptive (applied) knowledge that turned positive during the eighteenth century ``Industrial Enlightenment''. This paper presents the first quantitative test of whether such a feedback loop existed. For this, it introduces new text-based measures for 1) the innovativeness of publications and 2) knowledge spillovers. It then applies them to ca. 300,000 English publication titles and short patent descriptions, 1600--1800. The paper finds that a feedback loop between propositional and prescriptive knowledge became positive by the mid-eighteenth century, supporting Mokyr's original hypothesis.
		\vspace{12pt} \newline
		\thispagestyle{empty} \noindent \textbf{Keywords}: \textsc{Economic Growth, Innovation, Feedback Loop Processes, Knowledge Spillovers, Natural Language Processing} \newline
	\end{abstract}
	\newpage

	\setcounter{page}{1}
	
	\section{Introduction}
	
	What explains the onset of modern economic growth at the end of the eighteenth century? Joel Mokyr (\citeyear{Mokyr2002}) has argued that self-sustained modern economic growth originated from structural change in the knowledge economy following the Scientific Revolution and Enlightenment. In the centuries before, waves of invention were rare and short-lived. After 1760, however, England entered a regime of steady growth in knowledge production, patents, and economic output \citep{Mokyr2002, Mokyr2016,Clark2007,baten2008book,Allen2009,Nuvolari2011}. According to \cite{Mokyr2002}, the transition to modern economic growth can be explained by a feedback loop between propositional (theoretical) and prescriptive (applied) knowledge. He argues that this feedback loop only became positive in the eighteenth century following changes in the production and sharing of ideas during the Enlightenment. While the feedback loop thesis has been widely influential \citep{Jacob2014,Squicciarini2015,Kelly2022a,prize2025scientific}, we still lack quantitative estimates of feedback loop processes that could either support or falsify the theory.

	This paper introduces the first estimates of feedback loop processes between propositional and prescriptive knowledge for seventeenth and eighteenth century England by utilizing recent advances in natural language processing. In doing so, the paper is the first to quantitatively test \cite{Mokyr2002}'s feedback loop hypothesis. It is also the first paper to trace long-run patterns of innovation and spillovers in Britain's knowledge economy before and during the Industrial Revolution.

	To do so, the paper introduces two new measures of 1) the innovativeness of publications and 2) spillovers between subject fields that are entirely based on information from text data. The measures are based on comparisons between the past and future of different fields while operationalizing a semantically deep and historically fine-tuned BERT model. With these two measures, it becomes possible to derive measures of innovation and knowledge spillovers i) in the absence of citation data and ii) in the presence of historical and context-sensitive text. The paper successfully validates the new measures against historical patent citations, number of editions, and authors publishing in multiple disciplines. It then applies the innovation and spillover measures to the universe of ca. 300,000 publication titles (historically equivalent to short abstracts) and patent descriptions in England, 1600--1800.\footnote{During this time period, English publishers used the full space of a book's cover to publish short abstracts of the book with an average count of 55 words. Moreover, the paper validates that titles are strong predictors of actual book content by drawing on a sample of 214 full-text books from Early English Books Online (EEBO).}

	This toolset makes it possible to directly test Mokyr's feedback loop thesis. According to \cite{Mokyr2002}, throughout most of human history, feedback loop processes were either non-existent or negative. Only with the ``Industrial Enlightenment'' of the eighteenth century did the feedback loop process turn positive, as propositional and prescriptive knowledge began to reinforce one another, ``eventually tipping the balance of the feedback mechanism from negative to positive.'' \citep[p. 33]{Mokyr2002}. To test this prediction, the paper estimates whether, over time, spillovers from one type of knowledge increased the innovativeness of the other. This follows Mokyr's definition of a positive feedback loop where ``growth in one increases the marginal product of the other'' \citep[p. 21]{Mokyr2002}. 
	
	First, the paper produces associational evidence on the development of the relationship between received knowledge spillovers and innovativeness at the title level.\footnote{To classify titles as propositional or prescriptive knowledge, the paper follows the classification system from \cite{Koschnick2025}. The set of propositional knowledge includes the subject fields of \textit{applied physics, astronomy, mathematics, chemistry, and encyclopedias} and the set of prescriptive knowledge includes the subject fields of \textit{technical publications, navigation, scientific instruments, and patents}. This  closely follows \cite{Mokyr2002}'s original definition of propositional knowledge being broader than just science, but also comprising simple facts and explanations. The approach of classifying titles into propositional and prescriptive knowledge is explained in detail in section~\ref{sec:assigning_subject_classes_prop_prescr}.}   Accounting for subject- and publication-year fixed effects, it finds evidence of neutral or negative knowledge spillovers at the beginning of the seventeenth century. Then, over the next 140 years, negative spillovers phased out until around 1740 when the relationship became positive for the first time. These findings directly correspond to the predictions from \cite{Mokyr2002}. The paper further links these results to the real economy by estimating the association between spillovers from propositional knowledge and patent citations in 1700--1800. As before, we find evidence of a positive relationship for the second half of the eighteenth century. 
	
	To rule out that the observed relationship is driven by shifts in language or stylistic conventions, the paper conducts a series of placebo tests. First, the paper presents evidence that spillovers from 14 plausibly unrelated fields, incl. \textit{poetry, drama, foreign languages, biographies,} and \textit{travel descriptions} did not exhibit any positive pattern. Second, it repeats the analysis for closely related fields that would nevertheless be unlikely to affect propositional or prescriptive knowledge, incl. \textit{classical education, philosophy}, and \textit{law}. Moreover, the paper also presents evidence from estimating spillovers from all 44 subject classes in the ESTC that were not part of propositional or prescriptive knowledge, incl. possibly related fields such as \textit{business} or \textit{exploration}. Even here, spillovers from propositional and prescriptive knowledge remain larger than any of the other spillover coefficients, further highlighting the importance of spillovers between propositional and prescriptive knowledge within the knowledge economy.
	To further test robustness to changes in style, the paper uses an LLM to rewrite all titles in a simplified style. Despite substantial changes in wording, the main results remain unchanged.
	
	Overall, these findings constitute strong evidence of the emergence of a structural break in Britain's knowledge economy in the seventeenth and eighteenth century. Growth (spillovers) in one type of knowledge indeed seems to have increased the marginal product of the other type of knowledge (innovation). Moreover, the timing of the flipping of the relationship from negative to positive spillovers follows exactly the narrative account of \cite{Mokyr2002}.

	To further understand the mechanism underlying the positive knowledge spillovers emerging at the end of the eighteenth century, the paper tests whether upper-tail human capital facilitated knowledge spillovers, as predicted by \cite{Mokyr2002,Mokyr2016}. Extracting information on authors' careers from the ESTC, we find that fellows of the Royal Society, the Society of Arts \citep{Howes2020}, and authors with a university education \citep{Koschnick2025} were more likely to incorporate spillovers from other fields. We interpret this as evidence that, consistent with recent literature \citep{Squicciarini2015,Kelly2020,hanlonengineer}, upper-tail knowledge elites were instrumental in bridging the gap between propositional and prescriptive knowledge.
	
	Moreover, the paper goes beyond \cite{Mokyr2002}'s theory and investigates how revolutions in methods that occurred during the seventeenth and eighteenth century could have acted as complements to feedback loop processes. The importance of revolutions in methods has been stressed by scholars such as Landes, Wootton, Kelly, and Ó Gráda. \cite{Landes1969,Landes1998} and \cite{Wootton2015} have placed special emphasis on the new scientific method that consisted of observation, formalization, and experiment. \cite{Jacob1997,Jacob2014} has stressed the importance of the Newtonian Revolution and Newtonian mechanics for the British Industrial Revolution. Moreover, \cite{OGrada2016} and \cite{Kelly2020} have highlighted the importance of precise measurement as a spillover from scientific instruments that enabled a new degree of complexity in machinery.

	The paper tests these channels by calculating embedding space similarities to a set of descriptive terms of these methods \citep[similar to][]{Garg2018,ash2025ideas}. It then interacts the spillover measure with the similarity-to-methods measure to estimate complementarities between the two. The paper finds that for the adoption of propositional knowledge in new prescriptive knowledge, precise measurement acted as a key catalyst. Vice versa, for the adoption of prescriptive knowledge in new propositional knowledge, Newtonian mechanics and the scientific method were key catalysts.

	Overall, the paper contributes to the literature  across four key dimensions. First, it provides a new approach to estimate structural change in the knowledge economy and contributes to the debate on the origin of modern economic growth \citep[see e.g.][]{Romer1986,Romer.1990,pomeranz2000great,Mokyr2002,Clark2007,Allen2009}. Here, it provides the first quantitative test of \cite{Mokyr2002}'s theory of a feedback loop process between propositional and prescriptive knowledge. The evidence found offers strong support for \cite{Mokyr2002}'s account of ideas and knowledge production as first-order drivers of the Industrial Revolution. Altogether, the findings of the paper place additional weight on the importance of the knowledge economy for the Industrial Revolution, in contrast to e.g. natural resources \citep{Wrigley2012} or factor prices \citep{Allen2009}, although these accounts can clearly be seen as complementary.\footnote{Here the printing press \citep{dittmar2011information,Dittmar2019} and institutions \citep{Mokyr2025_CultureVsInstitutions} feature as an important precondition for the European divergence in knowledge production and science.} 
	
	Second, the paper moves beyond \cite{Mokyr2002}'s theory and investigates the mechanisms behind the feedback loop mechanism. The paper places new emphasis on the role of revolutions in methods that occurred during the seventeenth and eighteenth century, such as the scientific method, the Newtonian Revolution, and precise measurement. These channels have been highlighted by \cite{Landes1969,Landes1998}, \cite{Wootton2015}, \cite{Jacob1997,Jacob2014}, \cite{OGrada2016}, and \cite{Kelly2020} as potential first-order determinants of changes in the knowledge economy before the Industrial Revolution. This paper provides the first quantitative evidence how these revolutions in methods entered the feedback loop process. Additionally, the paper shows that upper-tail human capital and networks through knowledge sharing societies contributed to the creation of knowledge spillovers. With this, the paper complements a long-standing literature on upper-tail human capital \citep{Meisenzahl2011,Squicciarini2015,hanlonengineer,Maloney2022,Kelly2022a,Kelly2023,cinnirella2025flow}. 
	
	Third, the paper relates to the long-standing debate on the influence of science on the Industrial Revolution. Here, scholars have been divided on whether science was a key driver of the Industrial Revolution \citep{Schofield1957,Schofield1963,Musson1969,Stewart1986,Stewart1995,Jacob1997,Jacob2014} or too underdeveloped to have had a direct effect \citep{Mathias1972,Hall1974,OGrada2016}. The paper contributes to the literature by providing quantitative evidence of innovation-enhancing spillovers from science. However, these occur in a setting that allows for a multitude of indirect effects, such as the adoption of the scientific method or a quantitative mindset of precise measurement. Thereby, it reconciles both positions and contributes to a literature that has studied the diffusion of technical knowledge \citep{hanlon2022penny,rosenberger2024innovation,Chiopris2024} and scientific knowledge \citep{Dittmar2021,zanardello2024early,curtis2023seeds,delacroix2025flora,Koschnick2025,cervellati2026knowledge,melillo2026republic} during the early modern period and Industrial Revolution. 
	
	Fourth, the paper contributes to the literature on innovation and the historical knowledge economy by introducing a new method to measure innovation and spillovers based on text data. It further develops an approach, using a time-locked BERT model, to apply these measures to historical text data. The measures are successfully validated against modern and historical data. Hence, they provide a new way to proxy innovation and spillovers, complementary to existing bag-of-words approaches \citep{Iaria2018,Kelly2021a,almelhem2023enlightenment,grajzl2024quiet,grajzl2026less,Chiopris2024}. Because they rest on sentence embeddings rather than word counts, these measarues are likely to work better with semantically complex text, and in particular to historical text that usually exhibits inconsistent spelling conventions and complex content.

	The paper proceeds in the following way. Section \ref{sec:literature} provides a conceptual overview of knowledge spillovers between propositional and prescriptive knowledge and illustrates them with the historical case of the improvement of water power.  Section \ref{sec:data} introduces the text data used in this paper, including the universe of printed titles in Britain between 1600 and 1800, as well as patent short descriptions.  Section \ref{sec:spillovers} introduces the innovation and spillover measure. The measure uses the embedding space of a historically trained and fine-tuned \textit{SteamBERTh} model. Section \ref{sec:validation_section} validates the innovation measure with historical patent citations. Next, section \ref{sec:empirical-framework} introduces the empirical framework for estimating the relationship between knowledge spillovers between propositional and prescriptive knowledge and innovation over time. Section \ref{sec:results} presents results. Lastly, section \ref{sec:mechanism} explores multiple mechanisms, including upper-tail human capital as well as the introduction of the scientific method. Section \ref{sec:conclusion} concludes.

	\FloatBarrier
	
	\section{\label{sec:literature}Conceptual and historical background}
	
	\subsection{Conceptual background}
	\label{sec:coneptual_background}

	This paper quantitatively tests Joel Mokyr's feedback loop hypothesis. In his \textit{Gifts of Athena}, \cite{Mokyr2002} argues that throughout most of history, humanity's stock of knowledge remained constant. Although there were few waves of macro inventions, such as mechanical clocks in the late medieval period or the casting of iron in the fifteenth century, all of these ebbed off quickly and were again followed by long-run stagnation. In contrast, according to Mokyr, in the eighteenth century, a positive feedback mechanism between theoretical and applied knowledge created self-sustained technological growth for the first time in history.

	To precisely capture the spheres of theoretical and applied knowledge, Mokyr introduces the terms of propositional knowledge, knowledge about how the (natural) world works and prescriptive knowledge, knowledge of how to manipulate the (natural) world. Propositional and prescriptive knowledge include science and technology respectively, but they are wider concepts going e.g. beyond the strict domain of science.

	Mokyr argues that the way both spheres operated changed fundamentally in the early modern period as a product of the Industrial Enlightenment. 
	First, knowledge became tighter and more useful to practical applications, not least because of the adoption of an experimental and empirical paradigm \citep[p.38 f.]{Mokyr} as a legacy of the Scientific Revolution \citep[for further evidence on the Scientific Revolution and Enlightenment in economic history, see][]{Squicciarini2015, Koschnick2025, delacroix2025flora}. Second, propositional and technical knowledge started to spread beyond their own domains. Advances in propositional knowledge were incorporated in practical technologies, while the application of technologies created useful prescriptive facts that re-informed propositional knowledge. This reciprocal relationship ultimately created a feedback loop mechanism where one type of knowledge increased the productivity of the other \citep[see][p. 21]{Mokyr2002}.

	 Mokyr argues that before the eighteenth century, feedback loops were either neutral or negative \citep[p. 33]{Mokyr2002}, constrained by their epistemic base and economic and social factors (ibid.). Yet, with the Industrial Enlightenment \citep[pp. 28--77]{Mokyr2002}, the spheres of propositional and prescriptive knowledge were increasingly integrated and, for the first time in human history, created a positive feedback loop.

			\begin{figure}[htp]
				\centering
				\begin{tikzpicture}[
					line width=0.5pt, >=Stealth,
					xscale=1.0, yscale=0.65
					]
					
					\node[draw,circle,minimum size=2.7cm] (Omega) at (-2.4,0) {\Large$\Omega$};
					\node[draw,circle,minimum size=2.7cm] (Lambda) at (2.4,0) {\Large$\lambda$};
					
					\def\R{1.9}
					
					\draw[->] (140:\R) arc[start angle=140, end angle=40, radius=\R]
					node[midway, above=3pt, font=\footnotesize] {$\beta^{\Omega \to \lambda} = \dfrac{\partial \ln \text{Innovation}^{\lambda}}{\partial \ln \text{Spillover}^{\Omega \to \lambda}}$};
					\draw[->] (-40:\R) arc[start angle=-40, end angle=-140, radius=\R]
					node[midway, below=3pt, font=\footnotesize] {$\beta^{\lambda \to \Omega} = \dfrac{\partial \ln \text{Innovation}^{\Omega}}{\partial \ln \text{Spillover}^{\lambda \to \Omega}}$};
					
				\end{tikzpicture}
				\caption{A feedback loop between propositional ($\Omega$) and prescriptive knowledge ($\lambda$) \vspace{4pt} \newline \footnotesize \emph{Notes:} The figure illustrates a feedback loop between propositional ($\Omega$) and prescriptive knowledge ($\lambda$) as two cross-elasticities: The response of a title's innovativeness to the knowledge spillovers it received from the other type of knowledge. The loop is positive when both $\beta^{\Omega \to \lambda}$ and $\beta^{\lambda \to \Omega}$ are positive. See equation~\ref{eq:cross_elasticities} and \cite[p. 21]{Mokyr2002}.}
				\label{fig:illustration_feedback_loop}
			\end{figure}

	The paper illustrates the functioning of this feedback loop in figure~\ref{fig:illustration_feedback_loop}. A feedback loop consists of two elements, first spillovers and secondly increased innovation. As illustrated, spillovers between both propositional ($\Omega$) and prescriptive knowledge ($\lambda$) increase innovation in the receiving field. New knowledge in the receiving field in turn creates new spillovers back. Mokyr's definition of a positive feedback loop, ``growth in one increases the marginal product of the other'' \citep[p. 21]{Mokyr2002}, can be written as two cross-elasticities for each direction of the loop. For each period $p$, we get
	\begin{equation}
		\label{eq:cross_elasticities}
		\beta^{\Omega \to \lambda}_{p} = \frac{\partial \ln \text{Innovation}^{\lambda}}{\partial \ln \text{Spillover}^{\Omega \to \lambda}}
		\qquad \text{and} \qquad
		\beta^{\lambda \to \Omega}_{p} = \frac{\partial \ln \text{Innovation}^{\Omega}}{\partial \ln \text{Spillover}^{\lambda \to \Omega}},
	\end{equation}
	where $\beta^{\Omega \to \lambda}_{p}$ and $\beta^{\lambda \to \Omega}_{p}$ are the elasticities of innovation in one type of knowledge with respect to spillovers from the other. We can estimate them from printed titles belonging either to propositional or prescriptive knowledge. Then, for a given title, a positive $\beta$ means that, within a field and year, titles that drew more on the other type of knowledge were more innovative than the other titles in the field.\footnote{Note that by construction the coefficients are cross-effects, not growth rates. They are agnostic about how much knowledge was produced, but capture whether the knowledge that crossed the boundary between $\Omega$ and $\lambda$ was productive.}
	
	Mokyr's account yields three predictions for these coefficients. 
	\begin{enumerate}
		\item A negative or neutral relationship between knowledge spillovers and innovation before the eighteenth century;
		\item A gradual ``tipping of the balance'' toward a positive relationship during the early eighteenth century;
		\item A positive relationship between knowledge spillovers and innovation, in both directions, by the middle of the eighteenth century.
	\end{enumerate}
	
	The timing prediction is based on Mokyr's mechanism of the ``Industrial Enlightenment'' that connected both spheres of knowledge and started the feedback loop mechanism. When did the ``Industrial Enlightenment'' begin? Being a gradual process, it is impossible to pin an exact year on it, yet we identify a set of important demarcation marks that all point to the middle of the eighteenth century. Lecture series on Newtonian science had already started by the early 1700s \citep{Stewart1992}. The publication of Voltaire's \textit{Letters Concerning the English Nation} in 1733 is often used to demarcate the beginning of the Enlightenment \citep{Wootton2015}. The 1750s and 1760s seem to have been a point of increased acceleration, marked by the publication of Diderot and d'Alembert's \textit{Encyclop\'edie} between 1751 and 1772 \citep{Squicciarini2015}, the foundation of the Society for the Encouragement of Arts, Manufactures and Commerce in 1754 \citep{Howes2020}, the Lunar Society of Birmingham in 1765 \citep{Schofield1963}, and a series of economic societies across the continent \citep{Stapelbroek2012,cinnirella2025flow}.
	
	\FloatBarrier
	\subsection{How did the feedback loop function in practice?}
	
	Importantly, spillovers in the feedback loop process could operate through a multitude of channels. The paper identifies a set of key spillovers, $A_1$--$A_4$ and $B_1$--$B_3$, that are either explicitly discussed in \cite{Mokyr2002} or prominently feature in the literature on knowledge, innovation, and technology \citep{Landes1969,Landes1998,stokes1997pasteur,Reyonolds1983,Stewart1986,Stewart2007,Jacob1997,Jacob2014,Mokyr2002}.

	\begin{figure}[htp]
		\centering
		\begin{tikzpicture}[line width=0.5pt, >=Stealth,
			xscale=0.8, yscale=0.52]  

			\def\R{3.3}

			\node[draw,circle,minimum size=3.2cm] (Omega) at (-4,0)
			{\Large$\Omega$};
			\node[draw,circle,minimum size=3.2cm] (S) at (4,0)
			{\Large$\lambda$};

			\node[inner sep=1pt] (A) at (0, 3.2)
			{\scriptsize$\begin{bmatrix}
					A_{1}\\[2pt]
					A_{2}\\[2pt]
					A_{3}\\[2pt]
					A_{4}
				\end{bmatrix}$};

			\node[inner sep=1pt] (B) at (0,-3.2)
			{\scriptsize$\begin{bmatrix}
					B_{1}\\[2pt]
					B_{2}\\[2pt]
					B_{3}
				\end{bmatrix}$};

			\draw[->] (150:\R)  arc[start angle=150,end angle=110,radius=\R];
			\draw[->] (70:\R)   arc[start angle=70, end angle=30, radius=\R];
			\draw[->] (-30:\R)  arc[start angle=-30,end angle=-70,radius=\R];
			\draw[->] (-110:\R) arc[start angle=-110,end angle=-150,radius=\R];

		\end{tikzpicture}
		\caption{A feedback loop between propositional ($\Omega$) and prescriptive knowledge ($\lambda$) \vspace{4pt} \newline \footnotesize \emph{Notes:} The figure illustrates a feedback loop between propositional  ($\Omega$) and prescriptive knowledge ($\lambda$). Innovation-enhancing spillovers from $\Omega$ to $\lambda$ operate through channels $A_1$--$A_4$. Vice versa, innovation-enhancing spillovers from $\lambda$ to $\Omega$ operate through channels $B_1$--$B_3$. See also \cite[p. 17, 22]{Mokyr2002}.}
		\label{fig:feedback_loop_channels}
	\end{figure}

	First, $A_1$--$A_4$ indicate channels for spillovers from propositional to prescriptive knowledge ($\Omega \rightarrow \lambda$). Most directly, a theoretical understanding of the laws of nature could lead to exact predictions of optimal engineering techniques ($A_1$). For example, early eighteenth century engineers could use Hooke's law to design springs, or model how bridges or tunnels would deform under load. Moreover, even without exact predictions, theories and observational knowledge could improve the search function over the technological space ($A_2$) \citep{kauffman2000optimal,acemoglu2016innovation}. Improvers could also profit from the adoption of the scientific toolset, such as mathematics or precise measurement ($A_3$) \citep{kellywatches2016}. Lastly, knowledge about limits or inconsistencies in existing knowledge could also help to set the agenda for practical inquiry ($A_4$). Conflicts between theories often served as a starting point for successful experimentation \citep{Reyonolds1983}.
	
	Notably, the presence of false predictions from faulty or underdeveloped scientific theories can also lead to negative spillovers into the technological space. Examples of this in the seventeenth century are manifold, such as the use of linear scaling laws before Galileo's formulation of the square-cube law in 1638. The use of linear scaling in the extension of bridge constructions or ship design often led to structural failure \citep{valleriani2009transformation}. Moreover, the Aristotelian belief that nature abhorred a vacuum led to the prediction that water could be raised indefinitely by suction pumps. In fact, it failed above 10 meters, surprising contemporaries and early scientists. Lastly, negative spillovers can also be explained by economic and social resistance to new knowledge or technologies (\citealp[pp.~218 f.]{Mokyr2002}; \citealp{caprettini2020rage}).
	
	Second, $B_1$--$B_3$ indicate channels through which spillovers from prescriptive to propositional knowledge ($\lambda \rightarrow \Omega$) operated. First, the practical implementation of theoretical predictions could help to reject false scientific theories ($B_1$). Second, research could be guided by puzzling facts ($B_2$). For example, Pasteur's theoretical research was prompted by remarkable observations and puzzles, such as anomalies in fermentation that finally led to his discovery of the germ theory of disease \citep{stokes1997pasteur}. Lastly, unsolved technological problems in commerce and trade could likewise help to set research agendas ($B_3$) \citep{stokes1997pasteur}. 
	
	We should also note here that false or incomplete empirical observations could lead to the development of incorrect and unsuccessful scientific theories. For example, early telescopes produced small discs around planets as optical artifacts. As these were mistaken for the true size of stars, this was used as an argument against heliocentrism, as star sizes appeared implausibly large  \citep{graney2010telescope,graney2011telescopic}. Likewise, optical inspection of early dissections seemed to suggest that blood in arteries ebbed and flowed, supporting the Galenic theory that blood is produced in the liver and therefrom flows outwards into the body to be consumed by tissues. Only later would venous valves be discovered, clearing the ground for William Harvey's discovery of pulmonary and systemic circulation of blood \citep{SCULTETUS2001435}.

	The literature on science and technology has identified the presence of such feedback loop processes for a variety of mid-eighteenth century industries. First, \cite{Reyonolds1983} has studied how a wide range of mathematical theorists, such as Parent, Bernoulli, and Euler, as well as practical engineers, such as Polhem, de Parcieux, and Smeaton, devoted themselves to the improvement of waterwheels, a major source of energy in the eighteenth century. Here, \cite{Reyonolds1983} documents how the integration of theory and practice changed the nature of innovation. Inventors stopped using stochastic principles based on questions such as ``Will it work if I build it this way? or, If I change this element, will it work any better?'' \cite[p. 232]{Reyonolds1983}. Instead, engineers started to rely on ``(1) systematic methods of experimentation, (2) the use of working models, and (3) the application of quantitative measurements to key variables'' ($A_2, A_3, B_1$) \cite[p. 232]{Reyonolds1983}.\footnote{A key input to quantitative experimentation was also the new design of testing devices \citep{Constant1983}.} Now, engineers like Polhem, de Parcieux, and Smeaton started from specific questions on the efficacy of designs that came from unsettled debates in physics ($A_2, A_4$).\footnote{Here, \cite{Reyonolds1983} argues that lacking a full theory of hydrodynamics, initial theoretical predictions were usually wrong. Yet, they provided useful and more directed starting places for practical engineers that experimentally tested their predictions and started to collect systematic quantitative data. In the end, the agenda was highly successful and led to significant increases in the efficiency of the water wheel. While the efficiency of traditional waterwheels lay between ca. 30--40\% \citep{Viollet2017}, Smeaton's breastshot wheel, as well as overshot and Poncelet undershot wheels (based on theoretical work by Borda in the 1760s) delivered efficiencies of 60--80\% (ibid.).} Thus, it was often the unsolved questions rather than successful predictions that gave engineers a specific starting place for experimentation.  In return, the scientific theory of hydrodynamics improved with each new practical insight ($B_1, B_2$) \citep{Reyonolds1983}.

	The same dynamic also applied to other fields. One can think of Josiah Wedgwood and the influence of his early chemistry studies on his pottery products. Here, early chemistry inspired the development of new materials and production methods even when the exact workings of the science were not yet clear. Similarly, dyeing at the end of the eighteenth century seems to have increasingly relied on chemical theory and controlled experimentation \citep[pp. 338--351]{Musson1969}. \cite{Schofield1963}, \cite{Stewart1992,Stewart2007}, \cite{Stewart1995} and \cite{Jacob2014} have further argued that knowledge flows between science and technological innovators in industries such as steam, cotton, weaving, metallurgy, and pottery steadily increased throughout the eighteenth century.
	
	Yet, despite this qualitative evidence of increasingly positive spillovers between propositional and prescriptive knowledge at the end of the eighteenth century, it remains hard to assess how representative these cases were for overall knowledge production \citep{OGrada2016}. This is where quantitative evidence can significantly contribute to our understanding of the emergence of feedback loops.  Therefore, the next sections will introduce data on the universe of published titles in Britain and present an empirical framework to estimate knowledge spillovers between prescriptive and propositional knowledge.

	\section{\label{sec:data}Data}

	\subsection{\label{sec:assigning_subject_classes}Publication titles 1600--1800}

	To capture the content of the British stock of knowledge, the paper uses the universe of all 285,985 unique printed titles from the English Short Title Catalogue (ESTC) between 1600 and 1800 from \cite{Koschnick2025}. The ESTC contains all works printed in England as well as all works in English. With an average length of ca. 55 words, these titles had the form of short abstracts, usually covering the full front page of a book.\footnote{The format was unique to England and was not adopted on the continent, e.g. in Germany or France. The format was furthermore consistent over time, with static word count levels between 1600 and 1800, see appendix figure~\ref{fig:estc_words_over_time}.} Hence, they contain information on both the broad topic covered by a title as well as a short abbreviated summary of its content. 
	
	To classify the ESTC, the paper uses subject classes from \cite{Koschnick2025}. \cite{Koschnick2025} assigned higher-order categories to librarian-assigned subject labels and then used a BERT model trained on historical titles to assign subject fields to unclassified titles. The dataset was further cleaned to remove duplicates. Appendix table~\ref{tab:summary_training_corpus_by_group} provides descriptive statistics of the ESTC catalogue. A typical example of an ESTC title reads:
	\begin{quote}
		``A compendium of algebra. Containing the principles of the analytick art, with rules for solving simple, quadratick, and cubick, \&c. equations; together with the method of converging series's; after so plain a method, that any one who understands numbers, may learn the solution of the said equations without a master. By George Gordon, assistant to the Reverend Dr. Desaguliers.'' (George Gordon, 1728)
	\end{quote}
	
	Additionally, the paper provides a validation exercise that tests whether ESTC titles provide meaningful summaries of the content of books. For this, we draw on full-text data from Early English Books Online (EEBO) that is available for the seventeenth century and match them to ESTC titles. We document an EEBO coverage of 30.34\%. Based on this subset, we establish that book titles are a strong predictor of actual content using a tf--idf approach (term frequency–inverse document frequency). Appendix~\ref{sec:title_full_text_sim} documents the validation exercise.
	
	Notably, the ESTC also covers a large number of foreign works that were reprinted in England, thereby also accounting for knowledge inflows into the British knowledge economy. This is important, since it has been argued that the continent possessed an advantage in codifying knowledge \citep{mokyr2021holy}. Figure~\ref{fig:languages} reports the number of all titles printed in foreign languages. For the NLP analysis, titles in other languages were translated using Google Translate \citep[see][]{Koschnick2025}. 
	
	Finally, the paper extracts information on authors' occupations from the ESTC. Seventeenth and eighteenth century authors usually added prestigious occupations and affiliations with societies to their names on the title page. The paper extracts information on the occupational categories of \textit{engineer, medical career}, and \textit{academic career}, as well as memberships in the \textit{the Royal Society} or other \textit{enlightenment societies}. Appendix section~\ref{sec:variable_descriptions} describes this procedure in further detail.

	\subsection{\label{sec:data_patents}Patent data}

	Beyond technical publications, a significant amount of societal technical knowledge is also stored in patents. To provide short summaries of patents that parallel the titles from the ESTC, the paper introduces a novel data source, the \textit{Chronological Index of Patents Applied for and Patents Granted} compiled by Bennett Woodcroft in \citeyear{Woodcroft1854a}. It covers the full time period of 1617--1852.\footnote{The description of patents is similar to Woodcroft's \textit{Subject-matter index of patents of invention} (\citeyear{Woodcroft1854a}) used by \cite{Billington2019} and \cite{Billington2020}, but has the advantage of containing the original full summaries, while the \textit{Subject-matter index} only includes extracts and summaries from the original source. Compare e.g. patent number $1812$, listed as ``engine for lessening the consumption of steam and fuel in steam or fire engines, and gaining a considerable effect in time \& force'' in the \textit{Chronological Index} \citep{Woodcroft1854a} and listed as ``Engine for saving fuel in steam-engines'' in the \textit{Subject-matter index} \citep{Woodcroft1854}.} The paper follows the literature \citep{Billington2020,Billington2021} by excluding patents before 1700 as some of these are closer to monopoly grants than modern ``patents for invention''.  Patents have an average word count of ca. 25 words (see appendix figure~\ref{fig:patent_words_over_time}). To extract the technical content from the \textit{Chronological Index}, the paper uses an LLM-based approach to exclude legal or procedural text passages.

	A typical example of a patent short description title reads:
	
	\begin{quote}
		``Machine which performs its opperacions either by fire or fall of water, or both together, \& the friction is thereby reduced SO as to have no solid bodys to rub but the injecting vapour \& water cocks or sluices'' (William Blakey, 1766)
	\end{quote}
	
	Note that this paper necessarily has to omit valuable inventions that were kept secret instead of being patented \citep{moser2005patent}.

	\subsection{Defining propositional and prescriptive knowledge}
	\label{sec:assigning_subject_classes_prop_prescr}
	\textbf{Conceptual classification}
	\cite{Mokyr2002} defines \textit{propositional knowledge} ($\Omega$) as the set of knowledge that describes \textit{how nature functions}. This includes laws of nature as formulated in science but also a collection of all empirical data about nature.\footnote{Note that according to \cite{Mokyr2002} propositional knowledge excludes knowledge that is about interactions between humans, e.g. the arts, literature, or the social sciences.} Next, \textit{prescriptive knowledge} ($\lambda$) is the set of knowledge that describes \textit{how to change nature} \citep{Mokyr2002}. It therefore encompasses all knowledge about technologies, whether in industry, trades, or agriculture. 
	
	The distinction is similar to science vs. technology. Indeed, propositional knowledge contains all knowledge from science.\footnote{We should note that \textit{scientific practice} relies on a significant amount of prescriptive knowledge, e.g. how to operate technical instruments or how to set up experiments.} Yet, it also includes other activities aimed at collecting, describing, and classifying natural phenomena \citep[see][pp. 4--6]{Mokyr2002}. Propositional knowledge is therefore wider than science. Likewise, prescriptive knowledge contains the widest array of possibilities of how to interact with or change nature, e.g. building machines or instruments, useful ways of navigating a ship, as well as new ways of breeding cattle.

	\begin{figure}[h]
		\centering
		\begin{tikzpicture}
			
			\node[
			draw,
			circle,
			minimum width=5cm,
			align=center,
			font=\small
			] (Omega) at (0,0) {
				\Large$\Omega$\\[5pt]
				Applied physics\\
				Astronomy\\
				Mathematics\\
				Chemistry\\
				Encyclopedias
			};
			\node[
			draw,
			circle,
			minimum width=5cm,
			align=center,
			font=\small
			] (Lambda) at (6,0) {
				\Large$\lambda$\\[5pt]
				Technology in trades\\
				Technology in agriculture \\
				Navigation\\
				Scientific instruments\\
				Patents
			};
		\end{tikzpicture}
		\caption{Assigning subject classes to propositional ($\Omega$) and prescriptive knowledge ($\lambda$) \newline \footnotesize \emph{Notes:} The figure reports the ESTC subject classes that form the set of propositional and prescriptive knowledge. Patents are added \cite{Nuvolari2011} and only available post 1700.}
		\label{fig:prop_and_prescr_fields}
	\end{figure}

	For the purpose of this paper, we aim to identify the set of propositional and prescriptive knowledge that was relevant for the Industrial Enlightenment and Industrial Revolution. As we use these categories in the downstream analysis of estimating feedback loops between related fields, we add two additional inclusion criteria: A) We require that spillovers should have been relevant to all receiving fields, and B) that the fields themselves should have been clearly distinct. For example, we plausibly expect e.g. discoveries in \textit{applied physics} $\in \Omega$ to have been relevant to \textit{technology in trades} $\in \lambda$ while also being distinct fields at the same time. On the other hand, we would not expect knowledge in \textit{medicine}  $\in \Omega$ to have been relevant to \textit{technology in trades}. Hence, one implication of this criterion is that, for this paper, we focus on the hard sciences and exclude the life sciences.\footnote{This decision is purely practical and follows a narrow interpretation of the Industrial Revolution with mechanical innovation at its heart. A study of the existence of feedback loops in the life sciences would be a promising route for future research.} 
	
	Figure~\ref{fig:prop_and_prescr_fields} shows the selected subject classes of propositional and prescriptive knowledge. They include the core fields of hard science and technical subjects. Given that encyclopedias were an important means for accessing society's stock of knowledge \citep{Squicciarini2015}, the paper further includes encyclopedic works within the group of propositional knowledge.\footnote{Encyclopedias were furthermore an important source for what \cite[p. 5]{Mokyr2002} terms technological science that includes e.g. knowledge about the properties of materials or knowledge about simple chemical reactions.} Additionally, we add patents to $\lambda$ as another important source for technological instructions. Here we caution that eighteenth century patents were public in principle but access to patent specifications was costly and uneven in practice \citep{MacLeod1989}.\footnote{First, before 1734 patent specifications were not requested for the registration of a patent, although the practice became more common during the early eighteenth century \citep[p. 49]{MacLeod1989}. After 1734, it became standard practice to require written patent specifications for registration (ibid), after which specifications were randomly enrolled at any of three separate offices \citep{MacLeodChristine1991}. Access was only granted if one knew the name of the patentee, but not provided for general technology classes (ibid). Important specifications were only released in print as late as 1794 through \textit{The Repertory of Arts and Manufactures} (ibid). Also, note that the British patent system operated purely through registration. Therefore, neither the functioning of the invention nor the completeness of patent specifications for replicability was extensively checked. Altogether, access to patents was significantly more costly and uneven than access to books that could circulate in the country.} Therefore, in the following baseline specifications, we exclude patents from the construction of the spillover index. Here we assume that the access cost to patents was much higher than to printed works from the ESTC. Appendix section~\ref{sec:robustness_patent_discont} reports robustness to either including patents in the spillover measure or excluding patents altogether.\footnote{Also note that since patents are only available after 1700, using patents as part of the spillover measure would introduce a discontinuity.}  Appendix table~\ref{tab:summary_training_corpus_by_group} presents summary statistics for the subject classes of $\Omega$ and $\lambda$ as shown in figure~\ref{fig:prop_and_prescr_fields}. Overall, the group of $\Omega$ includes 6,266 titles and the group of $\lambda$ includes 4,784 titles between 1600 and 1800. Lastly, appendix section~\ref{sec:timetrends} reports time trends.

\FloatBarrier
	
	\section{\label{sec:spillovers}Measuring knowledge spillovers and innovation}
	\subsection{Conceptual framework}
	
	\textbf{Innovation}

	How can we measure innovation and knowledge spillovers in historical data? Generally, citations are not available for the seventeenth and eighteenth century. Therefore, we have to rely on the information contained in the text data itself that can be quantified using recent advances in natural language processing.
	
	The innovation and knowledge spillover measures introduced in this paper exploit the temporal dimension of innovation. They make use of the fundamental insight from \cite{Kelly2021a} that states:

	A title is innovative if:
	\begin{enumerate}
		\item It is similar to publications in the future in its field
		\item It is dissimilar to publications in the past in its field
	\end{enumerate}
	
	\begin{figure}[!ht]
		\centering	
		\begin{tikzpicture}[scale=0.72, transform shape]
			\node[title] (T) at (0,0) {Title $v_{it_0}$};
			\node[doc] (p1) at (-4.6, 1.3) {};  \node[lab, left=4pt of p1] {$v_1$};
			\node[doc] (p2) at (-4.6, 0.0) {};  \node[lab, left=4pt of p2] {$v_2$};
			\node[doc] (p3) at (-4.6,-1.3) {};  \node[lab, left=4pt of p3] {$v_3$};
			\node[doc] (f1) at ( 4.6, 1.3) {};  \node[lab, right=4pt of f1] {$v_4$};
			\node[doc] (f2) at ( 4.6, 0.0) {};  \node[lab, right=4pt of f2] {$v_5$};
			\node[doc] (f3) at ( 4.6,-1.3) {};  \node[lab, right=4pt of f3] {$v_6$};
			\draw[sim] (T.west) -- (p1.east) node[alab, pos=0.5, above=2pt] {$\delta_1$};
			\draw[sim] (T.west) -- (p2.east) node[alab, pos=0.5, above=2pt] {$\delta_2$};
			\draw[sim] (T.west) -- (p3.east) node[alab, pos=0.5, above=2pt] {$\delta_3$};
			\draw[sim] (T.east) -- (f1.west) node[alab, pos=0.5, above=2pt] {$\gamma_1$};
			\draw[sim] (T.east) -- (f2.west) node[alab, pos=0.5, above=2pt] {$\gamma_2$};
			\draw[sim] (T.east) -- (f3.west) node[alab, pos=0.5, above=2pt] {$\gamma_3$};
			\node[win] at (-4.6,-2.25) {$t < t_0$};
			\node[win] at ( 4.6,-2.25) {$[t_0+1,\; t_0+\tau]$};
			\legendline{leg}{0,-2.35}{Innov: if $\bar{\gamma} > \bar{\delta}$}
		\end{tikzpicture}
		\caption{Calculating the innovation index \vspace{4pt} \newline \footnotesize \emph{Notes:}   A title, $v_{it}$, is innovative if its average similarity to titles in the future, $\bar{\gamma}$, is larger than its average similarity to titles in the past, $\bar{\delta}$.  }\label{fig:innovation}
	\end{figure}

	Altogether, a publication is innovative if it is both \textit{impactful} (1) and \textit{novel} (2). We can formalize this by dividing a title's forward similarity (1) by its backward similarity (2). Concretely, for a set of titles $S$, and title $v_{it_0} \in S$, we calculate:
	\begin{equation}
		\text{Forward similarity}_{it_0}  \;=\; f^{k}\!\bigl(v_{it_0},\, \mathcal{F}_{t_0,\tau}\bigr)
	\end{equation}
	\begin{equation}
		\text{Backward similarity}_{it_0}  \;=\; f^{k}\!\bigl(v_{it_0},\, \mathcal{B}_{t_0,\tau}\bigr)
	\end{equation}
	\begin{equation}
		\label{eq:innovation}
		\mathrm{Innovation}_{it_0}  \;=\; 
		\frac{f^{k}\!\bigl(v_{it_0},\, \mathcal{F}_{t_0,\tau}\bigr)}{f^{k}\!\bigl(v_{it_0},\, \mathcal{B}_{t_0,\tau}\bigr)},
	\end{equation}
	where $f^{k}(.)$ is the similarity function defined in Section~\ref{sec:defining-similarity}: The average cosine similarity, in the embedding space of the language model, between the title and the $k$ most similar titles of a pool. $\mathcal{F}$ is the pool of future titles of the field, $\mathcal{F}^{}_{t_0,\tau} \;:=\; \bigl\{\,v_{j t}\in S \;\bigm|\; t_0+1 \le t \le t_0+\tau \bigr\}$, and $\mathcal{B}$ is the pool of past titles of the field, $\mathcal{B}^{}_{t_0} \;:=\; \bigl\{\,v_{j t}\in S \;\bigm|\; t < t_0 \bigr\}$. In the baseline, the forward window is $\tau = 20$ years and the backward pool includes all earlier titles of the field, so that novelty is measured against the field's accumulated stock and impact against the next twenty years.\footnote{Appendix~\ref{sec:different_time_windows} reports versions with forward windows of 10 and 30 years. The backward window of the received spillover index in equation~\ref{eq:received_spillover} is $\tau = 20$ years on both sides. Because the innovation index requires a forward window, it is defined for titles published up to $1800-\tau$.} Figure~\ref{fig:innovation} illustrates the logic. The index is greater than one if the average of the $\gamma$ similarities to the future exceeds the average of the $\delta$ similarities to the past. Intuitively, a title is innovative if its field's future looks more like it than its field's past did. Novelty is a title's inverse backward similarity and impact is its forward similarity. Multiplying the two yields equation~\ref{eq:innovation}.\footnote{See \cite{Koschnick2025} for a similar BERT-based innovation index. However, the \cite{Koschnick2025} model differs in the use of the similarity function. It further uses a BERT model trained on contemporary data to derive similarities.}

	\noindent \textbf{Creating knowledge spillovers}
	
	\begin{figure}[!ht]
		\centering	
		\begin{tikzpicture}[scale=0.72, transform shape]
			\node[title] (T) at (0,3.3) {Title $v_{it_0}$};
			\draw[fieldline] (-6.2,2.0) -- (6.2,2.0);
			\node[lab, anchor=south west] at (-6.2,2.08) {Field $A$};
			\node[lab, anchor=north west] at (-6.2,1.92) {Field $B$};
			\node[doc] (p1) at (-4.2, 0.9) {};  \node[lab, left=4pt of p1] {$v_1$};
			\node[doc] (p2) at (-4.2,-0.4) {};  \node[lab, left=4pt of p2] {$v_2$};
			\node[doc] (p3) at (-4.2,-1.7) {};  \node[lab, left=4pt of p3] {$v_3$};
			\node[doc] (f1) at ( 4.2, 0.9) {};  \node[lab, right=4pt of f1] {$v_4$};
			\node[doc] (f2) at ( 4.2,-0.4) {};  \node[lab, right=4pt of f2] {$v_5$};
			\node[doc] (f3) at ( 4.2,-1.7) {};  \node[lab, right=4pt of f3] {$v_6$};
			\draw[sim] (T.south west) -- (p1.north east) node[alab, pos=0.18, left=1pt] {$\delta_1$};
			\draw[sim] (T.south west) -- (p2.north east) node[alab, pos=0.42, left=1pt] {$\delta_2$};
			\draw[sim] (T.south west) -- (p3.north east) node[alab, pos=0.52, left=1pt] {$\delta_3$};
			\draw[sim] (T.south east) -- (f1.north west) node[alab, pos=0.18, right=1pt] {$\gamma_1$};
			\draw[sim] (T.south east) -- (f2.north west) node[alab, pos=0.42, right=1pt] {$\gamma_2$};
			\draw[sim] (T.south east) -- (f3.north west) node[alab, pos=0.52, right=1pt] {$\gamma_3$};
			\node[win] at (-4.2,-2.65) {$[t_0-\tau,\; t_0-1]$};
			\node[win] at ( 4.2,-2.65) {$[t_0+1,\; t_0+\tau]$};
			\legendline{leg}{0,-2.75}{Spillover: if $\bar{\gamma} > \bar{\delta}$}
		\end{tikzpicture}

		\caption{Calculating spillovers between fields \newline \footnotesize \emph{Notes:}  A title, $v_{it}$, in field $A$ creates a spillover into field $B$ if its average similarity to titles in the future of $B$, $\bar{\gamma}$, is larger than its average similarity to titles in the past of $B$, $\bar{\delta}$.  }\label{fig:spillovers}
	\end{figure}

	We can easily extend the innovation concept to knowledge spillovers: Let $A,B \subset \mathcal{S}$ be two subject classes with $A\neq B$. Now, a title in field $A$ creates a knowledge spillover into field $B$ if:
	\begin{enumerate}
		\item It is similar to publications in the future in field $B$
		\item It is dissimilar to publications in the past in field $B$
	\end{enumerate}
	
	Figure~\ref{fig:spillovers} illustrates this logic. In the same way that title $v_{it_0}$ shifted its own field in figure~\ref{fig:innovation}, title $v_{it_0} \in A$ now shifts field $B$. Thus, title $v_{it_0} \in A$ creates a spillover if it is closer to $v_4, v_5, v_6$ in the future of field $B$ than to $v_1, v_2, v_3$ in the past of field $B$. We can formalize this as:
	\begin{equation}
		\label{eq:spillover}
		\mathrm{Spillover}_{i , A \!\to\! B}
		\;=\;
		\frac{
			f^{k}\!\bigl(v_{it_0},\;\mathcal{F}^{B}_{t_0,\tau}\bigr)
		}{
			f^{k}\!\bigl(v_{it_0},\;\mathcal{B}^{B}_{t_0,\tau}\bigr)
		},
	\end{equation}
	where $\mathcal{F}^{B}_{t,\tau}$ is the forward pool in field $B$, 
	$\mathcal{F}^{B}_{t_0,\tau} \;:=\; \bigl\{\,v_{j t}\in B \;\bigm|\; t_0+1 \le t \le t_0+\tau \bigr\}$, and $\mathcal{B}^{B}_{t,\tau}$ is the backward pool in field $B$, $\mathcal{B}^{B}_{t_0,\tau} \;:=\; \bigl\{\,v_{j t}\in B \;\bigm|\; t_0 - \tau \le t \le t_0-1 \bigr\}$.
	Figure~\ref{fig:spillovers} illustrates this logic.

	\noindent \textbf{Receiving knowledge spillovers}
	
	Next, we move from titles \textit{creating} spillovers to titles \textit{receiving} spillovers.	
	A title in field $B$ \emph{received} a spillover from field $A$ if it resembles the recent output of $A$ more closely than its own predecessors did. This means that we divide the similarity between title $v_{it}$ in field  $B$ and the past of field $A$ by the similarity between the past of field $B$ and the past of field $A$.  To ensure title $i$ and the past of field $B$ are comparable groups, we perform the comparison between $B$ and $A$ only for the $k$ most similar titles in $B$ to $i$. This yields:

	\begin{equation}
		\text{Received spillover}_{A \to B}(v_{it})
		:= \frac{f^{k}\bigl(v_{it},\,\mathcal{B}^{A}_{t,\tau}\bigr)}{f^{k}\bigl(\mathcal{P}^{k}_{it},\,\mathcal{B}^{A}_{t,\tau}\bigr)},
		\label{eq:received_spillover}
	\end{equation}
	where $\mathcal{B}^{A}_{t,\tau}$ is the backward pool of $A$, $\mathcal{P}^{k}_{it}$ are the peers of $v_{it}$ in the backward pool of $B$, and $f^{k}$ is the similarity function of Section~\ref{sec:defining-similarity}, the mean of the $k$ largest cosine similarities between the titles on its left and the pool on its right. The index exceeds one if the title came closer to the past of $A$ than its peers had been. Figure~\ref{fig:received_spillover} illustrates this: Title $v_4$ received a spillover if its links $\gamma$ to the recent past of $A$ are tighter than the links $\delta$ of its peers $v_1, v_2, v_3$. In the baseline we define $k = 20$ and $\tau = 20$ years.

	\begin{figure}[!ht]
		\centering	
		\begin{tikzpicture}[scale=0.72, transform shape]
			\node[doc] (a1) at (-1.2, 3.0) {};  \node[lab] (la1) at (-1.2, 3.75) {$v_5$};
			\node[doc] (a2) at ( 0.0, 3.0) {};  \node[lab] (la2) at ( 0.0, 3.75) {$v_6$};
			\node[doc] (a3) at ( 1.2, 3.0) {};  \node[lab] (la3) at ( 1.2, 3.75) {$v_7$};
			\begin{scope}[on background layer]
				\node[pool, fit=(a1) (a2) (a3) (la1) (la2) (la3), inner xsep=10pt, inner ysep=6pt] (A) {};
			\end{scope}
			\node[lab, left=8pt of A] {$\mathcal{B}^{A}_{t,\tau}$};
			\node[win, right=8pt of A] {$[t-\tau,\; t-1]$};
			\draw[fieldline] (-6.2,2.0) -- (6.2,2.0);
			\node[lab, anchor=south west] at (-6.2,2.08) {Field $A$};
			\node[lab, anchor=north west] at (-6.2,1.92) {Field $B$};
			\node[doc] (p1) at (-4.2, 0.4) {};  \node[lab] (lp1) at (-4.95, 0.4) {$v_1$};
			\node[doc] (p2) at (-4.2,-0.9) {};  \node[lab] (lp2) at (-4.95,-0.9) {$v_2$};
			\node[doc] (p3) at (-4.2,-2.2) {};  \node[lab] (lp3) at (-4.95,-2.2) {$v_3$};
			\begin{scope}[on background layer]
				\node[pool, fit=(p1) (p2) (p3) (lp1) (lp2) (lp3), inner xsep=8pt, inner ysep=8pt] (P) {};
			\end{scope}
			\node[lab, left=8pt of P] {$\mathcal{P}^{k}_{it}$};
			\node[win, below=3pt of P] {$[t-\tau,\; t-1]$};
			\node[doc, minimum size=1.1cm] (v4) at (4.2,-0.9) {};  \node[lab, right=4pt of v4] {$v_4$};
			\node[win, below=4pt of v4] {$[t]$};
			\draw[sim] ($(A.south west)+(0.3,0)$) -- (p1.north east) node[alab, pos=0.45, left=1pt] {$\delta_1$};
			\draw[sim] ($(A.south west)+(1.0,0)$) -- (p2.north east) node[alab, pos=0.45, left=1pt] {$\delta_2$};
			\draw[sim] ($(A.south west)+(1.7,0)$) -- (p3.north east) node[alab, pos=0.45, left=1pt] {$\delta_3$};
			\draw[sim] ($(A.south east)+(-1.7,0)$) -- (v4.west) node[alab, pos=0.30, right=1pt] {$\gamma_1$};
			\draw[sim] ($(A.south east)+(-1.0,0)$) -- (v4.north) node[alab, pos=0.30, right=1pt] {$\gamma_2$};
			\draw[sim] ($(A.south east)+(-0.3,0)$) -- (v4.north east) node[alab, pos=0.30, right=1pt] {$\gamma_3$};
			\draw[sel] (v4.south west) -- (P.east |- 0,-2.35) node[alab, pos=0.5, below=2pt, font=\normalsize\bfseries\boldmath, text=black] {$k$ most similar to $v_4$};
			\legendline{leg}{1.0,-3.65}{Received spillover: if $\bar{\gamma} > \bar{\delta}$}
		\end{tikzpicture}
		\caption{Calculating receiving knowledge spillovers (2) \newline \footnotesize \emph{Notes:}  The figure illustrates the calculation of the \textit{received spillover} measure from equation~\ref{eq:received_spillover}.  Title $v_4$ in field $B$ received a spillover from field $A$ if its average similarity $\bar{\gamma}$ to the recent past of $A$, $\mathcal{B}^{A}_{t,\tau}$, exceeds the average similarity $\bar{\delta}$ between that past and its peers $\mathcal{P}^{k}_{it}$, the $k$ titles in the recent past of $B$ most similar to $v_4$ ($v_1, v_2, v_3$).}\label{fig:received_spillover}
	\end{figure}

	\subsection{Defining the similarity function}
	\label{sec:defining-similarity}
	For calculating similarity, we rely on the semantic content of a historically trained BERT model, introduced in section~\ref{sec:hist-bert-model}. This model yields semantic text-vector representations in a 768-dimensional embedding space. Let $v_{it}$ be a single title embedding. Let $sim(x,y)=\frac{x\cdot y}{\lVert x\rVert\,\lVert y\rVert}$ be the cosine similarity between two titles in the embedding space $x,y \in \mathbb{R}^d$.\footnote{Embeddings are normalized so that cosine similarities are equivalent to the dot product.} Lastly, let $Y=\{y_1, \dots, y_m\} \in \mathbb{R}^d$ be a set of embeddings. Then, we define similarity as the average cosine similarity between title $i$ and the top-$k$ most similar titles:

	\begin{equation}
		f(v_{it},\,Y;\,k\bigr)
		\;=\;
		\frac{1}{k}
		\sum_{\,y\in\operatorname{Top}_{k}
			\{\!sim(v_{it},y') : y'\in Y\}}
		\!sim(v_{it},y),
		\qquad
		k:=\min\{k,|Y|\}
		\label{eq:similarity-ktop}	
	\end{equation}
	
	Calculating similarity only to the most similar $k$ titles in the comparison group directly addresses two key biases in the calculation of spillovers:

	First, the spillover and innovation measure can be sensitive to the inclusion of irrelevant titles. Adding irrelevant titles (e.g. imagine the emergence of a new sub-field) decreases average forward similarity. Yet, by only calculating similarity to the $k$-most similar titles, we ensure that irrelevant titles are not part of the comparison group as long as $k$ is sufficiently small.
	
	Second, the spillover and innovation measure can also be sensitive to titles assigned to the wrong subject class. By definition, titles from another subject class are highly dissimilar to other titles in this class and therefore could bias the spillover and innovation measures. By only calculating similarity to the $k$-most similar titles, we can exclude wrongly assigned titles from the comparison group.

	For this paper, we choose a top-$k$ value of 20. Appendix section~\ref{app:sec:robustness_k_rho} shows robustness to different values of $k$. Appendix section~\ref{sec:app:comparing-sim} further shows that the k-top similarity measure (equation~\ref{eq:similarity-ktop}) outperforms the average similarity measure in predicting patent citations using the innovation measure from equation~\ref{eq:innovation}.  	
	The next section introduces a historical BERT model for the calculation of the embedding space.

	\subsection{\label{sec:hist-bert-model}A BERT similarity model for historical science and patent data}
	\label{sec:steamberth}
	To measure text similarity, the paper draws on context-sensitive BERT models \citep{devlin2019bert} that yield a multidimensional embedding space representation of the semantic content of a title. However, a key concern is that models trained on modern text data may induce two forms of anachronistic bias:
	\begin{enumerate}
		\item Bias from shifts in word-object representations in language. For example, a model might miss that historically a ``fire engine'' would refer to a steam engine or that ``philosophical instruments'' would refer to any scientific instruments.
		\item Bias from a modern representation of concepts within the embedding space. For example, models trained on modern data would position ``phlogiston'' or ``ether'' as semantically distant from other scientific terms, thereby failing to reproduce the internal logic of historical scientific texts, in which ``phlogiston'' or ``ether'' were treated as equally valid scientific concepts.
	\end{enumerate}
	
	To directly address these biases, the paper draws on a \textit{MacBERTh} model that was pre-trained on a historical corpus \citep{MacBERTh_2021}. The main weight of the corpus lies on the early modern period, 1600--1700, with \textit{Early English Books Online} (EEBO) as one of its main sources. Yet, given a general lack of large-scale full-text data for the eighteenth century, the authors decided to extend their corpus until 1950. Nonetheless, the pre-1800 period still captures ca. 70\% of the entire dataset \citep[][p. 25]{MacBERTh_2021}, making \textit{MacBERTh} a good candidate model for embedding early modern text data.
	
	To further align the \textit{MacBERTh} model with seventeenth and eighteenth century publications, this paper fine-tunes the \textit{MacBERTh} model on the ESTC and patent corpus. This step also addresses problems in using base models, such as \textit{MacBERTh}, for semantic comparison that arise from a) CLS layers trained for classification tasks and b) embedding space anisotropy. The result is a new fine-tuned \textit{MacBERTh} model that we name \textit{SteamBERTh}. 
	
	Appendix section~\ref{sec:fine-tuning-evaluation} documents the fine-tuning procedure and presents evaluative statistics. For fine-tuning, the paper employs a \textit{simple contrastive learning of sentence embeddings} approach (SimCSE) \citep{gao2021simcse}. The training data covers the universe of short patent descriptions from \cite{Woodcroft1854a} and the universe of titles within \textit{applied physics, astronomy, chemistry, mathematics} and \textit{technical publications in trades and agriculture, navigation, and scientific instruments} from the ESTC.
	
	Evaluative statistics from appendix section~\ref{sec:fine-tuning-evaluation} show that the fine-tuned \textit{SteamBERTh} model significantly improves on the base model.  However, to assess the quality of the model as an input for historical innovation and spillover measures, we still need to validate the model with real economic data: In the next section, we assess the relationship between the innovation measure and historical patent citations. We then compare the received spillover index to the share of authors publishing in multiple disciplines.

	\section{Validation}	
	\label{sec:validation_section}
	\subsection{Validation on modern data}
	\label{sec:validation_modern}
	
	The paper first validates the innovation and received spillover index against modern data where we can draw on best-quality validation data. For this, the paper draws on the abstracts of $545,505$ journal articles in OpenAlex in \textit{physics and astronomy, chemistry}, and \textit{mathematics} (a subset of propositional knowledge), \textit{engineering, computer science}, and \textit{materials science} (a subset of prescriptive knowledge) as well as the abstracts of the 2.8 million US utility patents of listed firms filed 1985--2024. The patents are linked to \cite{kogan2017technological} who provide forward citations, market values and technology classes. Since the modern corpus does not need a historical embedding model, we apply a general-purpose sentence transformer \texttt{all-mpnet-base-v2} model. For time windows, we select $\tau = 5$ years, since we assume that modern knowledge cycles are faster.\footnote{The innovation index is computed within technology class for patents and within field for articles, mirroring the within-field construction of Section~\ref{sec:empirical-framework}. Appendix~\ref{sec:app:modern} describes the corpus, cleaning and presents descriptive statistics.}

	\begin{figure}[htp]
		\centering
		\begin{subfigure}[t]{.32\textwidth}
			\includegraphics[width=\linewidth]{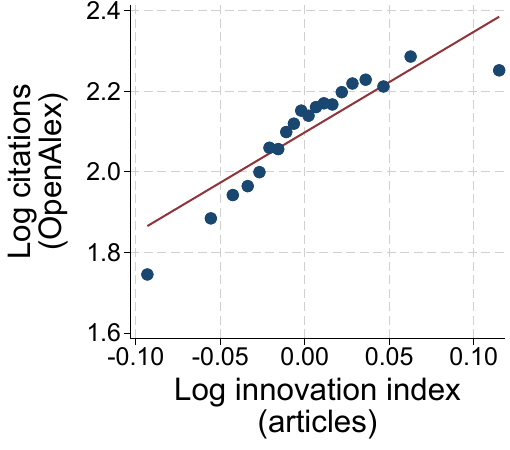}
			\subcaption{Articles: OpenAlex citations}
		\end{subfigure}\hfill
		\begin{subfigure}[t]{.32\textwidth}
			\includegraphics[width=\linewidth]{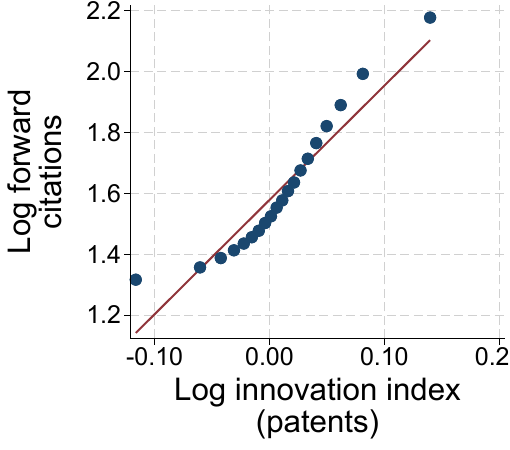}
			\subcaption{Patents: forward citations}
		\end{subfigure}\hfill
		\begin{subfigure}[t]{.32\textwidth}
			\includegraphics[width=\linewidth]{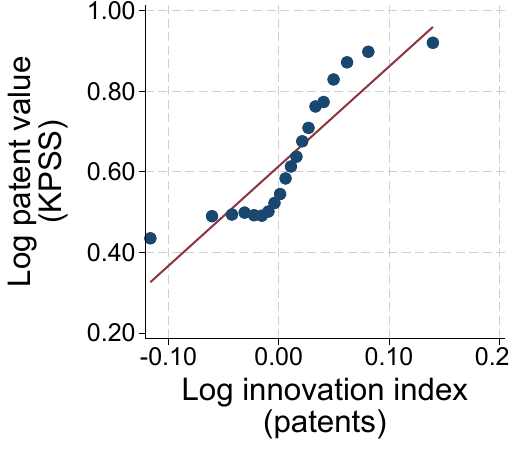}
			\subcaption{Patents: market value}
		\end{subfigure}
		\caption{Validation of the modern innovation index \vspace{4pt} \newline \footnotesize \emph{Notes:}   The figure presents binscatter plots (20 bins) for the \textit{all-mpnet-base-v2}-based innovation measure from equation~\ref{eq:innovation} ($\tau=5$, $k=20$) and modern validation measures, residualized for year and field fixed effects in panel a) and for year and CPC-section fixed effects in panels b) and c). Panel a) presents validation results for OpenAlex articles in \textit{physics and astronomy, chemistry, and mathematics}, 1990--2019, using citation counts from OpenAlex. Panels b) and c) present validation results for US utility patents of listed firms filed 1990--2019, using forward citations and the market value of the patent from \cite{kogan2017technological}. All measures are transformed using the natural logarithm. The estimated elasticity is $2.57$ for panel a) ($N=157,085$), $3.74$ for panel b) ($N=2,348,639$), and $2.49$ for panel c) ($N=2,348,639$). Please refer to appendix section~\ref{sec:app:modern} for formal regression results with additional control variables, such as the length of text.}
		\label{fig:modern_validation}
	\end{figure}
	
	First, the paper validates the innovation measure from equation~\ref{eq:innovation} using scientific article citations as well as patent forward citations and patent value. We find raw correlations of 0.08 for scientific citations, 0.25 for forward patent citations, and 0.04 for patent market value. The numbers are of a similar magnitude as correlations between modern proxies of innovation, such as 0.11 for patent citations and patent value.\footnote{The correlation between patent citations and patent value is calculated using data from \cite{kogan2017technological} and further corresponds to the findings in the innovation literature \citep{harhoff1999citation,lanjouw2004patent}.} As shown in Figure~\ref{fig:modern_validation}, the association between the innovation measure from equation~\ref{eq:innovation} and the other proxies is stable and monotonically positive. We find that for scientific articles, a one per cent higher innovation index is associated with 2.6 per cent more citations for articles. For patents, a one per cent higher innovation index is associated with 3.7 per cent more forward citations as well as 2.5 per cent higher market value for patents.

	\begin{figure}[htp]
	\centering
	\begin{subfigure}[t]{.32\textwidth}
		\includegraphics[width=\linewidth]{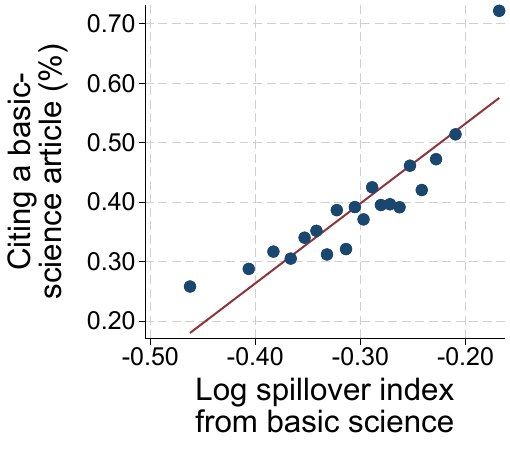}
		\subcaption{Received spillovers from physics, astronomy, chemistry, and math}
	\end{subfigure}\hspace{0.03\textwidth}%
	\begin{subfigure}[t]{.32\textwidth}
		\includegraphics[width=\linewidth]{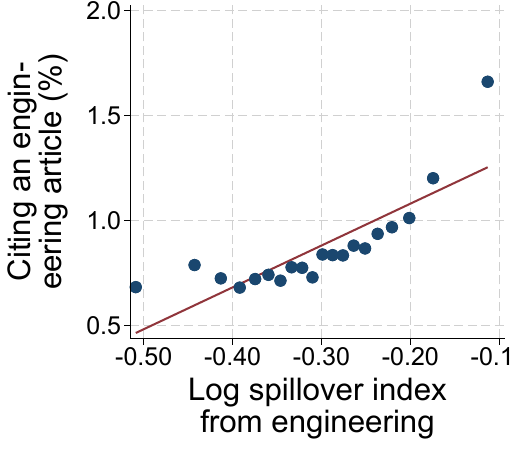}
		\subcaption{Received spillovers from engineering, computer science, and materials science}
	\end{subfigure}
	\caption{Validation of the modern received spillover index \vspace{4pt} \newline \footnotesize \emph{Notes:}   The figure presents binscatter plots (20 bins) for the \textit{all-mpnet-base-v2}-based received spillover measure from equation~\ref{eq:received_spillover} ($\tau=5$, $k=20$) and the share of patents citing at least one article in the source field, in per cent, residualized for CPC class $\times$ filing year fixed effects. Panel a) presents validation results for received spillovers from \textit{physics and astronomy, chemistry, and mathematics} and panel b) for received spillovers from \textit{engineering}. Patent-to-article citations are taken from \cite{marx2020reliance} and matched to the articles in the paper's OpenAlex dataset. The received spillover measure is transformed using the natural logarithm. A one log point higher index raises the share of citing patents by $1.68$ percentage points in panel a) and by $1.54$ percentage points in panel b) ($N=1,624,402$). Please refer to appendix section~\ref{sec:app:modern} for formal regression results with additional control variables, such as the length of text.}
	\label{fig:modern_spillover_validation}
\end{figure}

	Next, the paper validates the received spillover index from equation~\ref{eq:received_spillover} against the citation flows. For patents, \cite{marx2020reliance} link the front-page references of US patents to the scientific articles they cite. We find a raw correlation of 0.02 for both citation flows in both directions --- here the low correlation is mainly driven by citation flows to science being a rare binary event. Figure~\ref{fig:modern_spillover_validation} plots the share of patents that cite an article in the source field against the received spillover index from that field while residualizing for technology class and filing year. Both shares rise almost monotonically with the index. A one percent increase in the received spillover index from \textit{physics, astronomy, chemistry}, and \textit{math},  is associated with a 4.3\% increase in citation to these fields, while a one percent increase in the received spillover index from \textit{engineering, computer science, and materials science} is associated 1.8\% increase in citations to the fields.

	For articles, the reverse flow can be derived from the rare references that articles make to patents. Residualizing for topics and publication years, a one per cent higher received spillover index from patents is associated with a 2.8 per cent increase in engineering articles citing a patent (see appendix section~\ref{sec:validation_modern_reverse_flow}). The spillover index thus tracks the direction of actual knowledge flows in both directions, from patents to scientific articles and vice versa.
	
	\FloatBarrier
	\subsection{Validation on historical data}
	\label{sec:validation}
	
	This section validates the innovation index from equation~\ref{eq:innovation} and the received spillover index from equation~\ref{eq:received_spillover} against historical data. The main challenge for historical validation is the absence of historical citation data from the majority of articles and publications. However, for patents we can draw on the Woodcroft reference index  from \cite{Nuvolari2011}. Moreover, we can use the number of editions a book went through as a measure of its value or success.

\begin{figure}[h]
		\centering
		\begin{subfigure}[t]{.32\textwidth}
			\includegraphics[width=\linewidth]{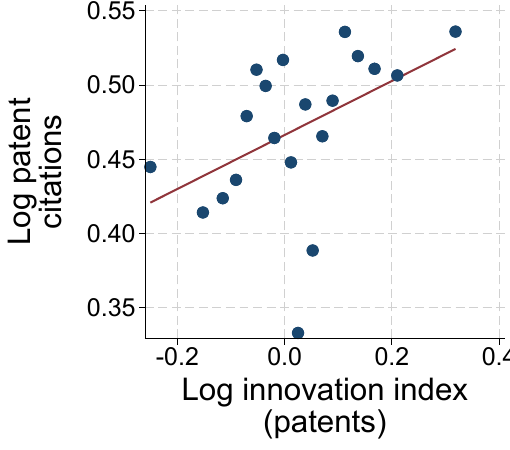}
			\subcaption{Patent citations, 1720--1799}
		\end{subfigure}\hfill
		\begin{subfigure}[t]{.32\textwidth}
			\includegraphics[width=\linewidth]{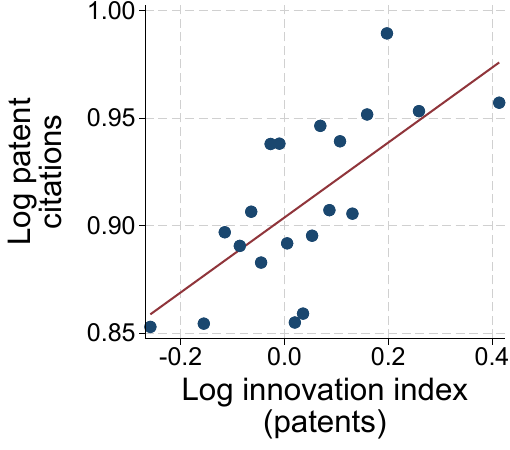}
			\subcaption{Patent citations, 1800--1841}
		\end{subfigure}\hfill
		\begin{subfigure}[t]{.32\textwidth}
			\includegraphics[width=\linewidth]{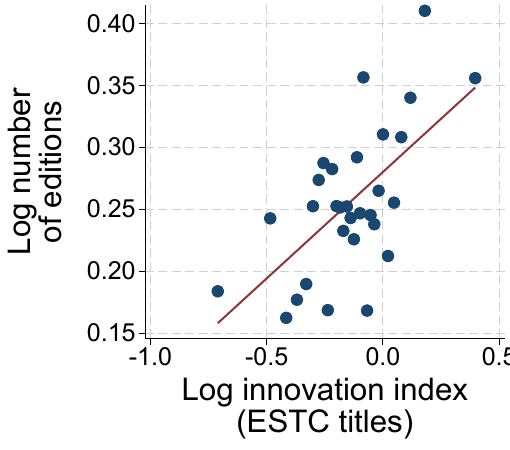}
			\subcaption{ESTC number of editions, 1620--1780}
			\label{fig:estc_editions}
		\end{subfigure}
		\caption{Validation of the historical innovation index \vspace{4pt} \newline \footnotesize \emph{Notes:}   The figure presents a binscatter plot for the SteamBERTh-based innovation measure and historical validation measures for year fixed effects. Panel a) and b) present validation results for historical patents using patent citations from \cite{Nuvolari2011}. Panel c) presents results for historical books from the ESTC database using a count of published editions for propositional and prescriptive knowledge. Propositional knowledge is defined as \textit{applied physics, astronomy, mathematics, chemistry, and encyclopedias} and prescriptive knowledge is defined as \textit{technical publications, navigation, scientific instruments, and patents}. All measures are transformed using the natural logarithm. The estimated elasticity is  $0.18$ for panel a) ($N=1,922$),  $0.17$ for panel b) ($N=6,684$), and $0.17$ for panel c) ($N=5,496$).  Please refer to appendix section~\ref{sec:validation_of_innovation} for formal regression results with additional control variables, such as the length of text.}
		\label{fig:binscatter}
	\end{figure}

	For constructing the historical innovation index we use embeddings from the \textit{SteamBERTh} model. We find a raw correlation of 0.02 for patent citations 1720--1799 and 0.07 for 1800--1841 (when citations become more common and represent a stronger signal). For the number of editions we find a raw correlation of 0.07. Figure~\ref{fig:binscatter} then plots the relationship between the \textit{SteamBERTh}-based innovation measure and historical patent citation counts residualized for publication years. For patents, we observe a within-year elasticity of $0.18$ for the time period 1720--1799 and $0.17$ for 1800--1841. For books from the ESTC, we also observe a within-year elasticity of $0.17$. The strength of the association is reassuring and shows that the innovation measure also picks up a sufficient signal from the less structured and more complex historical data.

	As an additional validation exercise, appendix table \ref{tab:top-20-innov-phys} lists the 20 most innovative ESTC titles in \textit{applied physics}. Reassuringly, the list is composed of the most famous scientists of the day, including Robert Boyle, Thomas Hobbes, Johannes Kepler, René Descartes, Gassendi, and Robert Hooke. It further includes a large number of seminal and groundbreaking works in the history of physics, including Francis Bacon's \textit{Great Instauration} and Robert Boyle's air pump experiments \citep[see e.g.][]{shapin1985leviathan}.\footnote{When going through this list note the following. By construction, the innovation measure identifies prominent works whose topics were novel and influential. This does not necessarily mean that the positions argued in these works were either correct or influential. The current approach would not be able to distinguish between writers arguing on the same subject, topic, and content but having differential views, nor to identify which writer won the debate. For example, appendix table \ref{tab:top-20-innov-phys} includes a title by Alexander Ross where he disapprovingly comments on Sir Kenelm Digby's natural philosophy. The impact of this title is most likely due to Sir Kenelm Digby's influence on natural science rather than Alexander Ross's position. Also note that the ESTC sometimes contains different versions of the same publication with differing titles. Reassuringly, these are treated consistently in the innovation measure.}  
	
	Additionally, appendix section~\ref{sec:validation_of_innovation} explores whether \textit{SteamBERTh} constitutes an improvement over standard BERT models trained on modern data. It runs a set of regressions with different BERT models as an input into the innovation measure from equation~\ref{eq:innovation} as competing predictors for patent citations. We find that the \textit{SteamBERTh} based innovation measure has the highest $R^2$ and strongly outperforms the other BERT models in a horse race specification for the time period 1720--1800.\footnote{We further document that the other BERT models are performing slightly better on nineteenth century patent citation data, 1800--1849. This is expected, the fine-tuning procedure offers time-period specific performance gains at the loss of generality for more recent time periods.} We also report a comparison between the performance of the bag-of-words based innovation measure from \cite{Kelly2021a} and the embedding space based measures from equation~\ref{eq:innovation} in appendix section~\ref{sec:validation_of_innovation} and~\ref{sec:app:modern}. While we find that the \cite{Kelly2021a} index performs well on modern data (it performs better on patents than scientific articles), it loses all predictive power when applied to historical data.\footnote{Here, the main problem for bag-of-words based methods is the varying spelling standards in historical text as well as the use of more abstract terms and concepts in general.}

	Lastly, validating the received spillover measure with historical data poses an especially hard problem since we lack any kind of citation data outside of patents. The paper therefore adopts a more indirect validation approach where we draw on the share of authors publishing in multiple fields. The intuition is simple. Authors who published in multiple fields would have been more likely to have access to ideas from multiple fields. Hence, we also expect them to have been more likely to incorporate those ideas into their work, effectively constituting knowledge spillovers.

	\newcommand{\pub}{\mathrm{pub}} 	\label{sec:validation_spillover}
	
	We operationalize this by using author $j$’s publication share in field $B$:
	\begin{equation}
		s_{jB}\equiv \frac{n_{jB}}{n_{j,A\cup B}},
		\label{eq:author_proxy}
	\end{equation}
	
	where $n_{jB}$ denotes the number of publications by author $j$ in field $B$ and $n_{j,A\cup B}\equiv n_{jA}+n_{jB}$ denotes total publications in $A \cup B$.
	We expect that for any author publishing in field $A$ their publication share in field $B$ should positively correlate with the amount of ideas they are exposed to from B and hence to the strength of knowledge spillovers from $B$. 
		
	\begin{figure}[h]
		\centering
		\includegraphics[width=0.32\linewidth]{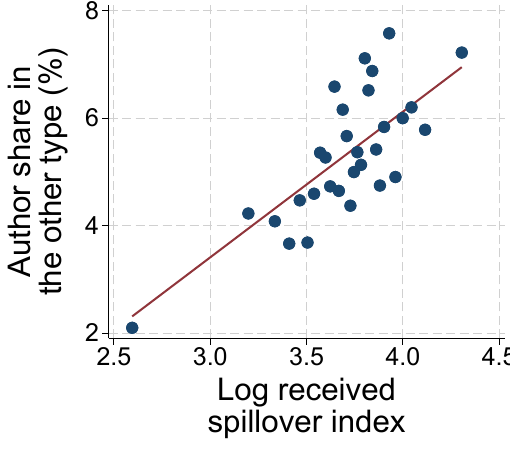}
		\caption{SteamBERTh-based received spillover index and author share proxy --- binscatter plot, residualized for year fixed effects \vspace{4pt} \newline \footnotesize \emph{Notes:}   The figure presents a binscatter plot (30 bins) for the SteamBERTh-based received spillover measure and the author proxy from equation~\ref{eq:author_proxy}, in per cent, residualized for year fixed effects. The sample period is 1620--1780. $N=5,827$. Please refer to appendix section~\ref{sec:validation_of_innovation} for formal regression results with additional control variables, such as the length of text.}
		\label{fig:binscatter_author_proxy}
	\end{figure}

	We then apply this proxy to the fields of propositional ($\Omega$) and prescriptive ($\lambda$) knowledge. For each title in $\Omega$, we measure the share of author $j$'s publications in $\lambda$ and vice versa. We find a raw correlation between the received spillover index and the author-share proxy of 0.04. Figure~\ref{fig:binscatter_author_proxy} plots the relationship between both measures while residualizing for publication year. We find a strong positive association, where a 1\% increase in the received spillover index is associated with a 0.027 percentage point increase in the author share, and a 100\% increase is associated with a 1.9 percentage point increase, constituting a 35\% increase at the mean. We further report formal regression results in Appendix~\ref{sec:validation_of_received_spillover_appendix} together with separate estimation results for spillover measures from $\Omega$ to $\lambda$ and $\lambda$ to $\Omega$. Overall, this section has shown that both the innovation and spillover measures based on \textit{SteamBERTh} can also be successfully validated against historical proxies of innovation and knowledge spillovers. 

	\section{Empirical framework}
	\label{sec:empirical-framework}

	The previous section has established a new method to measure innovation and knowledge spillovers from historical text data. It has further validated the innovation and spillover measures using modern and historical data. Next, this section develops a framework to estimate the relationship between knowledge spillovers and innovation at the title level. Observing a positive relationship between knowledge spillovers and innovation would be a key characteristic of a positive feedback loop process between propositional and prescriptive knowledge as outlined by \cite{Mokyr2002}:
	
	According to Mokyr, feedback loop processes are characterized by the presence of \textit{innovation-inducing} knowledge spillovers between the two types of knowledge \citep{Mokyr2002}. We would expect that ``growth in one increases the marginal product of the other'' \citep[p. 21]{Mokyr2002}.\footnote{\citep[see also][]{milgrom1991complementarities}} Here, an increased marginal product corresponds to innovation as captured through the innovation measure from equation~\ref{eq:innovation}. Likewise, we can directly capture spillovers from new knowledge through the received spillover measure from equation~\ref{eq:received_spillover}. Hence, with these two measures, it becomes possible to estimate whether titles that received knowledge spillovers were more likely to be innovative. If this was the case, growth in either $\Omega$ or $\lambda$ would have increased the marginal product of the other, thereby creating the foundations for a positive feedback loop.
	
	Section~\ref{sec:coneptual_background} has shown how we can formalize the feedback loop as two cross-elasticities:
	\begin{equation}
		\label{eq:cross_elasticities_framework}
		\beta^{\Omega \to \lambda}_{p} = \frac{\partial \ln \text{Innovation}^{\lambda}}{\partial \ln \text{Spillover}^{\Omega \to \lambda}}
		\qquad \text{and} \qquad
		\beta^{\lambda \to \Omega}_{p} = \frac{\partial \ln \text{Innovation}^{\Omega}}{\partial \ln \text{Spillover}^{\lambda \to \Omega}},
	\end{equation}	
	
	To capture them empirically, we test whether a given title that drew more on knowledge from the other sphere was more innovative than other titles within its field and same year. For this, we draw on the innovation and received spillover index from equation~\ref{eq:innovation} and~\ref{eq:received_spillover}:

	\begin{equation}
		\label{eq:ols_spillover}
		\text{Innovation}^{A}_{ijt} = \sum_{p=1600-1619}^{1760-1789} ( \beta_p^{B \to A} \cdot 	\text{Received spillover}_{B \to A}(v_{ijt}) \times \eta_p) + \boldsymbol{X'_{ijt}}\zeta	+ \gamma_j + \alpha_t + \varepsilon_{ijt}
	\end{equation}

	The dependent variable, $\text{Innovation}^{A}_{ijt}$, is the innovation index from equation~\ref{eq:innovation} for title $i$ in field $j$ at time $t$ in knowledge field $A$. The main explanatory variable, $\text{Received spillover}_{B \to A}(v_{ijt})$, captures the strength of knowledge spillovers that title $i$ in time $t$ received from knowledge field $B$. The measure is defined in equation~\ref{eq:received_spillover}. We use it to capture spillovers between $\lambda$ and $\Omega$ as well as $\Omega$ and $\lambda$. The fields contained in $\Omega$ are \textit{applied physics, astronomy, mathematics, chemistry,} and \textit{encyclopedias} and the fields contained in $\lambda$ are \textit{technical publications, navigation, scientific instruments,} and \textit{patents}. To estimate the strength of the association between these spillovers and innovation over time, the spillover coefficient is interacted with twenty-year period indicator variables $\eta_p$. Both the dependent and independent variables are transformed using the natural logarithm. $\boldsymbol{X'_{ijt}}$ is a vector of control variables, incl. a title's level and quadratic word count. Lastly, $\gamma_j$ and $\alpha_t$ capture subject and year fixed effects.

	Note that mechanically the calculation of the innovation index from equation~\ref{eq:innovation} for title $i$ in year $t$ requires a pre- and post-period of length $\tau$. Therefore, the ESTC sample is mechanically limited to $[1600+\tau,1800-\tau]$. For this paper, we choose $\tau=20$ years. Additionally, we restrict the start of the sample period to 1640 --- due to the scarcity of publications before that date, see appendix~\ref{fig:all_estc}. Robustness to alternative $\tau$ and to including the full 1620--1780 period are shown in appendix section~\ref{sec:10-year-interval}. 

	As argued in section~\ref{sec:literature}, we expect to find:
	\begin{enumerate}
		\item A negative or neutral relationship between knowledge spillovers and innovation before the eighteenth century
		\item A gradual ``tipping of the balance'' toward a positive relationship during the early eighteenth century
		\item A positive relationship between knowledge spillovers and innovation by the middle of the eighteenth century
	\end{enumerate}

	Equation~\ref{eq:ols_spillover} estimates the association between a title's innovativeness and it receiving knowledge spillovers. Reading $\beta_p^{B \to A}$ as the cross-elasticity of equation~\ref{eq:cross_elasticities_framework}, it captures the productivity payoff to drawing on the other type of knowledge. For this reading to hold, three conditions must be met.
	
	\begin{enumerate}
		\item A positive coefficient must reflect the use of knowledge rather than the convergence of vocabularies.
		\item The measures must not generate patterns mechanically.
		\item No common shock may raise both a title's spillover index and its innovativeness within field and year.
	\end{enumerate}
	
	The paper carefully ensures that all three conditions are met. First, the previously shown validation results (patent and article citations, patent value, citation flows between fields, editions, and cross-field authorship) document that the measures pick up economic signals and not just changes in vocabulary. Moreover, all coefficients are estimates within a field and year --- simple corpus changes as documented by \cite{almelhem2023enlightenment} are absorbed by the fixed effects. Lastly, a rewriting exercise that unifies the style of all titles leaves the results unchanged (see robustness section~\ref{sec:robustness}).

	Second, we ensure that the index is not mechanically driven by artefacts such as corpus growth. While year-specific changes are absorbed by year fixed effects, we might still worry about heterogeneous effects of pool size on individual titles. To account for such mechanical effects, the paper performs two analyses in appendix~\ref{sec:mechanical}. First, it artificially balances the subject-pools by repeatedly drawing from each subject's own pool until subjects are balanced over years. Results are almost identical to the baseline. Additionally, it performs a placebo test where the growing pool size of the real data is preserved but individual titles are shuffled randomly between years. The resulting year-coefficients are precisely estimated at zero and do not exhibit any trends. 
	
	Third, a natural worry is whether results are driven by common shocks raising both innovation and spillovers simultaneously. We show that results are robust to including to subject-by-year fixed effects. Therefore, a shock such as a naval war, which could raise the demand for both navigation manuals and astronomical tables, affects a field-year as a whole and is absorbed. Yet, other shocks might affect different fields simultaneously --- for example, the adoption of a common writing style could create spurious spillovers. To account for this, the paper adopts a placebo approach where we estimate equation~\ref{eq:ols_spillover} with spillovers from unrelated fields. For example, spillovers from logic or philosophy should not have affected innovation in practical fields. Throughout all placebo field specification we do not find evidence of spurious estimates. Nonetheless, we stress that estimates remain associations at the title level. They capture fundamental characteristics of the knowledge economy, not the causal effect of title-specific spillovers.

	\section{\label{sec:results}Empirical results}
	\label{sec:empirical_results}
	\subsection{Knowledge spillovers and innovation}
	\label{sec:results_spillovers_innovation}

\begin{figure}[H]
		\centering
		\tiny
		\begin{subfigure}[t]{0.49\textwidth}
			\includegraphics[width=1\linewidth]{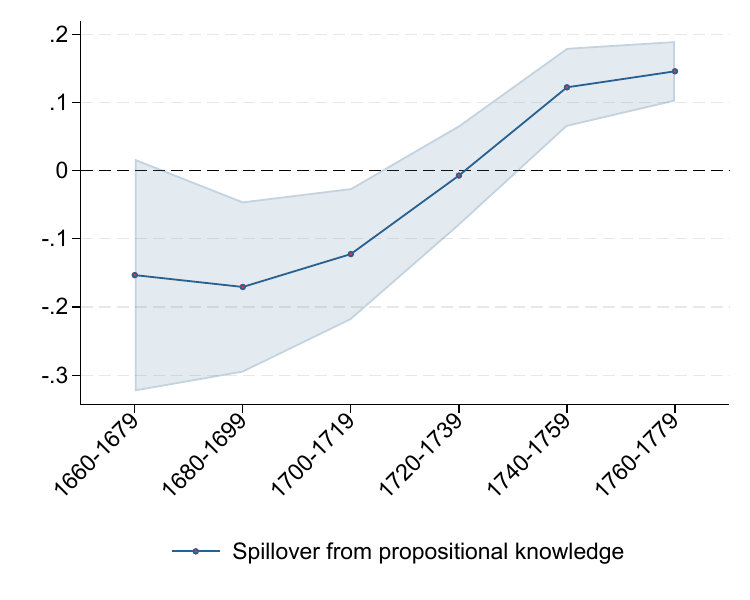}
			\subcaption{Spillovers from $\Omega$ $\rightarrow$ innovation in $\lambda$}
		\end{subfigure}
		\begin{subfigure}[t]{0.49\textwidth}
			\includegraphics[width=1\linewidth]{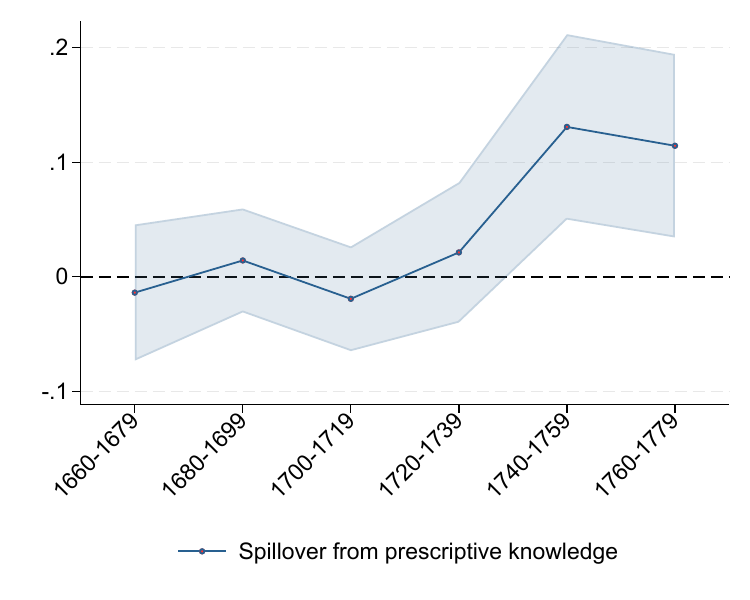}
			\subcaption{Spillovers from $\lambda$ $\rightarrow$ innovation in $\Omega$ }
		\end{subfigure}
		\caption{Spillovers from prescriptive ($\lambda$) and propositional knowledge ( $\Omega$) $\rightarrow$ innovation  \vspace{4pt} \newline \footnotesize \emph{Notes:}   The figure reports the estimated coefficients from equation \ref{eq:ols_spillover} estimated via OLS. The dependent variable is the innovation index from equation~\ref{eq:innovation}. The graph plots the coefficients from interacting the received spillover index from equation~\ref{eq:received_spillover} with twenty-year time periods. Dependent and independent variables are transformed using the natural logarithm. The model also controls for the level and quadratic count of words and includes year fixed effects. $N=$ 3,119 (a) and 2,940 (b). 90\% confidence intervals estimated using standard errors clustered at the publication year level.}
		\label{fig:results_prop_prescr}	
	\end{figure}

	Figure~\ref{fig:results_prop_prescr} reports the main results from equation~\ref{eq:ols_spillover}. Panel a) reports estimated coefficients for the relationship between spillovers from propositional to prescriptive knowledge, $\Omega \rightarrow \lambda$, and innovation over time. Panel b) reports estimated coefficients for the relationship between spillovers from prescriptive to propositional knowledge, $\lambda  \rightarrow \Omega$, and innovation over time.

	In panel a), we first find a negative relationship between knowledge spillovers from propositional ($\Omega$) to prescriptive knowledge ($\lambda$) and innovation before 1720. Next, we find that the relationship ``flipped'' in 1720--1739 and became positive in 1740--1759 and 1760--1779. For 1760--1779, we find that an increase in knowledge spillovers from $\Omega$ to $\lambda$ by 100\% was associated with a 10.6\% increase in innovation.
	
	In panel b), we first find a neutral relationship between knowledge spillovers from prescriptive ($\lambda$) to propositional knowledge ($\Omega$) before 1740. Then in 1740--1759 and 1760-1779, the relationship turned positive. In 1760--1779, a 100\% increase in knowledge spillovers from $\lambda$ to $\Omega$ was associated with an 8.3\% increase in innovation. 
	
	Hence, in line with Mokyr, we find evidence of an absence of a positive feedback loop before the middle of the eighteenth century. Even more, the relationship between spillovers from propositional knowledge and innovation in prescriptive knowledge was negative. This finding illustates the obstacles to the successful application of early theory. As argued before in section~\ref{sec:literature}, negative spillovers could have arisen from relying on the predictions from immature theories ($A_1$) or an underdeveloped and imprecise toolset ($A_3$). Moreover, if early applications of science did function, they were often economically non-viable. Appendix section~\ref{sec:app:negative_coefficient} decomposes the contribution of individual titles to the negative coefficient and discusses individual titles. We find evidence that titles often integrated traditional knowledge, e.g. from medicine or agriculture, rather than from e.g. physical experiments. Moreover, we find cases that were clearly not economically viable such as a burning glass to melt metals or a distillation machine for drinking water on ships.
	
	Next, as predicted by \cite{Mokyr2002}'s theory, the results show that the relationship started to flip in the eighteenth century and became positive by 1740--1759 and 1760--1779. As argued in section~\ref{sec:literature}, this could have been driven by a range of indirect channels ($A_1$--$A_4$ and $B_1$--$B_3$). Appendix section~\ref{sec:app:negative_coefficient} presents additional evidence, that by 1740, titles were drawing on a broader base of propositional knowledge, leading to the useful applications of subjects such as electricity, chemistry, and agriculture. Altogether, the empirical results support \cite{Mokyr2002}'s thesis.

	\noindent \textbf{Patents and the real economy}
	
	Next, the paper specifically focuses on patents as a subset of prescriptive knowledge. Applying the model from equation~\ref{eq:ols_spillover} to patents, the paper estimates whether knowledge spillovers from propositional knowledge were associated with increased innovation in patents, and hence with direct changes in the real economy. The paper further presents results for spillovers from applied physics, capturing the applied mechanics channel most commonly associated with knowledge spillovers during the Industrial Revolution.

	\begin{figure}[h]
		\centering
		\tiny
		\begin{subfigure}[t]{0.42\textwidth}
			\includegraphics[width=1\linewidth]{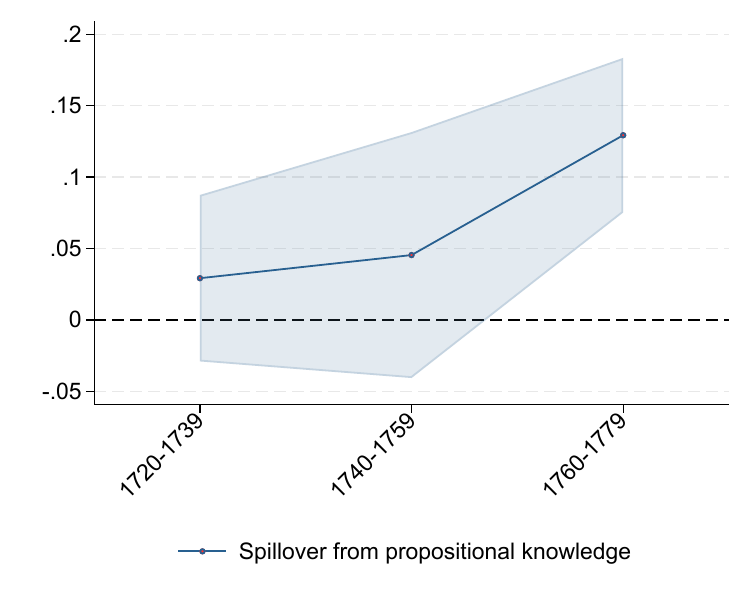}
			\subcaption{Spillovers from $\Omega$ $\rightarrow$ patent innovation}
		\end{subfigure}
		\begin{subfigure}[t]{0.42\textwidth}
			\includegraphics[width=1\linewidth]{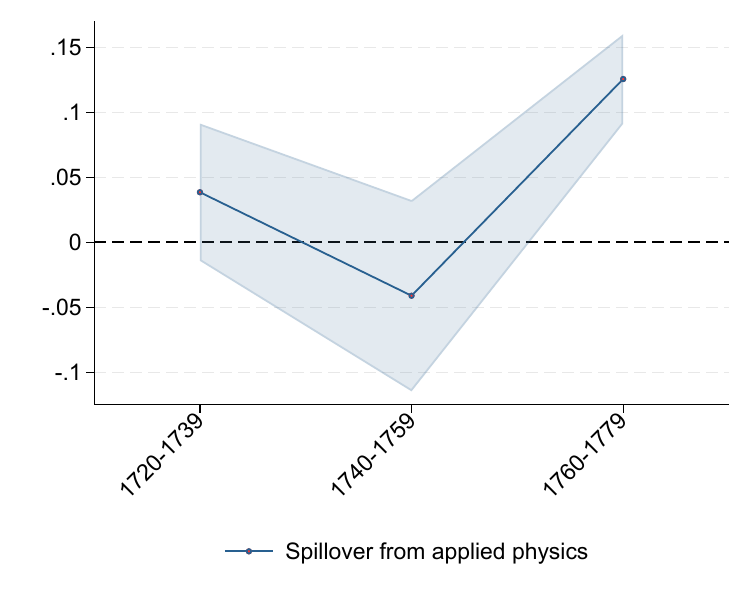}
			\subcaption{Spillovers from applied physics $\rightarrow$ patent innovation}
		\end{subfigure}
		\caption{Spillovers from propositional knowledge ($\Omega$) and applied physics $\rightarrow$ patent innovation \vspace{4pt} \newline \footnotesize \emph{Notes:}    The figure reports the estimated coefficients from equation \ref{eq:ols_spillover} estimated via OLS. The dependent variable is the innovation index from equation~\ref{eq:innovation}. The graph plots the coefficients from interacting the received spillover index from equation~\ref{eq:spillover} with twenty-year time periods. Panel a) shows results for spillovers from the full set of $\Omega$ and panel b) spillovers from applied physics. Dependent and independent variables are transformed using the natural logarithm. The model controls for the level and quadratic count of words and includes year fixed effects. Given that the patent data only starts in 1700 and that the innovation index uses a comparison period of $[-\tau,\tau]$, $\tau=20$ , the sample is 1719--1779. $N=$ 784 in both panels. 90\% confidence intervals estimated using standard errors clustered at the publication year level.}
		\label{fig:results_patents}	
	\end{figure}

	Figure~\ref{fig:results_patents} presents the results. Panel a) presents results for spillovers from propositional knowledge ($\Omega$) and panel b) presents results for spillovers from applied physics. Since the patent data only starts in 1700, we can only estimate effects on innovation post-1719.\footnote{This is due to the necessary comparison period of $[-\tau,\tau]$, $\tau=20$ for the innovation index from equation~\ref{eq:innovation}.} In panel a), we find that the relationship between knowledge spillovers from $\Omega$ to patents and innovativeness was neutral in 1720--1739 and 1740--1759 and then became positive and significant in 1760--1779. In 1760--1779, an increase in spillovers from $\Omega$ by 100\% was associated with an increase in innovativeness of 9.37\%. For spillovers from applied physics in panel b), we find a similar, if slightly smaller effect. Here, an increase in spillovers from applied physics was associated with an increase in innovativeness of 9.10\%. Hence, we find that knowledge spillovers from propositional knowledge were also associated with productivity increases in the real economy, as captured through patents.
	
	For additional robustness, panels c) and d) from appendix figure~\ref{fig:results_patents_robustness} show results with patent citations from \cite{Nuvolari2011} as the dependent variable. We find similar patterns.\footnote{We find positive post 1760 coefficients for both outcomes. Results diverge with respect to pre 1740 coefficients: For the \cite{Nuvolari2011} are negative, while they are neutral using the innovation index. Note here that patent citations before 1740 were extremely rare.} Moreover, appendix section~\ref{sec:by-industry} presents estimated spillover coefficients interacted by industry. We find that the relationship between spillovers from both $\Omega$ and applied physics was strongest for \textit{engines, food, glass, metallurgy, mining, pottery}, and \textit{ship} industries. Notably, we do not find a significant effect for \textit{textiles}, potentially reflecting the importance of tweaking and tinkering in textile innovations \citep{mokyr1992,OGrada2016} or the relative decentrality of textiles within Britain's innovation network \citep{rosenberger2024}.

	\subsection{Placebo spillovers}
	
	The previous results provide strong support for the Mokyrian hypothesis. A potential concern, however, is that the estimates may be confounded by general linguistic trends correlated with title innovativeness. These may include changes in linguistic style, more sophisticated language, or more frequent references to places and events.
	
	To address this concern, we report placebo spillovers from alternative counterfactual fields. We begin with spillovers from clearly unrelated fields to show that the results are not driven by broad linguistic trends. We then consider spillovers from clearly related fields that, however, would have been unlikely to be a productive input into either propositional or prescriptive knowledge. This analysis shows that even spillovers from related and linguistically similar fields cannot explain our earlier results. Finally, we account for the fact that the selection of these placebo groups allows for a certain degree of subjectivity, and therefore present placebo spillovers from all other fields in the ESTC. Even here, the effects from spillovers from propositional and prescriptive knowledge dominate spillovers from all other placebo fields.
	
	\noindent \textbf{Placebo spillovers from plausibly unrelated fields}

	\begin{figure}[h]
		\centering
		\tiny
		\begin{subfigure}[t]{0.49\textwidth}
			\includegraphics[width=1\linewidth]{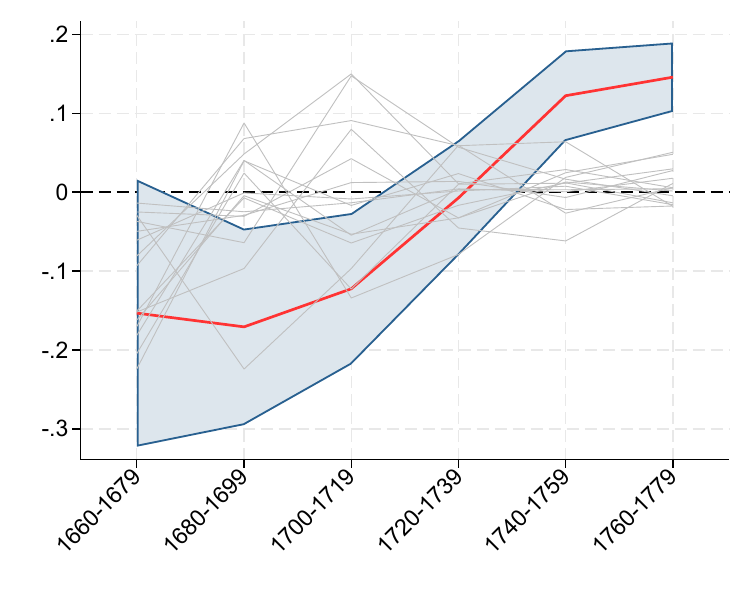}
			\subcaption{Spillovers from $\Omega$ $\rightarrow$ innovation in $\lambda$}
		\end{subfigure}
		\begin{subfigure}[t]{0.49\textwidth}
			\includegraphics[width=1\linewidth]{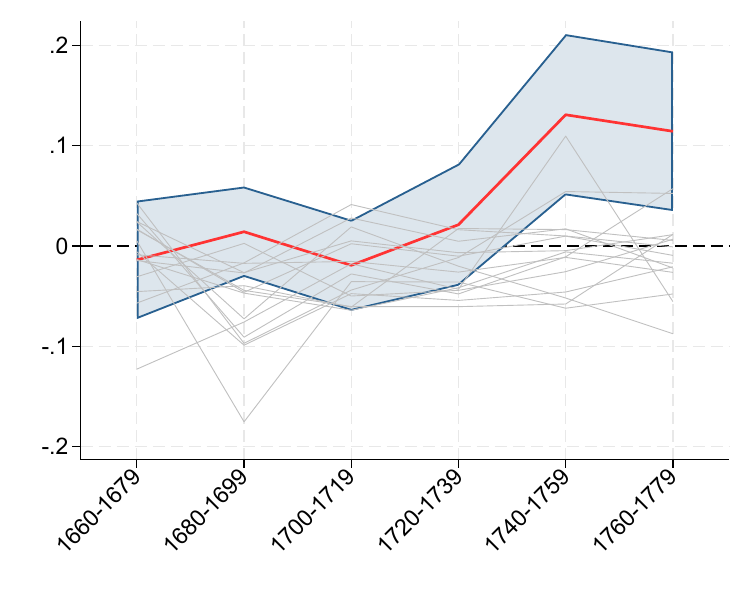}
			\subcaption{Spillovers from $\lambda$ $\rightarrow$ innovation in $\Omega$ }
		\end{subfigure}
		\caption{Placebo spillovers from plausibly unrelated fields \vspace{4pt} \newline \footnotesize \emph{Notes:}  The figure reports placebo results from plausibly unrelated spillovers. These are defined as \textit{drama, poetry, antiquities, amusements, foreign languages, printing, prophecies, stories, superstition, architecture, art, biographies, moral tales, state affairs}, and \textit{travel descriptions}. Placebo coefficients are reported as gray lines, while the spillovers from propositional and prescriptive knowledge are reported as red lines. Coefficients estimated from equation~\ref{eq:ols_spillover}. Panel a) shows the association between spillovers and innovation in prescriptive knowledge ($\lambda$) and panel b) shows the association between spillovers and innovation in propositional knowledge ($\Omega$).  90\% confidence intervals estimated using standard errors clustered at the publication year level for spillovers from propositional and prescriptive knowledge.}
		\label{fig:placebo_tests}	
	\end{figure}
	
\begin{figure}[H]
		\centering
		\tiny
		\begin{subfigure}[t]{0.49\textwidth}
			\includegraphics[width=1\linewidth]{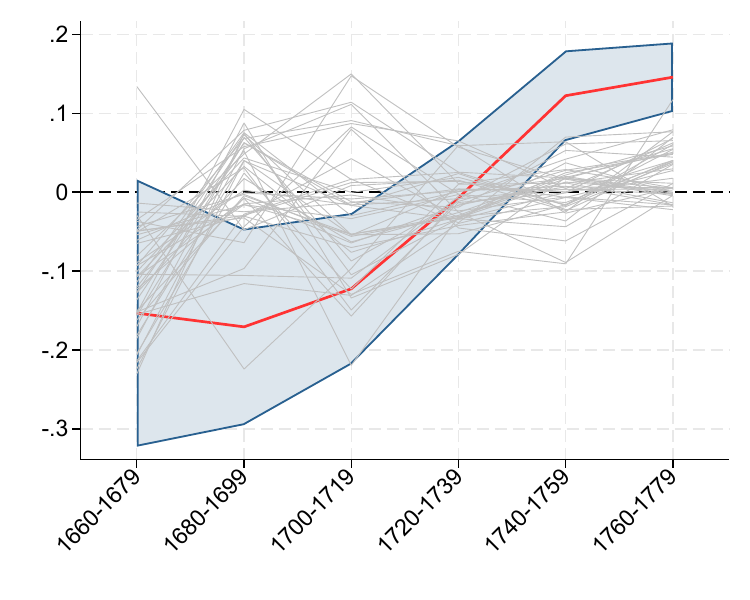}
			\subcaption{Spillovers from $\Omega$ $\rightarrow$ innovation in $\lambda$}
		\end{subfigure}
		\begin{subfigure}[t]{0.49\textwidth}
			\includegraphics[width=1\linewidth]{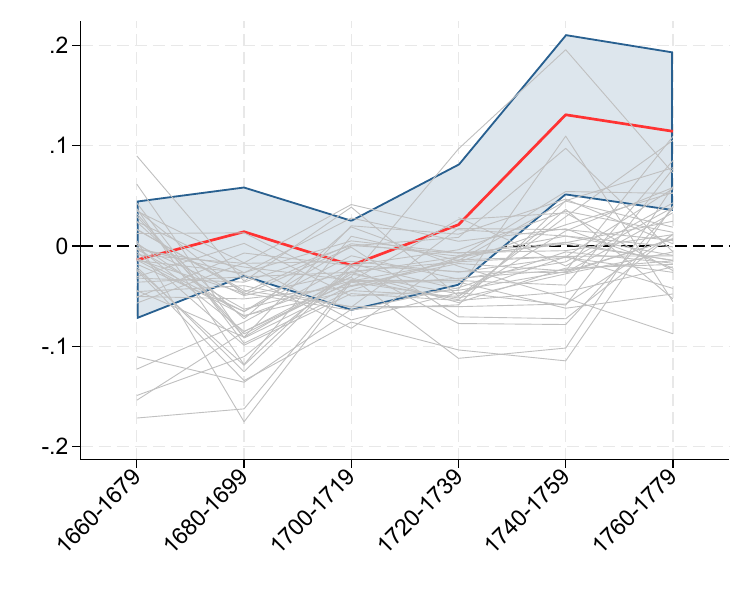}
			\subcaption{Spillovers from $\lambda$ $\rightarrow$ innovation in $\Omega$ }
		\end{subfigure}
		\caption{Placebo spillovers from all ESTC fields \vspace{4pt} \newline \footnotesize \emph{Notes:}  The figure reports placebo results from spillovers in all ESTC fields that are not part of prescriptive or propositional knowledge. Placebo coefficients are reported as gray lines, while the spillovers from propositional and prescriptive knowledge are reported as red lines. Coefficients estimated from equation~\ref{eq:ols_spillover}. Panel a) shows the association between spillovers and innovation in prescriptive knowledge ($\lambda$) and panel b) shows the association between spillovers and innovation in propositional knowledge ($\Omega$). 90\% confidence intervals estimated using standard errors clustered at the publication year level for spillovers from propositional and prescriptive knowledge.}
		\label{fig:placebo_tests_all}	
	\end{figure}

	We first start with placebo spillovers from plausibly unrelated fields. We define the set of plausibly unrelated fields as \textit{drama, poetry, antiquities, amusements, foreign languages, printing, prophecies, stories, superstition, architecture, art, biographies, moral tales, state affairs, and travel descriptions}. We argue that these fields cover both a broad range of potentially confounding topics and writing styles, while also being irrelevant to knowledge production in propositional and prescriptive knowledge.\footnote{Note that we do not include religion in the list of plausibly unrelated fields given the findings from \cite{almelhem2023enlightenment} that demonstrate a structural change in the relationship between science and religion throughout the seventeenth and eighteenth century. See also \cite{Hornung2014} and \cite{becker2024} for potential influences of religion or religious minorities on technological innovations. However, religion is included in the next placebo specification in figure~\ref{fig:placebo_tests_all}.} Figure~\ref{fig:placebo_tests} reports results.

	We find that for none of the plausibly unrelated fields does the relationship between spillovers and innovativeness increase over time. Furthermore, none of the plausibly unrelated fields yields positive coefficients of the same size as the positive spillovers from $\Omega \rightarrow \lambda$ and $\lambda \rightarrow \Omega$ for the period 1740--1759 and 1760--1779. We can also use this information for statistical inference by computing Fisher-exact p-values using \( p = \frac{1 + \sum_{j=1}^{J} \mathbf{1}(|\hat{\tau}_{j}| \geq |\hat{\tau}_{\text{prop$\mid$pres}}|)}{1 + J} \) where $J$ denotes the number of placebo coefficients. The p-value equals the share of placebo effects that are at least as extreme in absolute value as the estimated coefficient for propositional and prescriptive knowledge. The test is non-parametric and does not rely on standard errors or asymptotic approximations. Its validity rests on the exchangeability assumption—that under the null, propositional and prescriptive knowledge are indistinguishable from the placebo fields in the absence of a true spillover effect. For figure~\ref{fig:placebo_tests}, this yields a Fisher exact p-value of 0.067. 
	
	\clearpage
	\FloatBarrier
	
	\noindent \textbf{Placebo spillovers from related but ineffective fields}
	
	\begin{figure}[h]
		\centering
		\tiny
		\begin{subfigure}[t]{0.49\textwidth}
			\includegraphics[width=1\linewidth]{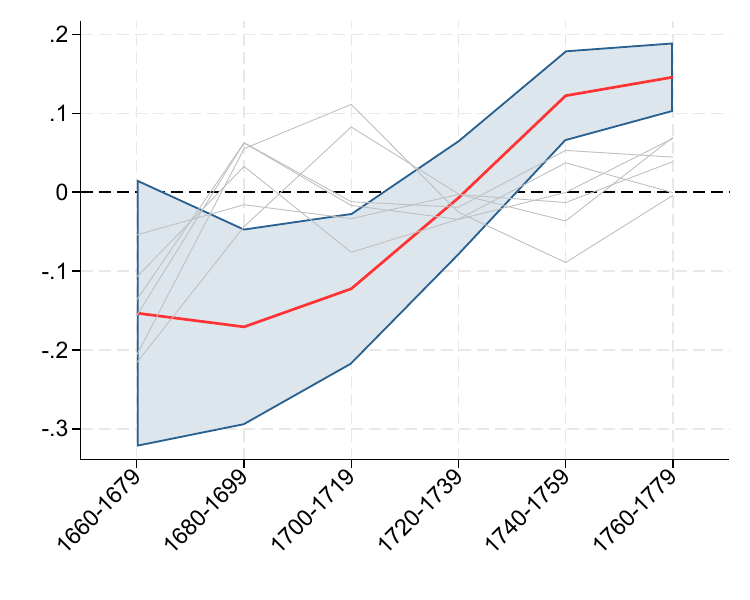}
			\subcaption{Spillovers from $\Omega$ $\rightarrow$ innovation in $\lambda$}
		\end{subfigure}
		\begin{subfigure}[t]{0.49\textwidth}
			\includegraphics[width=1\linewidth]{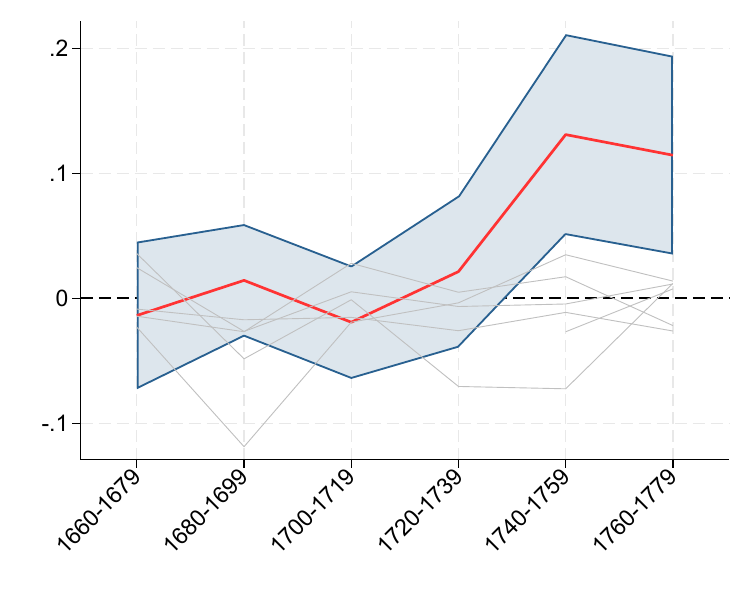}
			\subcaption{Spillovers from $\lambda$ $\rightarrow$ innovation in $\Omega$ }
		\end{subfigure}
		\caption{Placebo spillovers from related but ineffective fields \vspace{4pt} \newline \footnotesize \emph{Notes:}  The figure reports placebo results from related but ineffective fields. For propositional knowledge, we define the set of fields that are similarly academic but unrelated to technical work as \textit{Logics and rhetorics at university, classical education, university matters, philosophy} and \textit{political philosophy}. For prescriptive knowledge, we define the set of fields that are similarly applied but should not have contributed to the understanding of nature as \textit{law, architecture, antiquities, military, travel descriptions,} and \textit{economics}. Coefficients estimated from equation~\ref{eq:ols_spillover}. Panel a) shows the association between spillovers and innovation in prescriptive knowledge ($\lambda$) and panel b) shows the association between spillovers and innovation in propositional knowledge ($\Omega$). 90\% confidence intervals estimated using standard errors clustered at the publication year level for spillovers from propositional and prescriptive knowledge.}
		\label{fig:placebo_tests_related}	
	\end{figure}

	Next, we make the placebo test more demanding by using titles from fields that were conceptually closely related but which we believe to have been ineffective for improving the receiving field. For this, we create two sets of placebo fields. The first is a counterfactual to propositional knowledge and the second one a counterfactual to prescriptive knowledge. For propositional knowledge, we select a list of fields that are as academic and learned as the scientific fields but that we believe to be irrelevant for inventive technical activity. These fields are \textit{Logics and rhetorics at university, classical education, university matters, philosophy} and \textit{political philosophy}. Then, for prescriptive knowledge we select a list of fields that are as practical and applied as the fields in prescriptive knowledge, but that should not have informed the understanding of nature. These fields are \textit{law, architecture, antiquities, military, travel descriptions,} and \textit{economics}.\footnote{For economics, we observe publications only in the periods of 1740--1759 and 1760--1779, incl. works on physiocracy and mercantilism.}
	
	Figure~\ref{fig:placebo_tests_related} presents the results. We find that spillovers from the sets of related fields are estimated close to zero. Moreover, we do not find evidence of an increase in placebo spillover coefficients post 1740. We take this as additional evidence that our results are not driven by similar linguistic trends that would also have shown up in these related fields. Given that each placebo group in this analysis has fewer than 10 fields, Fisher exact p-values are not calculated.

	\FloatBarrier
	\clearpage
	
	\noindent \textbf{Placebos from all fields}

	Given that the earlier selection of plausibly unrelated placebos and related but ineffective placebos inevitably involves some subjective judgment, we next include all 44 ESTC subject classes as placebo fields to assess robustness. Note that this also includes subject classes that we expect to have been interacting with our fields of propositional and prescriptive knowledge such as \textit{medicine} or \textit{biology} (see appendix table~\ref{tab:all_subject_classes} for a list of all subject classes).\footnote{Note that many of these topics might have interacted with propositional and prescriptive knowledge in various ways. For example, topics such as \textit{music} or \textit{curiosities and wonders} might contain objects studied by science --- concepts such as vibration, pitch, and resonance were actively studied during the Scientific Revolution and Isaac Newton treated both sound and colors analogously in his \textit{Opticks}. Furthermore, other fields like \textit{legal} or \textit{administrative} might increase the value of e.g. patents through different channels.} Yet, as shown in figure~\ref{fig:placebo_tests_all}, spillovers from propositional and prescriptive knowledge still dominate spillovers from \textit{all other} fields.

	\begin{table}[h]
		\centering
		\begin{adjustbox}{max width=8.5\columnwidth}
			\begin{threeparttable}\fontsize{10}{13}\selectfont
				\caption{Fisher exact p-values from placebo tests}
				\label{tab:fisher_p_test}
				\begin{tabular}{p{4.5cm}p{2cm}p{1.5cm}p{1.5cm}p{1.5cm}}
\toprule
Period & Original spill. coeff. & Exact $p$-value & Placebo spill. $J$ & \# as extreme \\
\midrule
\textit{Panel A: Spillovers $\rightarrow$ innov. in $\lambda$ }& & & & \\
1740--1759 &     0.122 &     0.022 &        45 &         0 \\
1760--1779 &     0.146 &     0.022 &        45 &         0 \\
\bottomrule
\end{tabular}

				\begin{tabular}{p{4.5cm}p{2cm}p{1.5cm}p{1.5cm}p{1.5cm}}
\textit{Panel B: Spillovers $\rightarrow$ innov. in $\Omega$} & & & & \\
1740--1759 &     0.131 &     0.043 &        45 &         1 \\
1760--1779 &     0.114 &     0.022 &        45 &         0 \\
\bottomrule
\end{tabular}

				\begin{tablenotes}
					\item {\footnotesize \emph{Notes:} The table reports Fisher-style exact $p$-values comparing the estimated science--to--technology spillover coefficients with placebo spillover coefficients from all other 44 fields that are neither a subset of propositional nor prescriptive knowledge. 
						The $p$-value is calculated as 
						$p = \frac{1 + \sum_{j=1}^{J} \mathbf{1}(|\hat{\tau}_{j}| \geq |\hat{\tau}_{\text{prop$\mid$pres}}|)}{1 + J}$, 
						where $J$ is the number of placebo coefficients. Panel a) reports statistics for spillovers into $\lambda$ and panel b) reports statistics for spillovers into $\Omega$ corresponding to panel a) and b) in figure~\ref{fig:placebo_tests_all}. }
				\end{tablenotes}
			\end{threeparttable}
		\end{adjustbox}
	\end{table}

	In both graphs, none of the spillovers from all other fields is larger than the spillovers for propositional and prescriptive knowledge for 1760--1779. Fisher exact p-values are reported in table~\ref{tab:fisher_p_test}. The coefficients for the period 1760--1779 have a p-value of 0.022. For 1740--1759, there are, respectively, 0 and 1 placebo coefficients that are larger in absolute value than the original coefficients, which yields p-values of 0.022 and 0.043 respectively. Furthermore, note that the only positively larger coefficient is found in panel b) for $\Omega$. Moreover, this is a spillover from \textit{medicine} that could have been considered part of propositional knowledge, but was excluded as a life science (see section ~\ref{sec:assigning_subject_classes_prop_prescr}).\footnote{As eighteenth century medicine was heavily drawing on innovations in other fields of propositional knowledge, such as e.g. discoveries from electricity or the chemical analysis of mineral waters, the existence of spillovers is not surprising.}

	Altogether, the placebo results show that the relationship between spillovers from propositional and prescriptive knowledge cannot be explained by trends in language that also would have affected spillovers from other fields. The results from placebo spillovers from all fields in the ESTC further illustrate the importance of the propositional and prescriptive channel in comparison to other knowledge related channels that might impact innovation, e.g. originating from legal innovation or administrative practices.

	\FloatBarrier
	
	\subsection{Robustness}
	\label{sec:robustness}
	\noindent \textbf{LLM-based rewriting of original titles in a simplified style}

	A final concern is that changes in cosine similarities might also be driven by changes in style. To abstract away from the unique style component of individual titles, we draw on an LLM to rewrite titles in a simplified style.\footnote{We employ a \texttt{gpt-4o} model that is explicitly instructed to rewrite titles in a simplified style while preserving semantic content. The full approach is documented in Appendix section~\ref{appendix:section:llm_rewrite}.} This approach should a) remove individual style components and b) add a significant permutation to the style. Hence, if the previous results were primarily driven by style, it is unlikely results would hold based on the rewritten titles with simplified style.
	
	Appendix section~\ref{appendix:section:llm_rewrite} presents the rewriting exercise. First we evaluate the effectiveness of the rewriting exercise. We find that the rewritten texts differ strongly from the original, with more than 50\% of words in each title having been replaced and sentence complexity metrics indicating simplified grammatical and sentence structure. We conclude that the operation successfully altered the style.  Yet we also caution that the LLM rewriting operation likely changed some of the semantic content as well. Hence, this exercise should be seen as a demanding robustness check. Given the tendency of LLMs to hallucinate, interpolate, and project modern notions on historical text, the rewriting approach should not be employed for first order results.
	
	As a next step, we recalculate the innovation and received spillover measures from equation~\ref{eq:innovation} and~\ref{eq:received_spillover} with the rewritten text as input. As shown in appendix section~\ref{appendix:section:llm_rewrite}, using these measures in our baseline model from equation~\ref{eq:ols_spillover}, our results still remain highly similar. Most importantly, the spillover coefficients at the end of the eighteenth century remain positive and significant throughout all specifications.\footnote{Overall, we estimate a set of 6 specifications. We employ the new measures first for the innovation measure, then for the spillover measure, and then for the innovation and spillover measure together. We then apply these three specifications to the two directions of spillovers. Spillovers at the end of the eighteenth century remain significant throughout all of these three specifications. Here, the rewritten measures replicate the previously found trends, where for spillovers from propositional to prescriptive knowledge the coefficient is largest for the period of 1760--1779. For spillovers from prescriptive to propositional knowledge, the coefficient is largest for the period 1740--1760 for out of two specifications. For these specifications the coefficient is significant. For a third specification, the time period of 1760--1779 is largest and significant.} We find small changes for the beginning of the period where our sample of titles is also the smallest. Thus, in one single specification, the negative coefficient for spillovers from propositional to prescriptive knowledge between 1680 and 1699 becomes neutral. Moreover, in two specifications, we also find evidence of a one-time positive spillover in 1680--1699 from prescriptive to propositional knowledge. Despite these deviations, the overall trends remain stable, especially with respect to the emergence of positive knowledge spillovers at the end of the period.
	
	We take this demanding rewrite operation as additional evidence that our results capture spillovers in actual semantic content and are not purely driven by style. 
	
	\bigskip

	\noindent \textbf{Robustness for parameter and specification choices}
	
	Lastly, the paper conducts a series of robustness tests, showing that results hold irrespective of parameter choices in the innovation and spillover function from equation~\ref{eq:innovation} and~\ref{eq:spillover}. We further conduct a heterogeneity analysis for spillovers from different fields in $\Omega$ and $\lambda$ and distinguish between micro- and macro-inventions. Lastly, we also account for field size compositional bias. 
	\begin{enumerate}
		\item \textbf{Different parameters in innovation measure.} For the calculation of the innovation index, we compare each title to the $k$ most similar forward and backward titles. The innovation index for the baseline results was calculated with $k=20$. We assume that this corresponds to the size of individual topics/sub-fields that are relevant for comparison. To show that results are robust to other values of $k$, appendix figure~\ref{fig:robustness_k_p} shows the baseline results from figure~\ref{fig:results_prop_prescr} for $k=10$, $k=20$, and $k=30$. Its second and fourth rows further change $k$ and $\rho$ from received spillover equation~\ref{eq:received_spillover} simultaneously to $k,\rho=10$, $k,\rho=20$, and $k,\rho=30$. Results are highly robust to changing the parameters for the innovation and received spillover index. 
		\item  \textbf{Different length of time window for innovation and spillover index.} Additionally, appendix~\ref{sec:different_time_windows} reports robustness when changing the length of the time windows, $\tau$, for the calculation of the innovation and received spillover index from equation~\ref{eq:innovation} and \ref{eq:received_spillover}. Note that changing the time window also changes the interpretation of the innovation index. Shorter time windows place more weight on titles with a short-run impact, while longer time windows place more weight on inventions with a long-run impact. Following \cite{mokyr1992}, this corresponds to the distinction between micro- and macro-inventions. Also note that by construction, $\tau$ determines the length of the sample, $[1600+\tau, 1800-\tau]$, for which the innovation index is defined. Figure~\ref{fig:robustness_time_windows} reports results for $\tau=10$, $\tau=20$, and $\tau=30$. For spillovers from $\lambda$ to $\Omega$, we find that results are stronger for shorter $\tau$, indicating that these spillovers mainly led to innovation in micro-inventions. For spillovers from $\Omega$ to $\lambda$, the positive coefficients after 1740 are of similar size across all time windows.
		\item \textbf{Robustness to changing pool size and title length.} To ensure that the results of the paper are not mechanically driven by growth in the number of titles, appendix~\ref{sec:mechanical} resamples subject-year periods to balance pool size and performs a placebo simulation where publication years are randomly permutated between fields, while holding field size fixed. We do not find evidence that pool size had a mechanical effect. Moreover, appendix~\ref{sec:app:title_length} reports results for length-neutral embeddings and for trimming the shortest and longest 10\% of each group, with results barely changed.
		\item \textbf{Heterogeneity across outcome distribution} To test whether the results are driven by the upper quantiles of innovation or evenly distributed, we report results from a quantile regression in appendix figure~\ref{fig:results_quantile}. Coefficients appear relatively evenly distributed across the outcome distribution, although there is some evidence that coefficients are stronger at the tails of the distribution.
		\item \textbf{Heterogeneity analysis by spillovers from different fields.} The framework from equation~\ref{eq:ols_spillover} also allows for separately estimating coefficients for spillovers from individual fields in $\Omega$ and $\lambda$. Appendix figure~\ref{fig:results_individual_fields} reports results. We find that positive spillovers at the end of the eighteenth century were strongest for \textit{applied physics, astronomy, and chemistry} from $\Omega$ and \textit{navigation, and scientific instruments} from $\lambda$.
		\item \textbf{10-year coefficients.} To show that results are robust to shorter time intervals interacted with the spillover coefficients, panels a) and d) of appendix figure~\ref{fig:robustness_periods_fe} also report results for 10-year period coefficients for the baseline results. Panel a) and b) of 
		appendix figure~\ref{fig:results_patents_robustness} reports 10-year period results for patents.
		\item \textbf{Accounting for compositional bias and full 1620--1780 period.} A potential concern is compositional bias from the changing number of titles per field over time. This holds especially for earlier periods with fewer titles per field and higher compositional variation. To account for this, panels b) and e) of appendix figure~\ref{fig:robustness_periods_fe} report results with additional subject $\times$ year fixed effects. Trends do not change and the positive effects post--1740 remain significant. Panels c) and f) additionally report robustness for including the full time period of 1620--1780. Again, the trends do not change. Note, however, that given the few underlying observations for the period 1620--1660 (see figure~\ref{fig:prop_and_prescr}), results pre--1660 should be interpreted cautiously.
		\item \textbf{Excluding encyclopedias and dictionaries.} Given that \textit{encyclopedias and dictionaries} are the most ambiguous category, we demonstrate that results are robust even when excluding titles from this category in appendix~\ref{sec:excluding-ency}.
		\item \textbf{Robustness to excluding patent data.} As argued in section~\ref{sec:assigning_subject_classes_prop_prescr}, access to patent specifications in the eighteenth century was costly, uneven, and sparingly used. Therefore, patents are excluded from the calculation of the spillover measure for all baseline specifications.\footnote{Also note that patents are only available post-1700 (see section~\ref{sec:data_patents}). Therefore, including patents in the calculation of the spillover index would add an undesirable discontinuity to the spillover measure. Therefore, results from appendix figure~\ref{fig:results_prescr_prop_no_pat} should be interpreted carefully.} Figure~\ref{fig:results_prescr_prop_no_pat} reports robustness to including patents into the spillover measure. Trends remain similar, although the 1760--1779 coefficient becomes insignificant. Likewise, we show robustness to entirely excluding patents from our definition of prescriptive knowledge. Results for $\Omega \rightarrow \lambda$ spillovers are reported in figure~\ref{fig:results_prop_prescr_no_pat}. Again, trends remain similar.
	\end{enumerate}

	\section{Mechanism}
	\label{sec:mechanism}
	
	\subsection{Upper-tail human capital}
	\label{sec:uthc}
	To examine the mechanism behind the previous results, we begin by asking what made knowledge spillovers possible in the first place. A key point is that authors' expertise and specialization within narrow areas was not sufficient in itself. Incorporating knowledge spillovers required access across epistemic domains. For spillovers to occur, specialized artisans needed to be exposed to scholarly ideas, just as scholars depended on insights from artisanal techniques. Historically, these channels of exchange were limited.
	
	Here, \cite{Mokyr2002,Mokyr2016,mokyr2021holy} has highlighted the importance of upper-tail human capital and the networks of the Republic of Letters for bridging the gap between propositional and prescriptive knowledge. Upper-tail human capital would have been necessary to access knowledge far from one's expertise, while Enlightenment networks would have connected scholars or practitioners to experts in those areas. 
	The importance of this channel has been shown in a burgeoning literature on upper-tail human capital.\footnote{See e.g. \cite{Squicciarini2015}, \cite{Mokyr2016}, \cite{dittmar2020public} , \cite{Maloney2022}, \cite{Mokyr2022}, \cite{Kelly2023}, \cite{hanlonengineer}, and \cite{cinnirella2025flow}. Following \cite{Mokyr2009,Mokyr2016,MOKYR_2018,mokyr2021holy}, we define upper-tail human capital as an endowment of skills that could be used to advance the knowledge frontier:
		\begin{quote}
			`` `upper-tail human capital,' that is, the skills and knowledge of the best scientists, artisans, engineers, mechanics, and physicians (\dots) the idea that the envelope of useful knowledge is pushed forward by a relatively small number of people'' \citep{MOKYR_2018}
	\end{quote}} 
	Based on this argument, this section tests whether proxies for upper-tail human capital and enlightenment networks can explain the strength of knowledge spillovers in 1740--1779, the moment the spillover mechanism turned positive. As proxies for upper-tail human capital and enlightenment networks, we use authors' occupation and education, as well as authors' membership in academic and other enlightenment societies.
	
	We extract information on authors' education and occupation from ESTC book titles (see data section~\ref{sec:assigning_subject_classes} and appendix section~\ref{sec:variable_descriptions}). The approach builds on the logic that authors with prestigious occupations or memberships, such as the Royal Society, were likely to report these when publishing their books. Concretely, we extract information on authors' membership in the Royal Society, England's first and most prestigious scientific society, as well as membership in the \textit{Society for the Encouragement of Arts, Manufactures and Commerce} \citep{Howes2020}, England's first economic society founded to promote useful knowledge \citep[see also][]{cinnirella2025flow}. We further use information on authors who were engineers or had an academic career. Lastly, we use data on university students from \cite{Koschnick2025} to capture authors who enrolled at either the university of Oxford or Cambridge.\footnote{In the seventeenth and eighteenth century, these were the only English universities.} 
	
	To test how authors' networks, occupation, and education could have impacted spillovers, we formulate the following model:
	
	\begin{equation}
		\label{eq:ols_spillover_mechanism}
		\text{Received spillover}_{B \to A}(v_{ijt}) = \sum_{p \in \text{Occ}} \theta_p \mathbf{1} \{\text{Occ}_{ijt}=p\}  + \boldsymbol{X'_{ijt}}\zeta	+ \gamma_j + \alpha_t + \varepsilon_{ijt}
	\end{equation}

	where the dependent variable, $\text{Received spillover}_{B \to A}(v_{ijt})$, captures the strength of received spillovers from equation~\ref{eq:received_spillover}. It is transformed using the natural logarithm. The main explanatory variable, $Occ_{ijt}$, is the set of authors' affiliations, occupations, and education, provided as a set of indicator variables. It includes membership in the \textit{Royal Society}, or other \textit{enlightenment societies}, being an \textit{engineer}, having once been\textit{ enrolled at university}, or having had an \textit{academic career}. The model further includes word count controls, $\boldsymbol{X'_{ijt}}$, and subject and year fixed effects,  $\gamma_j$  and $\alpha_t$. Because we want to understand the determinants of spillovers during the period when spillovers were associated with innovation, we estimate the model for the time period 1740--1779.
	
	Overall, we expect that membership in the Royal Society or other enlightenment societies would have created access to more knowledge, improving the search function over knowledge domains ($A_2$ from section~\ref{sec:literature}).\footnote{See also \cite{Mokyr2005}, \cite{zanardello2024early}, \cite{cinnirella2025flow}, and \cite{delacroix2025flora} for the role of enlightened academies and societies in economic history.} Likewise, a university education would have endowed students with the skills to more easily access and apply formalized and complex knowledge ($A_1$ from section~\ref{sec:literature}). Moreover, engineers would have been able to access both theory and practice, helping them to combine ideas ($A_2$), and to apply theories with precise measurement tools and to set the research agenda through puzzling facts or practical needs ($B_2$, $B_3$ from section~\ref{sec:literature}).\footnote{See also \cite{depleijt2019}, \cite{mokyr2022wheels}, \cite{Maloney2022}, and \cite{hanlonengineer} for the role of the engineer in economic history.}

	Panel A of table~\ref{tab:mechanism} reports results for spillovers from propositional to prescriptive knowledge ($\Omega \rightarrow \lambda$) and panel B reports results for spillovers from prescriptive to propositional knowledge ($\lambda \rightarrow \Omega$). For panel A, we find largest effects for membership in the Royal Society and the Society of Arts, university enrollment and for academic careers. Being a fellow of the Royal Society is associated with an increase in authors' chances of receiving (incorporating) spillovers from propositional knowledge by 8.7\%. The coefficient for membership in the Society of Arts is even larger with a magnitude of 13.31\%. University enrollment at Oxford and Cambridge and being an academic are associated with a 4.4\% and 5.3\% increase respectively. The horse race regression in column~6 shows that the occupational channels mainly operated independently of each other.

	\begin{table}[!htbp]
		\centering
		\begin{adjustbox}{max width=0.85\columnwidth}
			\begin{threeparttable}\fontsize{9}{11}\selectfont
				\caption{Determinants of spillovers between propositional ($\Omega$) and prescriptive knowledge ($\lambda$)}
				\label{tab:mechanism}
				{
\def\sym#1{\ifmmode^{#1}\else\(^{#1}\)\fi}
\begin{tabular}{l*{6}{c}}
\hline\hline
                    &\multicolumn{6}{c}{Time frame: 1740--1779}                                                     \\\cline{2-7}
                    &\multicolumn{1}{c}{(1)}   &\multicolumn{1}{c}{(2)}   &\multicolumn{1}{c}{(3)}   &\multicolumn{1}{c}{(4)}   &\multicolumn{1}{c}{(5)}   &\multicolumn{1}{c}{(6)}   \\
\hline
\multicolumn{7}{l}{\textbf{Panel A: Spillovers from $\Omega$ to $\lambda$}} \\
Fellowship in Royal Society&      0.0834***&               &               &               &               &      0.0713***\\
                    &    (0.0222)   &               &               &               &               &    (0.0203)   \\
[0.5em]
Membership in Society of Arts&               &       0.125** &               &               &               &      0.0986*  \\
                    &               &    (0.0497)   &               &               &               &    (0.0502)   \\
[0.5em]
Engineer            &               &               &     -0.0180   &               &               &     -0.0211   \\
                    &               &               &    (0.0212)   &               &               &    (0.0226)   \\
[0.5em]
University enrollment&               &               &               &      0.0800***&               &      0.0434** \\
                    &               &               &               &    (0.0188)   &               &    (0.0183)   \\
[0.5em]
Academic career     &               &               &               &               &      0.0626***&      0.0519** \\
                    &               &               &               &               &    (0.0200)   &    (0.0199)   \\
[0.5em]
Word count controls &         Yes   &         Yes   &         Yes   &         Yes   &         Yes   &         Yes   \\
[0.5em]
Year fixed effects  &         Yes   &         Yes   &         Yes   &         Yes   &         Yes   &         Yes   \\
[0.5em]
Subject class fixed effects&         Yes   &         Yes   &         Yes   &         Yes   &         Yes   &         Yes   \\
\hline
Observations        &         985   &         985   &         985   &        1313   &         985   &         985   \\
R-squared           &        0.19   &        0.17   &        0.17   &        0.20   &        0.18   &        0.20   \\
\hline
\multicolumn{7}{l}{\textbf{Panel B: Spillovers from $\lambda$ to $\Omega$}} \\
Fellowship in Royal Society&      0.0354** &               &               &               &               &      0.0276*  \\
                    &    (0.0146)   &               &               &               &               &    (0.0147)   \\
[0.5em]
Membership in Society of Arts&               &       0.218***&               &               &               &       0.253***\\
                    &               &    (0.0117)   &               &               &               &    (0.0345)   \\
[0.5em]
Engineer            &               &               &    -0.00178   &               &               &     -0.0287   \\
                    &               &               &    (0.0350)   &               &               &    (0.0364)   \\
[0.5em]
University enrollment&               &               &               &      0.0295*  &               &      0.0152   \\
                    &               &               &               &    (0.0169)   &               &    (0.0165)   \\
[0.5em]
Academic career     &               &               &               &               &      0.0284** &      0.0203   \\
                    &               &               &               &               &    (0.0134)   &    (0.0127)   \\
[0.5em]
Word count controls &         Yes   &         Yes   &         Yes   &         Yes   &         Yes   &         Yes   \\
[0.5em]
Year fixed effects  &         Yes   &         Yes   &         Yes   &         Yes   &         Yes   &         Yes   \\
[0.5em]
Subject class fixed effects&         Yes   &         Yes   &         Yes   &         Yes   &         Yes   &         Yes   \\
\hline
Observations        &        1068   &        1068   &        1068   &        1358   &        1068   &        1068   \\
R-squared           &        0.35   &        0.35   &        0.35   &        0.33   &        0.35   &        0.35   \\
\hline\hline
\end{tabular}
}

				\begin{tablenotes}
					\item {\footnotesize \emph{Notes:} The table shows coefficients from estimating equation~\ref{eq:ols_spillover_mechanism} via OLS for 1740--1779. The dependent variable is the received spillover index from equation~\ref{eq:received_spillover}, for spillovers from propositional ($\Omega$) to prescriptive knowledge ($\lambda$) in panel A and from prescriptive ($\lambda$) to propositional knowledge ($\Omega$) in panel B. It is transformed using the natural logarithm. The main explanatory variables are a set of indicator variables capturing authors' membership, occupations, and education. Columns~1--5 present results for each indicator variable separately. Column~6 reports a horse race with all explanatory variables. The model further controls for the level and quadratic count of words in titles and patents and includes year and subject class fixed effects. Standard errors clustered by publication year. *** denotes statistical significance at the 1\% level, ** at the 5\% level, and * at the 10\% level.}
				\end{tablenotes}
			\end{threeparttable}
		\end{adjustbox}
	\end{table}

	The results point to the importance of both network connections and access to scholarly knowledge as facilitators of knowledge spillovers from propositional into prescriptive knowledge. We find that the Society of Arts \citep{Howes2020}, with its purpose of promoting useful knowledge, seems to have played a key role, alongside the Royal Society, Britain's most prestigious scientific society. We further document a positive and significant coefficient for engineers when extending the time frame to 1740--1800 in appendix section~\ref{sec:appendix_mechanism_robustness}.\footnote{Received spillovers are only defined backwards, in contrast to the innovation measure that is calculated by comparing the future and past. Hence, we can extend the time frame further than in section~\ref{sec:empirical_results} that estimates the association between innovation and spillovers.}

	Panel B of table~\ref{tab:mechanism} presents spillovers from prescriptive to propositional knowledge. Again, we find large coefficients for fellowship in the Royal Society and membership in the Society of Arts, university enrollment, and academic careers. However, in the horse race specification in column~6, only fellowship in the Royal Society and membership in the Society of Arts remain significant. This appears intuitive, since a university education or academic career would have been more important for the purpose of incorporating propositional knowledge than incorporating prescriptive, applied knowledge.

	Appendix section~\ref{sec:appendix_mechanism_robustness} further reports descriptive statistics for the overlap between occupational categories and the development of university enrollment among authors between 1600 and 1800. It shows a significant overlap between membership in Enlightenment societies, engineering, and university background. This reflects findings from \cite{Schofield1963}, \cite{Mokyr2002}, and \cite{Jacob1997,Jacob2014} who have highlighted the interconnectedness of enlightenment culture and applied science. Appendix section~\ref{sec:appendix_mechanism_robustness} also reports horse race results when excluding university enrollment and only focusing on career outcomes.

	\subsection{The scientific method}
	
	After having examined how upper-tail human capital helped facilitate knowledge spillovers, this section investigates whether there were complementary factors that were necessary to turn knowledge spillovers into innovation. Several scholars have argued that the scientific method \citep{Landes1969,Landes1998,Wootton2015}, the Newtonian Revolution \citep{Jacob1997,Jacob2014}, or precise measurement \citep{OGrada2016,Kelly2022a} were crucial for the creation of new knowledge during the Industrial Revolution. According to these accounts, such developments transformed the way knowledge was produced. Careful observation and measurement of nature, combined with the systematic testing of theories against empirical evidence, contributed to a more reliable knowledge base, the key to applying knowledge to innovation. 
	
	To test how these new methods aided the adoption of new knowledge in the innovation process, we interact the spillover measure from model~\ref{eq:ols_spillover} with embedding space similarity to these three revolutions in methods. Similar to the approach in \cite{Garg2018}  and \cite{ash2025ideas}, we capture each concept through a set of descriptive terms. For example, the scientific method is characterized through \textit{observation, measurement, mathematical formalization}, or \textit{experiment},  Newtonian mechanics is characterized through \textit{Newtonian mechanics, force, momentum}, or \textit{impulse}, and precise measurement is characterized through \textit{precise measurement, precise instruments, standardized scales} or \textit{calibration}. The full set of terms is documented in appendix table~\ref{tab:revolution_method_terms}. These terms are then projected into the \textit{SteamBERTh} embedding space. We then calculate the average cosine similarity between each title and each list of concept terms.
	
	To test whether these methods were complementary to knowledge spillovers in the innovation-generating process, we formulate the following model with $S_{ijt} := \text{Received spillover}_{B \to A}(v_{ijt})$ and $M_{ijt} := \text{Similarity to Method}_{ijt}$:
	
	\begin{equation}
		\label{eq:ols_method_complementarities_full}
		\text{Innovation}^{A}_{ijt}
		= \beta_{1} S_{ijt}
		+ \beta_{2} M_{ijt}
		+ \beta_{3} S_{ijt} M_{ijt}
		+ \beta_{4} \boldsymbol{X}'_{ijt}\boldsymbol{\zeta}
		+ \gamma_j{j} + \alpha_{t} + \varepsilon_{ijt}.
	\end{equation}
	
	The dependent variable is the innovation index from equation~\ref{eq:innovation}. Then, the main set of explanatory variables are the interaction terms between the spillover measure from equation~\ref{eq:received_spillover} and the similarity measures to the \textit{scientific method}, \textit{Newtonian mechanics}, and \textit{precise measurement}. As before, the innovation and spillover measures are transformed using the natural logarithm. To ease the interpretation of the coefficients, the similarity measures are standardized to z-scores. As before, the model also contains a vector of control variables, $\boldsymbol{X'_{ijt}}$ , incl. a title's level and quadratic word count as well as subject and year fixed effects, $\gamma_j$ and $\alpha_t$.

	\begin{table}[!htbp]
		\centering
		\begin{adjustbox}{max width=1\columnwidth}
			\begin{threeparttable}\fontsize{9}{11}\selectfont
				\caption{Complementary factors for spillovers between propositional ($\Omega$) and prescriptive knowledge ($\lambda$)}
				\label{tab:method_mechanism}
				{
\def\sym#1{\ifmmode^{#1}\else\(^{#1}\)\fi}
\begin{tabular}{l*{5}{c}}
\hline\hline
                    &\multicolumn{5}{c}{Time frame: 1740--1779}                                     \\\cline{2-6}
                    &\multicolumn{1}{c}{(1)}   &\multicolumn{1}{c}{(2)}   &\multicolumn{1}{c}{(3)}   &\multicolumn{1}{c}{(4)}   &\multicolumn{1}{c}{(5)}   \\
\hline
\multicolumn{6}{l}{\textbf{Panel A: Spillovers from $\Omega$ to $\lambda$}} \\
Spillover $\Omega \rightarrow \lambda$&       0.108***&       0.118***&       0.103***&       0.115***&      0.0948***\\
                    &    (0.0171)   &    (0.0200)   &    (0.0172)   &    (0.0183)   &    (0.0214)   \\
[0.5em]
Similarity to Newtonian mechanics terms&               &    0.000695   &               &               &     -0.0155   \\
                    &               &    (0.0109)   &               &               &    (0.0159)   \\
[0.5em]
Spillover $\Omega \rightarrow \lambda$ $\times$ Similarity to Newtonian mechanics terms&               &      0.0155   &               &               &     -0.0161   \\
                    &               &    (0.0138)   &               &               &    (0.0213)   \\
[0.5em]
Similarity to precise measurement terms&               &               &     0.00842   &               &      0.0126   \\
                    &               &               &   (0.00963)   &               &    (0.0143)   \\
[0.5em]
Spillover $\Omega \rightarrow \lambda$ $\times$ Similarity to precise measurement terms&               &               &      0.0291** &               &      0.0349*  \\
                    &               &               &    (0.0115)   &               &    (0.0180)   \\
[0.5em]
Similarity to scientific method terms&               &               &               &     0.00727   &     0.00470   \\
                    &               &               &               &    (0.0113)   &    (0.0216)   \\
[0.5em]
Spillover $\Omega \rightarrow \lambda$ $\times$ Similarity to scientific method terms&               &               &               &      0.0259*  &     0.00321   \\
                    &               &               &               &    (0.0137)   &    (0.0281)   \\
[0.5em]
Word count controls &         Yes   &         Yes   &         Yes   &         Yes   &         Yes   \\
[0.5em]
Year fixed effects  &         Yes   &         Yes   &         Yes   &         Yes   &         Yes   \\
[0.5em]
Subject class fixed effects&         Yes   &         Yes   &         Yes   &         Yes   &         Yes   \\
\hline
Observations        &        3047   &        3047   &        3047   &        3047   &        3047   \\
R-squared           &        0.20   &        0.20   &        0.21   &        0.21   &        0.21   \\
\hline
\multicolumn{6}{l}{\textbf{Panel B: Spillovers from $\lambda$ to $\Omega$}} \\
Spillover $\lambda \rightarrow \Omega $&       0.130***&      0.0676*  &      0.0836** &      0.0914** &      0.0656*  \\
                    &    (0.0328)   &    (0.0360)   &    (0.0350)   &    (0.0342)   &    (0.0376)   \\
[0.5em]
Similarity to Newtonian mechanics terms&               &      0.0258*  &               &               &      0.0308   \\
                    &               &    (0.0137)   &               &               &    (0.0203)   \\
[0.5em]
Spillover $\lambda \rightarrow \Omega $ $\times$ Similarity to Newtonian mechanics terms&               &      0.0581***&               &               &      0.0471*  \\
                    &               &    (0.0176)   &               &               &    (0.0259)   \\
[0.5em]
Similarity to precise measurement terms&               &               &      0.0109   &               &     -0.0567   \\
                    &               &               &    (0.0189)   &               &    (0.0418)   \\
[0.5em]
Spillover $\lambda \rightarrow \Omega $ $\times$ Similarity to precise measurement terms&               &               &      0.0484** &               &     -0.0353   \\
                    &               &               &    (0.0224)   &               &    (0.0526)   \\
[0.5em]
Similarity to scientific method terms&               &               &               &      0.0273   &      0.0476   \\
                    &               &               &               &    (0.0183)   &    (0.0475)   \\
[0.5em]
Spillover $\lambda \rightarrow \Omega $ $\times$ Similarity to scientific method terms&               &               &               &      0.0608***&      0.0497   \\
                    &               &               &               &    (0.0222)   &    (0.0608)   \\
[0.5em]
Word count controls &         Yes   &         Yes   &         Yes   &         Yes   &         Yes   \\
[0.5em]
Year fixed effects  &         Yes   &         Yes   &         Yes   &         Yes   &         Yes   \\
[0.5em]
Subject class fixed effects&         Yes   &         Yes   &         Yes   &         Yes   &         Yes   \\
\hline
Observations        &        1358   &        1358   &        1358   &        1358   &        1358   \\
R-squared           &        0.25   &        0.26   &        0.26   &        0.26   &        0.27   \\
\hline\hline
\end{tabular}
}

				\begin{tablenotes}
					\item {\footnotesize \emph{Notes:} The table shows coefficients from estimating equation~\ref{eq:ols_method_complementarities_full} via OLS. The dependent variable is the innovation index from equation~\ref{eq:innovation}. The main set of explanatory variables are the interaction terms between the spillover measure from equation~\ref{eq:received_spillover}, from $\Omega$ to $\lambda$ in panel A and from $\lambda$ to $\Omega$ in panel B, and the similarity measures to the \textit{scientific method}, \textit{Newtonian mechanics}, and \textit{precise measurement}. The innovation and spillover measures are transformed using the natural logarithm, similarity measures are transformed using z-scores. The model further controls for the level and quadratic count of words in titles and patents and includes year fixed effects. Standard errors clustered by publication year. *** denotes statistical significance at the 1\% level, ** at the 5\% level, and * at the 10\% level.}
				\end{tablenotes}
			\end{threeparttable}
		\end{adjustbox}
	\end{table}

	Panel A of table~\ref{tab:method_mechanism} reports results for spillovers from propositional to prescriptive knowledge and panel B reports results for spillovers from prescriptive to propositional knowledge. First, in panel A, we find a significant and positive coefficient for the interaction term between spillovers and similarity to \textit{precise measurement}. The coefficient also remains significant in the horse race between all similarity measures in column~5. This indicates that \textit{precise measurement} was a complementary factor to receiving knowledge spillovers - being able to use precision measurement in the implementation of technology made the adoption of propositional knowledge more productive ($A_3$ from section~\ref{sec:literature}) \citep[see also][]{OGrada2016,Kelly2022a}. The channels seem to be able to explain a relevant part of the original spillover effect, reducing the spillover coefficient by 12\% in the horse race specification in column~5. In contrast, \textit{Newtonian mechanics} and the \textit{ scientific method} seem to have been less important for the implementation of propositional knowledge into new techniques in $\lambda$.

	Next, panel B of table~\ref{tab:method_mechanism} reports interaction terms between spillovers from prescriptive to propositional knowledge. Here, we find that individually (column~1--5), all interaction terms are positive and significant. This indicates that all three methods, \textit{Newtonian mechanics}, \textit{precise measurement}, and the \textit{scientific method}, were important for integrating insights from $\lambda$ into the stock of knowledge in $\Omega$. Here it seems that Newtonian theory \citep[see][]{Jacob1997,Jacob2014} helped to integrate empirical observations into new productive theory. Likewise, the scientific method with its focus on experimentation as a means to reject false theories  seems to have been important ($B_1$ from section~\ref{sec:literature}) \citep[see also][]{Landes1998,Wootton2015}. We can think of the practical work on water wheels, discussed in section~\ref{sec:literature}, where engineers like John Smeaton conducted experimental work on water wheels that led to the rejection of predictions from theory, crucially helping to develop the theory of hydrodynamics. Altogether, all three channels reduce the spillover coefficient by 50\% in the horse race specification in column~5.

	Altogether, we find that for incorporating propositional into prescriptive knowledge, precise measurement seems to have played the largest role. In contrast, the process of incorporating prescriptive into propositional knowledge was strongly complemented by the scientific method and the Newtonian Revolution. This shows the complex operation of the feedback loop process, running through a range of channels  and resting on recent innovations in methods that originated from the Scientific Revolution.

	\FloatBarrier
	
	\section{Conclusion}
	\label{sec:conclusion}
	This paper has provided the first quantitative test of the feedback loop hypothesis proposed by \cite{Mokyr2002}, who has argued that the emergence of self-sustained modern growth was caused by a structural change in the relationship between propositional and prescriptive knowledge. Using two new innovation and spillover measures constructed from a semantically rich, historically fine-tuned BERT model, the paper has estimated the relationship between knowledge spillovers between propositional and prescriptive knowledge and innovation between 1600 and 1800 in England. The evidence presented shows that knowledge spillovers were negative or neutral before the eighteenth century but began to turn positive in the 1720s and were firmly positive by the 1760s. This timing fits well with the predictions from the Mokyrian account and supports the idea of a structural break in the knowledge economy during the Industrial Enlightenment.

	The paper has further shown that spillovers from propositional knowledge were positively associated with patent innovation and patent citations at the end of the eighteenth century. Thus, the paper documents that the positive knowledge spillovers also extended to the real economy. Moreover, it has shown that upper-tail human capital, especially for Royal Society fellows and members fo the Society of Arts, played an instrumental role in facilitating these spillovers. Additionally, revolutions in methods, such as \textit{Newtonian mechanics, precise measurement} and \textit{the scientific method} seem to have been complementary factors to knowledge spillovers. 
	
	Moreover, the paper has introduced a framework to estimate spillovers between different corpora of text that can be applied to a variety of other settings.  Overall, the new NLP measures of innovation and spillovers can help us gain a more structural understanding of the structure of either the historical or modern knowledge economy.

	All in all, the paper provides empirical support for the Mokyrian thesis. It presents evidence of the emergence of a positive relationship between knowledge spillovers and innovation at the end of the eighteenth century. It also presents evidence that upper-tail human capital played a facilitating role and has highlighted the importance of the revolution in methods produced as part of the Scientific Revolution. Altogether, these results provide crucial quantitative backing to \cite{Mokyr2002}'s account of the transition towards modern self-sustained growth --- and add to our understanding of the origin of modern economic growth.

	\small
	
	\bibliographystyle{aea}
	\bibliography{biblio}
	\inputencoding{utf8}
	
	\clearpage
	\begin{appendices}
		\renewcommand{\topfraction}{0.9}
		\renewcommand{\bottomfraction}{0.8}
		\renewcommand{\textfraction}{0.07}
		\renewcommand{\floatpagefraction}{0.7}
		\setcounter{topnumber}{4}
		\setcounter{bottomnumber}{3}
		\setcounter{totalnumber}{6}

		\section{Data}

		\subsection{\label{sec:variable_descriptions}Variable and source descriptions}
		
		\noindent \textbf{Text sources}
		
		\noindent\textbf{English Short Title Catalogue.} The English Short Title Catalogue (ESTC) kindly shared by the British Library with the author. The cleaned ESTC version is taken from \cite{Koschnick2025} and includes translated titles and has been cleaned for duplicates. Topics assignments are taken from \cite{Koschnick2025}. Topics are based on higher-order classes based on subjects assigned by expert librarians. \cite{Koschnick2025} further trained a BERT model on the labeled ESTC corpus to predict missing labels. Further descriptions and model evaluations are provided in \cite{Koschnick2025}. The ESTC catalogue includes 285,985 titles over the time period 1600--1800. These cover the universe of all printed works in England as well as all works printed in the English language elsewhere. \medskip
		
		\noindent\textbf{Patents.} Patent for the time period 1700--1851 are obtained from the \textit{Chronological Index of Patents Applied for and Patents Granted} compiled by Bennett Woodcroft in \citeyear{Woodcroft1854a}. They are merged with patent statistics from \cite{Nuvolari2011}. Overall, the patent dataset includes 12,722 patent short descriptions. \medskip

		\noindent \textbf{Variable descriptions}

		\noindent\textbf{Innovation index.}  
		The innovation index measures how much a title introduces new ideas relative to past publications. For title $v_{it_0}$, it is defined as:
		\[
		\mathrm{Innovation}_{it_0} = 
		\frac{f^{k}(v_{it_0},\, \mathcal{F}_{t_0,\tau})}{f^{k}(v_{it_0},\, \mathcal{B}_{t_0})},
		\]
		where $f^{k}(\cdot)$ is the mean cosine similarity between the title and the $k=20$ most similar titles of a pool, computed from embeddings of the fine-tuned \textit{SteamBERTh} model, $\mathcal{F}_{t_0,\tau}$ is the set of titles of the same field published in the following $\tau=20$ years, and $\mathcal{B}_{t_0}$ the set of all earlier titles of the field. Titles with $\mathrm{Innovation}_{it_0}>1$ are more similar to future than to past texts and hence classified as innovative.  \medskip
		
		\noindent\textbf{Received spillovers.}  
		The received spillover index quantifies the extent to which a title $v_{it}$ in field $B$ draws on prior work from field $A$:
		\begin{align*}
			\text{Received spillover}_{A \to B}(v_{it})
			&:= \frac{
				f^{k}\bigl(v_{it},\mathcal{B}^{A}_{t,\tau}\bigr)
			}{
				\displaystyle
				f^{k}\bigl(\mathcal{P}^{k}_{it},\mathcal{B}^{A}_{t,\tau}\bigr)
			}.
		\end{align*}
		where $\mathcal{B}^{A}_{t,\tau}$ is the set of titles of $A$ published in the previous $\tau=20$ years, and $\mathcal{P}^{k}_{it}$ are the peers of $v_{it}$: the $k=20$ titles of its own field $B$ published in the previous $\tau=20$ years that are most similar to $v_{it}$. In the denominator, $f^{k}$ is the mean of the $k$ largest similarities between any peer and the titles of $\mathcal{B}^{A}_{t,\tau}$. Values above one indicate that $v_{it}$ is closer to prior work in $A$ than its peers in its own field were. \medskip
		
		\noindent\textbf{Received spillovers from propositional knowledge.}  Received spillovers are calculated using the above formula from equation~\ref{eq:received_spillover}. For each title in $v_{it}$ in prescriptive knowledge, spillovers are calculated to the backward pool, $\mathcal{B}^{A}_{t,\tau}$, of $\mathcal{P}:=$\textit{applied physics, astronomy, mathematics, chemistry}, and \textit{encyclopedias}. Spillovers from propositional knowledge are then averaged as	$\frac{1}{|\mathcal{P}|}\sum_{A\in\mathcal{P}}	\mathrm{Received\ spillover}_{A \to B}(v_{it})$. \medskip

		\noindent\textbf{Received spillovers from prescriptive knowledge.}   Received spillovers are calculated using the above formula from equation~\ref{eq:received_spillover}. For each title $v_{it}$ in propositional knowledge, spillovers are calculated to the backward pool, $\mathcal{B}^{A}_{t,\tau}$, of $\mathcal{P}:=$\textit{technology in trades, technology in agriculture, navigation}, and \textit{scientific instruments}. Spillovers from prescriptive knowledge are then averaged as	$\frac{1}{|\mathcal{P}|}\sum_{A\in\mathcal{P}}	\mathrm{Received\ spillover}_{A \to B}(v_{it})$.  \medskip

		\noindent\textbf{Patent citations.} Woodcroft patent citations obtained from \cite{Nuvolari2011}. \medskip

		\noindent\textbf{Number of ESTC editions.} The ESTC lists individual editions per work. To identify the number of editions per work, the paper uses the same duplicate-clearing Jaccard distance based strategy from \cite{Koschnick2025}. Note that duplicates are excluded from the corpus throughout this paper and not used for the calculation for the innovation index. \medskip

		\noindent\textbf{Occupations of ESTC authors.} Occupational information derived from authorship details in ESTC titles. In a first step, information on authors was identified using a regex-routine. Overall, we identify author information for 47\% of all titles within prescriptive and propositional knowledge. Following this, post-comma additional information on authors (beyond their names) was extracted. This covers 32\% of all authors. Based on these strings, we identified the following groups: \textit{engineer, medical career, academic career, fellow of the Royal Society}, and \textit{membership in other enlightenment societies}. \textit{Fellow of the Royal Society} includes all mentions of \textit{F.R.S} or \textit{fellow of the Royal Society}. \textit{Member in the Society of Arts} includes all mentions of Society for the \textit{Society for the Encouragement of Arts, Manufactures and Commerce}, either in the form of \textit{Society of Arts}, \textit{Society for the Encouragement of ...}, or \textit{Society for Promoting Arts}. \textit{Engineer} includes all mentions of \textit{engineers, surveyors, millwrights, shipwrights, and instrument makers}. \textit{Academic} includes all mentions of \textit{Teacher, librarian, reader of, lecturer, professor, M.A., Fellow of...}.\footnote{An extra rule ensures that fellows at colleges are separated from fellows of the Royal Society. The approach further accounts for different capitalization} Finally, going beyond the title information, authors are matched to authors who, at some point, were enrolled at the universities of Oxford and Cambridge based on \cite{Koschnick2025}. \medskip
		
		\noindent\textbf{Similarity to new methods.} Similarities to the three methodological revolutions of \textit{Newtonian mechanics, precise measurement,} and the \textit{scientific method}. Each concept is captured through a list of essential terms that are presented in table~\ref{tab:revolution_method_terms}. Similarities are then calculated at the title level and averaged accross each term. 
		
		\subsection{\label{sec:sources}Training corpus}
		
		\FloatBarrier

		\begin{table}[!htbp]
			\centering
			\begin{adjustbox}{max width=1\columnwidth}
				\begin{threeparttable}\footnotesize
					\caption{Summary statistics for training corpus}
					\label{tab:summary_training_corpus}
					\label{tab:summary_training_corpus_by_group}
					\begin{tabular}{lrrr@{\hspace{2em}}lrrr}
\toprule
\multicolumn{4}{l}{\textit{Patents (1700--1850), by industry}} & \multicolumn{4}{l}{\textit{ESTC titles (1600--1800), by subject class}} \\
\cmidrule(r{2em}){1-4}\cmidrule{5-8}
 & Publications & Avg. words & Avg. characters & & Publications & Avg. words & Avg. characters \\
\midrule
Agriculture & 418 & 16.8 & 99.9 & Applied physics & 822 & 42.1 & 264.0 \\
Carriages & 789 & 17.4 & 104.9 & Astronomy & 907 & 56.5 & 344.8 \\
Chemicals & 1{,}089 & 18.1 & 110.5 & Chemistry & 370 & 43.2 & 269.5 \\
Clothing & 320 & 18.4 & 113.1 & Encyclopedias and dictionaries & 1{,}352 & 76.4 & 480.3 \\
Construction & 633 & 19.5 & 118.8 & Mathematics & 1{,}345 & 56.0 & 355.1 \\
Engines & 1{,}577 & 17.0 & 101.8 & Navigation & 964 & 53.7 & 329.8 \\
Food & 698 & 16.9 & 102.0 & Scientific instruments & 367 & 46.4 & 288.6 \\
Furniture & 642 & 16.0 & 97.1 & Technical instructions Agriculture & 1{,}288 & 54.4 & 335.1 \\
Glass & 120 & 17.3 & 105.0 & Technical instructions Trades & 1{,}632 & 52.5 & 321.9 \\
Hardware & 817 & 16.4 & 96.7 &  &  &  &  \\
Instruments & 582 & 18.6 & 114.4 &  &  &  &  \\
Leather & 214 & 17.3 & 102.5 &  &  &  &  \\
Manufacturing & 664 & 15.2 & 90.7 &  &  &  &  \\
Medicines & 281 & 20.0 & 123.1 &  &  &  &  \\
Metallurgy & 662 & 17.1 & 100.7 &  &  &  &  \\
Military & 242 & 17.5 & 101.3 &  &  &  &  \\
Mining & 76 & 20.2 & 116.4 &  &  &  &  \\
Paper & 465 & 14.9 & 90.8 &  &  &  &  \\
Pottery & 272 & 20.7 & 125.8 &  &  &  &  \\
Ships & 563 & 19.1 & 114.1 &  &  &  &  \\
Textiles & 1{,}598 & 17.5 & 107.4 &  &  &  &  \\
\midrule
All patents & 12{,}722 & 17.4 & 105.1 & All ESTC titles & 9{,}047 & 55.8 & 346.8 \\
\midrule
 &  &  &  & Combined corpus & 21{,}769 & 33.4 & 205.6 \\
\bottomrule
\end{tabular}

					\begin{tablenotes}
						\item {\footnotesize \emph{Notes:} The table reports summary statistics for the corpus used for fine-tuning the MacBERTh base model, by patent industry and ESTC subject class. Publications: number of publications; Avg. words and Avg. characters: average length in words and characters.}
					\end{tablenotes}
				\end{threeparttable}
			\end{adjustbox}
		\end{table}

		\begin{table}[!htbp]
			\centering
			\begin{adjustbox}{max width=1\columnwidth}
				\begin{threeparttable}\footnotesize
					\caption{Summary statistics of the variables for ESTC titles and patents}
					\label{tab:descriptive_estc}
					\label{tab:descriptive_patents}
					\begin{tabular}{lrrrrr}
\toprule
 & Mean & Std. dev. & Min & Max & Obs. \\
\midrule
\multicolumn{6}{l}{\textbf{Panel A: ESTC titles in $\Omega$ and $\lambda$} (N = 9{,}024)} \\
Innovation index & 0.90 & 0.31 & $-$2.17 & 12.41 & 5{,}835 \\
Spillover $\Omega \rightarrow \lambda$ & $-$0.77 & 0.27 & $-$2.95 & 2.02 & 4{,}224 \\
Spillover $\lambda \rightarrow \Omega $ & $-$0.83 & 0.42 & $-$5.63 & 4.41 & 4{,}757 \\
Fellowship in Royal Society & 0.11 & 0.31 & 0.00 & 1.00 & 7{,}006 \\
Medical career & 0.07 & 0.26 & 0.00 & 1.00 & 7{,}006 \\
Engineer & 0.04 & 0.19 & 0.00 & 1.00 & 7{,}006 \\
Academic career & 0.19 & 0.39 & 0.00 & 1.00 & 7{,}006 \\
Word count & 55.86 & 49.21 & 1.00 & 527.00 & 9{,}024 \\
\midrule
\multicolumn{6}{l}{\textbf{Panel B: Patents} (N = 11{,}755)} \\
Innovation index & 1.04 & 0.31 & $-$20.11 & 6.14 & 7{,}704 \\
Patent citations & 1.78 & 1.19 & 1.00 & 21.00 & 2{,}027 \\
Spillover $\Omega \rightarrow \lambda$ & $-$0.86 & 0.40 & $-$7.52 & $-$0.10 & 3{,}828 \\
Spillover $\lambda \rightarrow \Omega $ & $-$0.82 & 0.38 & $-$7.45 & $-$0.11 & 3{,}836 \\
Word count & 18.61 & 18.02 & 3.00 & 472.00 & 11{,}755 \\
\bottomrule
\end{tabular}

					\begin{tablenotes}
						\item {\footnotesize \emph{Notes:} The table reports summary statistics for titles in propositional ($\Omega$) and prescriptive knowledge ($\lambda$) in the English Short Title Catalogue (ESTC, panel A) and for patents (panel B). The innovation index is by definition only measured for the period 1620--1780. The spillover index $\Omega \rightarrow \lambda$ is only defined for subset $\lambda$ and the spillover index $\lambda \rightarrow \Omega$ only for subset $\Omega$. Patent citations obtained from \cite{Nuvolari2011}.}
					\end{tablenotes}
				\end{threeparttable}
			\end{adjustbox}
		\end{table}

		\begin{table}[!htbp]
			\centering
			\begin{adjustbox}{max width=1\columnwidth}
				\begin{threeparttable}\footnotesize
					\caption{List of all subject classes in the ESTC}
					\label{tab:all_subject_classes}
					\begin{tabular}{p{12cm}}
\toprule
Subject classes \\
\midrule
Administrative, Alchemy, Almanacs, Amusements, Antiquities, Applied physics, Architecture, Art, Astrology, Astronomy, Biography, Biology, Chemistry, Church administration, Classical education, Curiosities and wonders, Drama, Economic societies, Economics, Education, Encyclopedias and dictionaries, Exploration, Foreign languages, Geography, History, Legal, Logics and rhetorics at university, Mathematics, Medicine, Mercantile, Military, Military Wars, Moral tales, Music, Navigation, Philosophy, Poetry, Political philosophy, Printing and book trades, Prophecies, Religious, Religious Catholicism, Religious Judaism, Religious Sects, Religious Sermons, Scientific instruments, Societies, State affairs, Stories, Supernatural, Technical instructions Agriculture, Technical instructions Trades, Travel descriptions,  University matters \\
\bottomrule
\end{tabular}

					\begin{tablenotes}
						\item {\footnotesize \emph{Notes:} Subject classes predicted with the model from \cite{Koschnick2025} to align with ESTC subject classes.}
					\end{tablenotes}
				\end{threeparttable}
			\end{adjustbox}
		\end{table}

		\FloatBarrier

		\begin{figure}[!htbp]
			\centering
			\begin{minipage}[t]{0.48\textwidth}
				\centering
				\includegraphics[width=\linewidth]{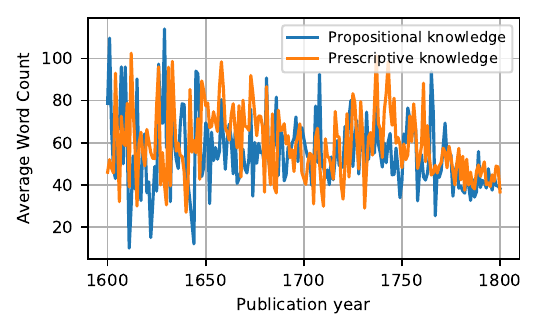}
				\caption{Average of word-count of ESTC titles over time \vspace{4pt} \newline \footnotesize \emph{Notes:} Titles classified as in main text.}
				\label{fig:estc_words_over_time}
			\end{minipage}\hfill
			\begin{minipage}[t]{0.48\textwidth}
				\centering
				\includegraphics[width=\linewidth]{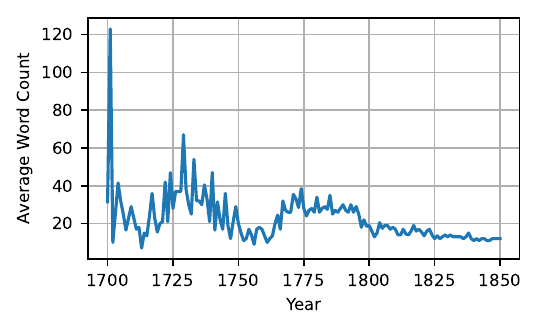}
				\caption{Average of word-count of short patent descriptions over time \vspace{4pt} \newline \footnotesize \emph{ Notes:} Patent descriptions taken from \cite{Woodcroft1854a}.}
				\label{fig:patent_words_over_time}
			\end{minipage}
		\end{figure}

		\begin{figure}[!htbp]
			\centering    
			\includegraphics[width=0.9\textwidth]{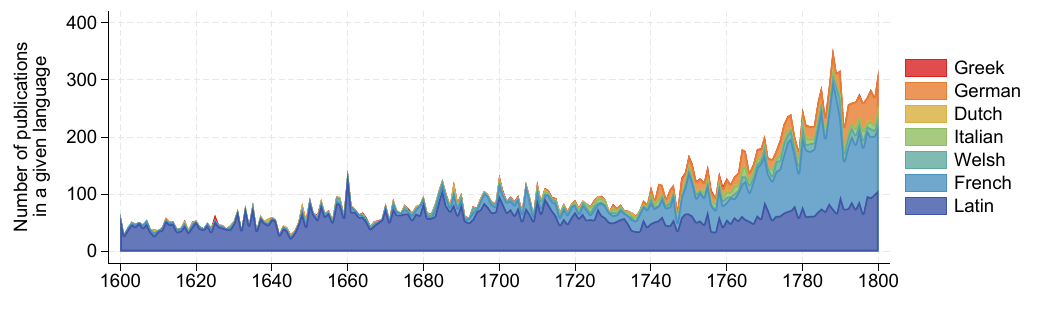}
			\caption{ESTC titles published in foreign languages over time \vspace{4pt} \newline \footnotesize \emph{ Notes:} Number of titles in the English Short Title Catalogue published in non-English languages. Figure reproduced from \cite{Koschnick2025}. In \cite{Koschnick2025}, languages are identified using fasttext and GoogleTranslate.}
			\label{fig:languages}
		\end{figure}

		\section{Time trends}
		\label{sec:timetrends}
		
		\begin{figure}[!htbp]
			\centering
			\begin{subfigure}[t]{0.32\textwidth}
				\includegraphics[width=\linewidth]{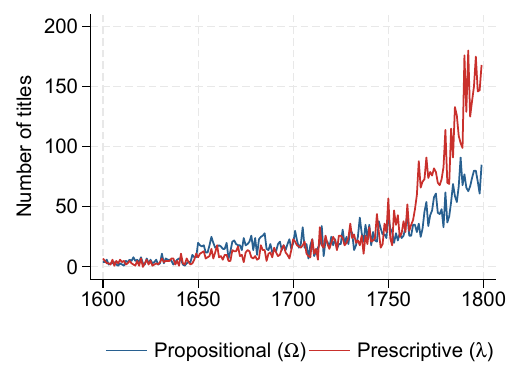}
				\subcaption{Titles in $\Omega$ and $\lambda$}
				\label{fig:prop_and_prescr}
			\end{subfigure}
			\begin{subfigure}[t]{0.32\textwidth}
				\includegraphics[width=\linewidth]{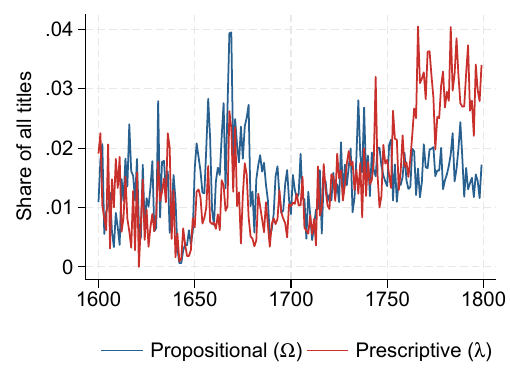}
				\subcaption{Share of $\Omega$ and $\lambda$}
				\label{fig:prop_and_prescr_share}
			\end{subfigure}
			\begin{subfigure}[t]{0.32\textwidth}
				\includegraphics[width=\linewidth]{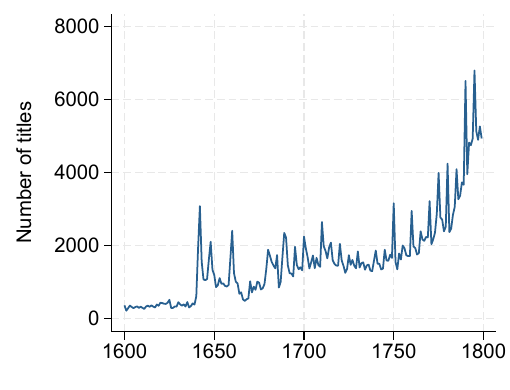}
				\subcaption{All English publications}
				\label{fig:all_estc}
			\end{subfigure}
			\caption{Propositional ($\Omega$) and prescriptive knowledge ($\lambda$) and all English publications over time \vspace{4pt} \newline \footnotesize \emph{Notes:} Panel a) plots the number of titles in propositional and prescriptive knowledge, panel b) their share of all published ESTC titles and patents, and panel c) the number of all English publications, composed of all printed titles listed in the ESTC as well as patents. Propositional and prescriptive knowledge defined as in main text.}
			\label{fig:time_trends}
		\end{figure}

		This section presents time trends for propositional and prescriptive knowledge. Figure~\ref{fig:prop_and_prescr} presents basic trends. Notably, both measures closely track the timing of the Industrial Revolution \citep{Clark2007,Allen2009,bouscasse2025did}. We can see that the by middle of the eighteenth century, the growth rate in both $\Omega$ and $\lambda$ increased markedly. Note that the increase in the growth rate is especially pronounced for $\lambda$. However, even before that, propositional and prescriptive knowledge started to slowly grow post 1700. For comparison, figure~\ref{fig:all_estc} plots the number of all English publications over time. Appendix figure~\ref{fig:prop_and_prescr_share} further reports the share of propositional and prescriptive knowledge over time. It shows that the share of prescriptive knowledge in Britain's knowledge production began to continuously increase from the early eighteenth century.

		Lastly, there are some salient patterns in the seventeenth century that need to be further discussed. First, we see a spike in all English publications during the period of the English Civil War, 1642--1649, and the interregnum 1649--1660.\footnote{On this, see \citet[pp. 163--165]{Raymond2003}.} We can also notice a jump in propositional publications post 1649. The jump likely reflects the beginning of the English Scientific Revolution in the 1650s that witnessed the invisible college as the origin of the Royal Society as well as the experimental philosophical club at the University of Oxford.\footnote{For further studies on science at the universities, see \cite{Koschnick2025}. \cite{Merton1938} argued that the Parliamentarian victory and Puritanism lay at the origin of the beginning of the Scientific Revolution in England. However, this has been controversially discussed in the historical literature \cite{Cohen1994}.} Additionally, the Civil War, 1642--1649 might have either suppressed or delayed scientific publications. 
		
		Therefore, given a) the distortionary effect of the Civil War, and b) the very low number of publications pre 1650, we decided to omit all pre-1650 publications from the main analyses in section~\ref{sec:results}. Section~\ref{sec:results} reports robustness to including this period.

		\FloatBarrier

		\FloatBarrier

		\section{Validation: Titles as a proxy for book content}
		\label{sec:title_full_text_sim}

		This section establishes that titles are actually good predictors of the full text of ESTC documents. For this, we draw on Early English Books Online that contains full-text corpora for seventeenth century English books. Based on EEBO and ESTC identifiers, we match titles in prescriptive and propositional knowledge, yielding a match rate of 30.34\%. To establish whether titles are good predictors of actual document content, we calculate a matrix of textual similarities between all titles and full text documents and estimate whether titles are good predictors of the similarity to their corresponding full text document. 
		
		\begin{table}[!htbp]
		\centering
			\begin{adjustbox}{max width=1\columnwidth}
				\begin{threeparttable}\fontsize{10}{13}\selectfont
					\centering
					\caption{Titles are a good predictor of similarity to full text documents}
					\label{tab:title_validation}
					{
\def\sym#1{\ifmmode^{#1}\else\(^{#1}\)\fi}
\begin{tabular}{l*{3}{c}}
\hline\hline
                    &\multicolumn{3}{c}{Cosine similarity, title and full text}\\\cline{2-4}
                    &\multicolumn{1}{c}{(1)}   &\multicolumn{1}{c}{(2)}   &\multicolumn{1}{c}{(3)}   \\
                    &        Sim.   &        Sim.   &        Sim.   \\
\hline
Title corresponds to full text&       2.291***&       2.295***&       2.323***\\
                    &    (0.0752)   &    (0.0751)   &    (0.0756)   \\
[1em]
ESTC subject fixed effects&          No   &         Yes   &         Yes   \\
[1em]
EEBO subject fixed effects&          No   &         Yes   &         Yes   \\
[1em]
ESTC year fixed effects&          No   &          No   &         Yes   \\
[1em]
EEBO year fixed effects&          No   &          No   &         Yes   \\
\hline
Observations        &       41807   &       41807   &       41807   \\
R-squared           &       0.017   &       0.060   &       0.204   \\
\hline\hline
\end{tabular}
}

					\begin{tablenotes}
						\item {\footnotesize \emph{Notes:} The table shows coefficients from estimating equation~\ref{eq:validate_titles} via OLS for the sample of 214 EEBO titles matched to ESTC titles within seventeenth century and subject class subset of prescriptive and propositional knowledge. The sample consists of all possible ESTC and EEBO title to full-text combinations. The dependent variable is tf--idf cosine similarity between titles and full text documents. It is transformed using the natural logarithm. The main explanatory variable captures whether a title corresponds to the full text. Standard errors two way clustered at the ESTC title and EEBO full text level. *** denotes statistical significance at the 1\% level, ** at the 5\% level, and * at the 10\% level.}
					\end{tablenotes}
				\end{threeparttable}
			\end{adjustbox}
		\end{table}

		To calculate similarity between titles and full documents, the paper uses a tf--idf (term frequency–inverse document frequency) representation that is well suited for comparisons between short texts (titles) and long documents (full texts).	Let $x$ denote a document (either a ESTC title or an EEBO full text) and let $w$ index terms in the corpus vocabulary. Each document is represented by a tf--idf vector with elements
		\[
		\text{tfidf}(w,x) = \text{tf}(w,x)\cdot \text{idf}(w),
		\]
		where $\text{tf}(w,x)$ is the term frequency of $w$ in document $x$, and
		\[
		\text{idf}(w) = \log\!\left(\frac{N}{\text{df}(w)}\right)
		\]
		downweights terms that frequently occur across the corpus. All vectors are $\ell_2$-normalized to ensure scale comparability between short and long texts.
		
		Textual similarity between ESTC title $i$ and full text document $j$ is then computed using cosine similarity
		\[
		\text{CosSim}_{ij}
		=
		\frac{\mathbf{v}_i \cdot \mathbf{v}_j}
		{\|\mathbf{v}_i\|_2 \, \|\mathbf{v}_j\|_2},
		\]
		where $\mathbf{v}_i$ and $\mathbf{v}_j$ denote the normalized TF--IDF vectors. Cosine similarity measures the alignment of term-weight distributions and ranges from 0 (no shared weighted terms) to 1 (identical term composition).

		To estimate whether titles are good predictors of the full text of a work, we estimate the following model:
		\begin{equation}
			\text{CosSim}_{ij} = \text{Same work}_{ij} + \zeta_i + \eta_j + \theta_i + \vartheta_j + \varepsilon_{ij} 
			\label{eq:validate_titles}
		\end{equation}
		where the unit of analysis are all possible combinations between ESTC title $i$ and EEBO full-text document $j$. The dependent variable, $\text{CosSim}_{ij}$ captures the similarity between ESTC title $i$ and full text document $j$ calculated using the above tf--idf approach. It is transformed using the natural logarithm. The main explanatory variable, $\text{Same work}_{ij}$, is an indicator variable that captures if the title and full-text data belong to the same work. Intuitively, we test whether information from a title can act as a predictor of full-text content information. $\zeta_i$, $\eta_j$ denote subject class- and $\theta_i$, $\vartheta_j$ year-fixed effects at the ESTC, $i$, and EEBO, $j$, level.

		Results are presented in table~\ref{tab:title_validation}. We find that titles are a very strong predictor of full text content. Increasing cosine similarity between titles and full text by 1\%, increases the chances of the title corresponding to the full text by ca. 2.3\%. Results are robust to including EEBO- and ESTC-level subject and year fixed effects.

		\FloatBarrier

		\section{Fine-tuning procedure for historical BERT model}
		
		\label{sec:app:fine-tuning}
		
		For fine-tuning, the paper uses a SimCSE approach. Intuitively, SimCSE works by treating each sentence as its own training signal: The model sees two copies of the same sentence and learns that they should be embedded close together. At the same time, all other sentences in the batch act as contrasts, so the model learns to push their embeddings farther apart. This makes the model learn a stable embedding space where sentences with similar meaning naturally cluster, even without labels.
		
		For fine-tuning, the paper proceed as follows. First, the training sample is constructed by combining titles in scientific and technical classes from table~\ref{tab:summary_training_corpus_by_group} with all patents. Next, patent descriptions with fewer than three words are dropped. The paper then performs an unsupervised SimCSE on the corpus of text. 
		
		The model is trained for two epochs with a multiple negatives ranking loss, a batch size of 32, a learning rate of $2\times10^{-5}$, and weight decay $0.01$. First, all random seeds are fixed, non-deterministic CUDA kernels disabled, and the full software environment (including \texttt{torch}, \texttt{ransformers}, and \texttt{sentence-transformers} versions) are logged to ensure exact replicability across runs. The resulting fine-tuned model, \textit{SteamBERTh}, provides domain-adapted embeddings for eighteenth-century scientific and technical language and is used throughout the paper to measure semantic similarity, and to construct spillover measures.
		The next appendix sections presents fine-tuning evaluative statistics.

		\section{Fine-tuning evaluation of \textit{SteamBERTh}}
		
		\label{sec:fine-tuning-evaluation}
		
		The paper employs a \textit{simple contrastive learning of sentence embeddings} approach (SimCSE) \citep{gao2021simcse} to fine-tune the \textit{MacBERTh} model on a set of exemplary texts. The training data covers the universe of short patent descriptions from \cite{Woodcroft1854a} and the universe of titles within \textit{applied physics, astronomy, chemistry, mathematics} and \textit{technical publications in trades and agriculture, navigation, and scientific instruments} from the ESTC. Table~\ref{tab:summary_training_corpus} provides an overview of the distribution of titles within these subsets. 
		SimCSE-style fine-tuning places semantically equivalent sentences close together while pushing unrelated sentences apart. It does so by treating two representations of the same sentence as a positive pair and uses other sentences in the batch as negatives, thereby producing a contrastive learning signal. For fine-tuning the model we choose a conservative set of hyperparameters, including an inverse temperature scale of 15, 2 epochs, a learning rate of $2e^{-5}$, a weight decay of 0.01, and warmup for 10\% of total steps.
		
		Following the fine-tuning operation, we evaluate the performance of the fine-tuned \textit{SteamBERTh} in comparison to the base \textit{MacBERTh} model by relying on descriptive statistics of the embedding space.	
		First, we present indicators of model anisotropy/isotropy. table \ref{tab:evaluative_statistics} reports \textit{mean nearest neighbor consine similarity}, \textit{principal component variance}, and \textit{rank overlap} of the embedding space of the training corpus for both the base \textit{MacBERTh} and \textit{SteamBERTh}. We can see that the embedding space of the base model shows severe signs of anisotropy, while the fine-tuned model is significantly more isotropic. Moreover, the k-rank comparisons show that the embedding spaces from both models are highly different.

\begin{figure}[!htbp]
			\centering
			\begin{minipage}{\textwidth}
				\begin{subfigure}[t]{.48\textwidth}
					\includegraphics[width=1\textwidth]{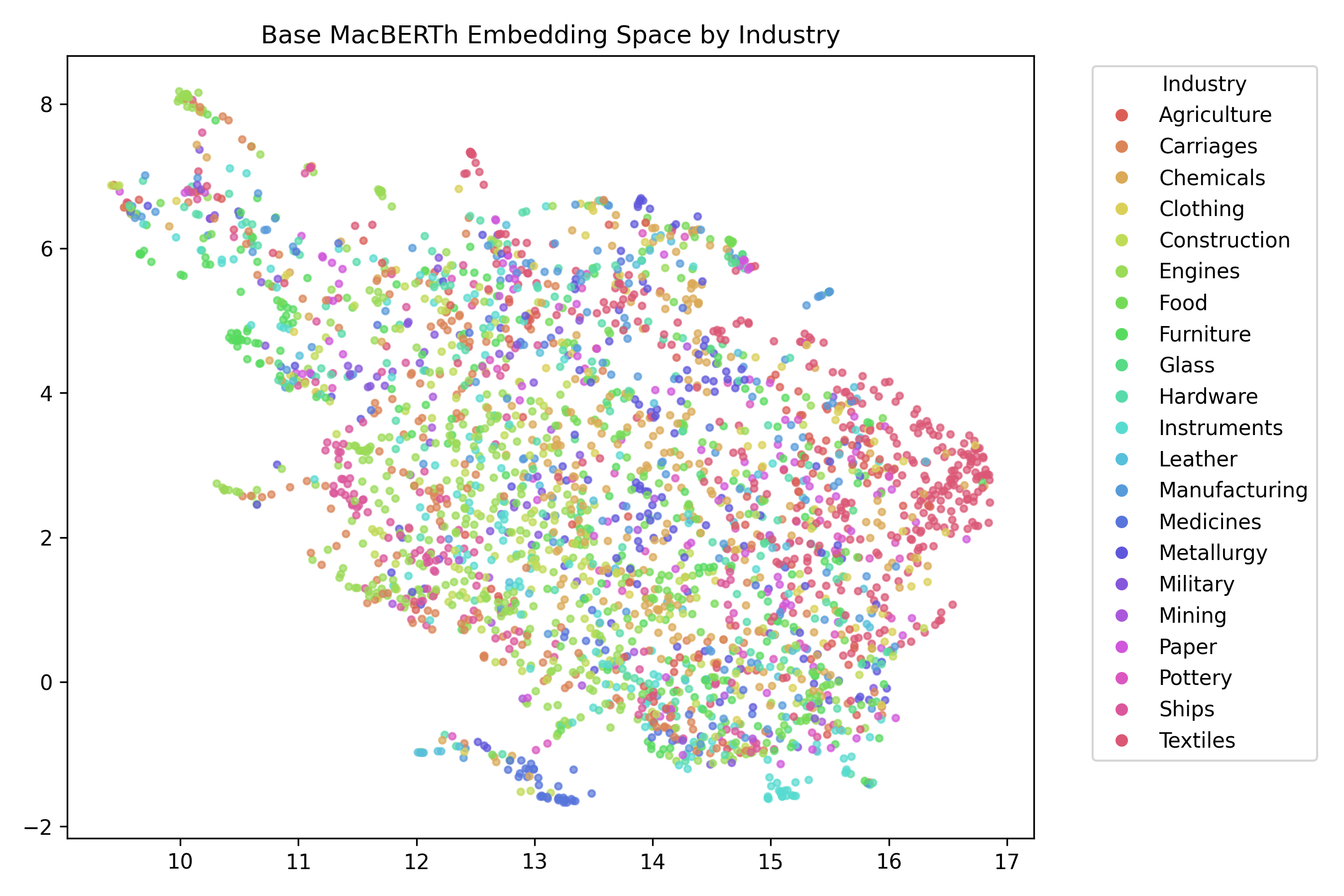}
					\subcaption{MacBERTh base model}
				\end{subfigure}
				\hfill
				\begin{subfigure}[t]{.48\textwidth}
					\includegraphics[width=1\textwidth]{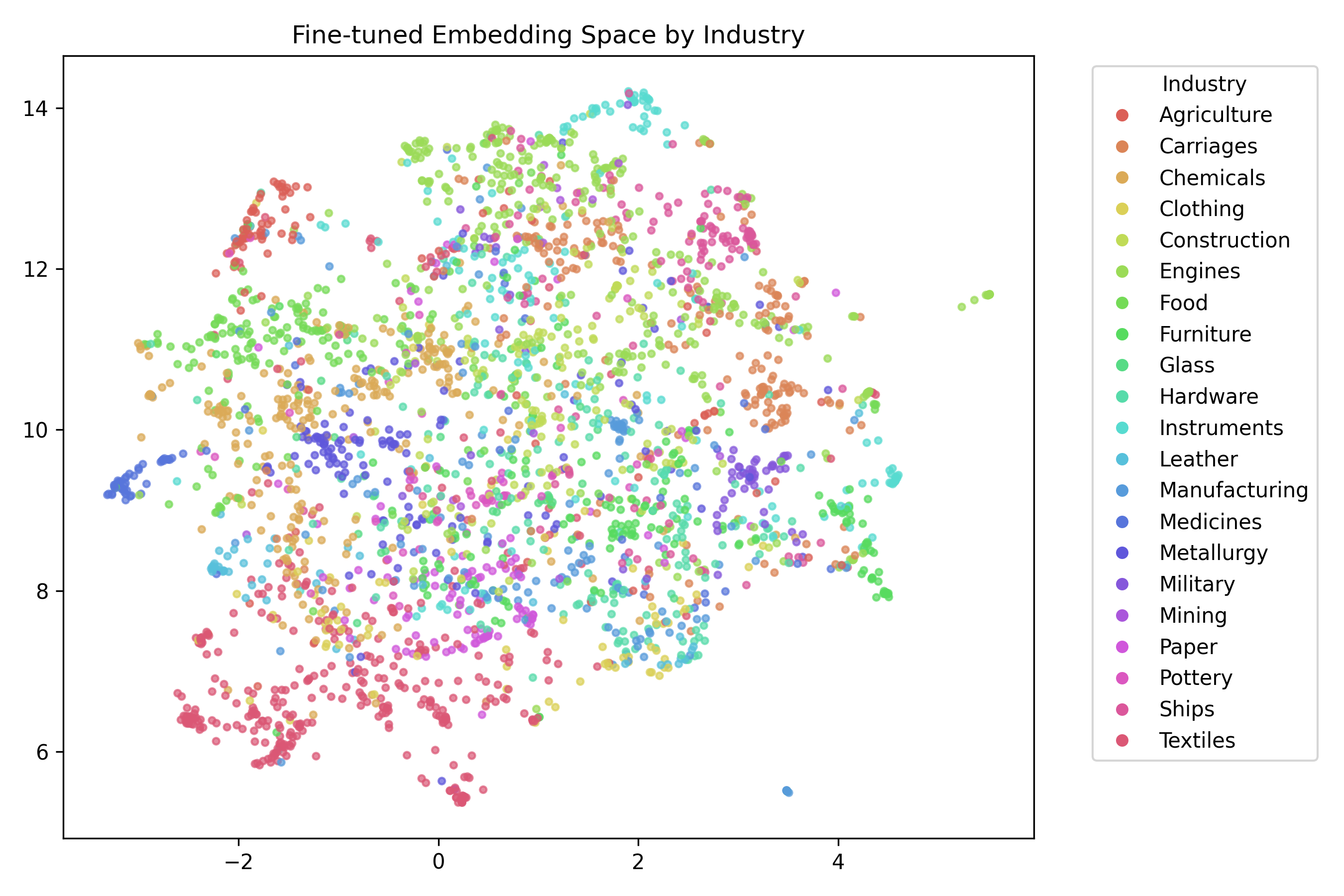}
					\subcaption{Fine-tuned model}
				\end{subfigure}
				\caption{Patents UMAP embeddings space projection by industry \vspace{4pt} \newline \footnotesize \emph{Notes:}   The figure compares a UMAP (Uniform Manifold Approximation and Projection) of the embedding space for patents between the MacBERTh base model and the fine-tuned SteamBERTh model. Different colors indicate industry classes from \cite{Nuvolari2011}. }
				\label{fig:embedding_umap_patent}
			\end{minipage}
			
			\vspace{1em}
			
			\begin{minipage}{\textwidth}
				\begin{subfigure}[t]{.48\textwidth}
					\includegraphics[width=1\textwidth]{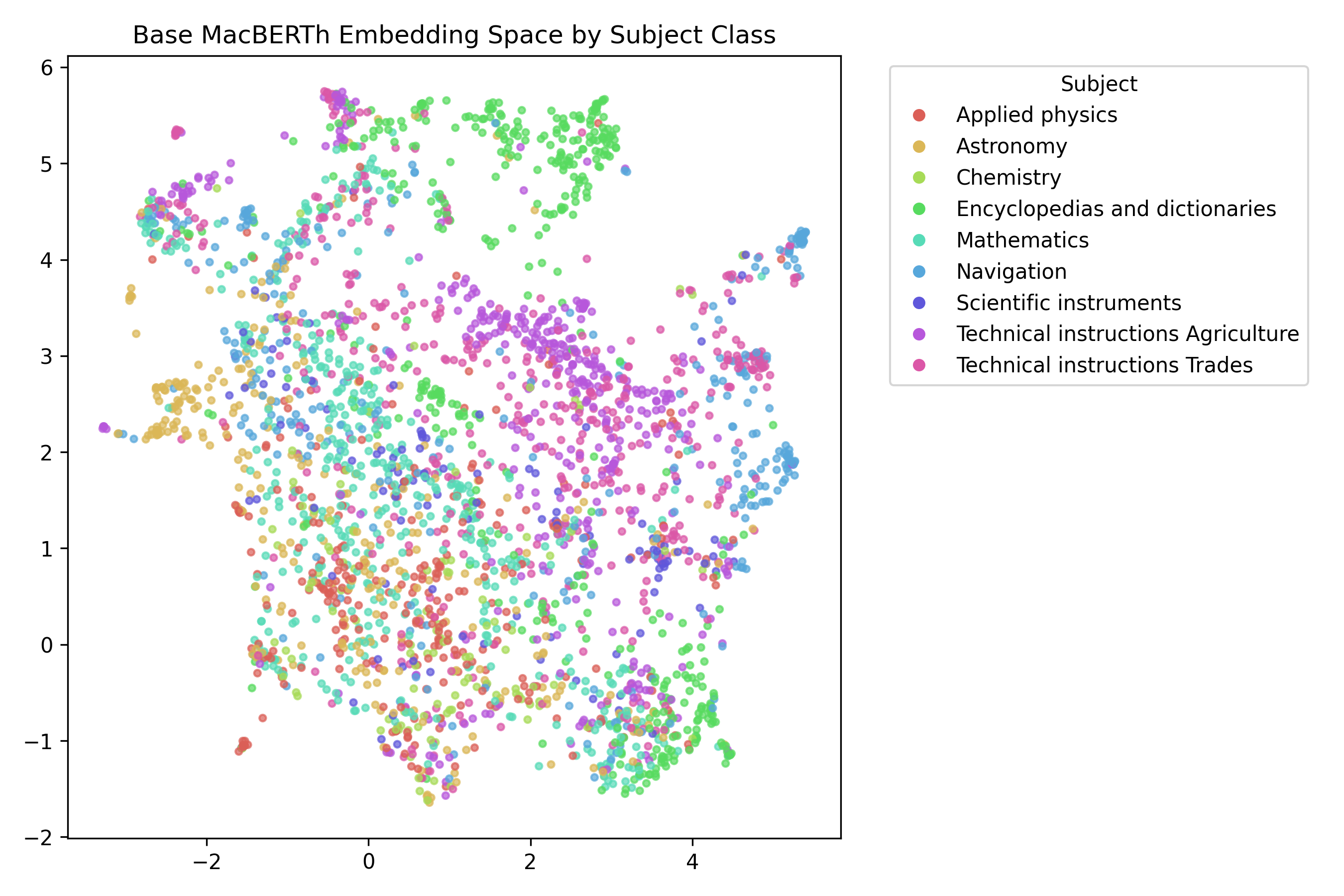}
					\subcaption{MacBERTh base model}
				\end{subfigure}
				\hfill
				\begin{subfigure}[t]{.48\textwidth}
					\includegraphics[width=1\textwidth]{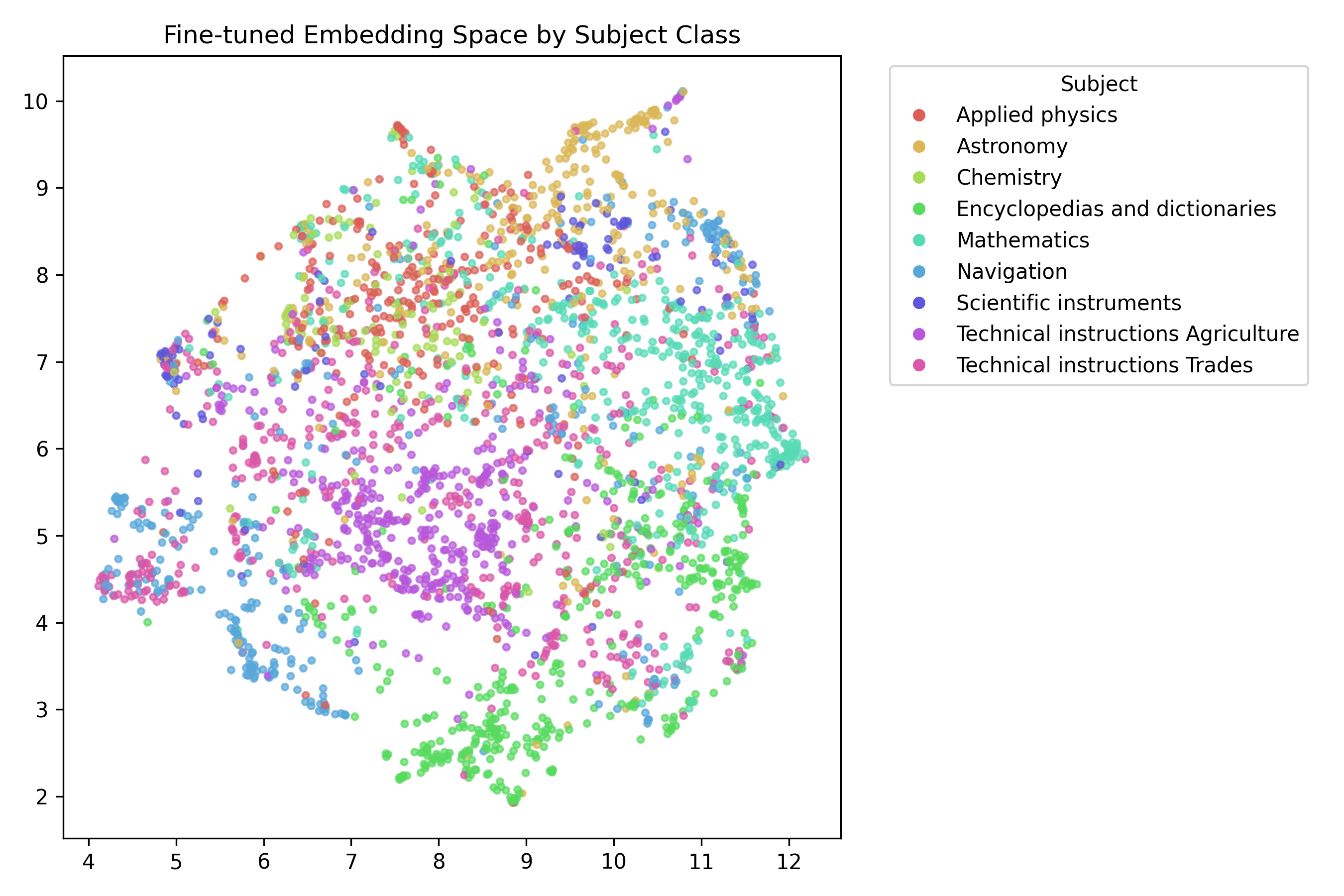}
					\subcaption{Fine-tuned model}
				\end{subfigure}
				\caption{ESTC title UMAP embeddings space projection by subject class \vspace{4pt} \newline \footnotesize \emph{Notes:}  The figure compares a UMAP (Uniform Manifold Approximation and Projection) of the embedding space for English Short Title Catalogue (ESTC) titles between the MacBERTh base model and the fine-tuned SteamBERTh model. Different colors indicate subject classes from \cite{Koschnick2025}.}
				\label{fig:embedding_umap_estc}
			\end{minipage}
			
			\vspace{1em}
			
			\begin{minipage}{\textwidth}
				\centering
				\captionsetup{type=table}
				\begin{adjustbox}{max width=0.7\columnwidth}
					\begin{threeparttable}\fontsize{10}{13}\selectfont
						\caption{Nearest neighbor cosine scores for base and fine-tuned models}
						\label{tab:umap_projection_evaluate_scores}
						\begin{tabular}{lll}
\hline
 Metric                  & Base MacBERTh model   & Fine-tuned model \\
\hline
\textit{Patents:}: & & \\
 5-NN accuracy  & 0.468\, & 0.599\,       \\[0.8ex] 
\textit{ESTC:} & & \\
 5-NN accuracy     & 0.619\, & 0.650\,      \\
\hline
\end{tabular}
						\begin{tablenotes}
							\item {\footnotesize \emph{Notes:} The table reports nearest neighbor cosine scores for base and fine-tuned models as measures of the semantic coherence within industry/subject class groups. Higher scores indicate increased semantic coherence within groups.}
						\end{tablenotes}
					\end{threeparttable}
				\end{adjustbox}
			\end{minipage}
		\end{figure}

		As further illustration of the efficacy of the fine-tuning process, figure \ref{fig:embedding_umap_patent}--\ref{fig:embedding_umap_estc} shows a UMAP embeddings space projection of the embedding space with subject class clusters denoted by different colors. We observe that the fine-tuning procedure significantly changes the positioning of subject classes and industry groups within the embedding space. Additionally, nearest neighbour cosine similarities by industry/subject class from table \ref{tab:umap_projection_evaluate_scores} show that the fine-tuning increased the semantic coherence within groups. 
		
		Overall, we find that the fine-tuned \textit{SteamBERTh} models seems to have higher internal consistency for the ESTC and patent data employed in this analysis.

		\begin{table}[!htbp]
			\centering
			\begin{adjustbox}{max width=0.7\columnwidth}
				\begin{threeparttable}\fontsize{10}{13}\selectfont
					\caption{Evaluative statistics}
					\label{tab:evaluative_statistics}
					\begin{tabular}{llr}
\hline
 Metric           & Base MacBERTh model  &   Fine-tuned model \\
\hline
 Mean NN-cosine similarity   & 0.974  &        0.655 \\
 PC-1 variance    & 0.198  &        0.033 \\
 Rank overlap $k=5$  & –      &        0.119 \\
 Rank overlap $k=10$ & –      &        0.071 \\
\hline
\end{tabular}
					\begin{tablenotes}
						\item {\footnotesize \emph{Notes:} The table reports a range of evaluative statistics comparing the embedding space for patents and ESTC training classes of the base and fine-tuned data. Mean NN-cosine reports the nearest neighbour cosine similarity and pc1-variance reports the fraction of total variance in the embedding matrix explained by the first principal component. Lower values in NN-cosine similarity and indicate a less collapsed or anisotropic embedding space. Likewise, the lower PC-1 variance indicates a more isotropic embedding space. Rank overlap measures are reported for top-5 and top-10 neighbors. The values indicate systematic changes in the embedding space after fine-tuning.}
					\end{tablenotes}
				\end{threeparttable}
			\end{adjustbox}
		\end{table}

		\section{Validation of innovation measure}
		\label{sec:validation_of_innovation}
		This section addresses the question whether using SteamBERTh, rather than BERT models that are trained on modern data, produces efficiency gains in capturing historical patterns of innovation. To test this, we formulate the following empirical model:
		\begin{equation}
			\text{Patent citations}_{it} = \beta \text{Innov meas}_{it} + X'_{it} \zeta + \alpha_t + \varepsilon_{it}
			\label{eq:validation_ols}
		\end{equation}
		The dependent variable is the count of patent citations, as an established measure of patent innovativeness in the literature \citep{trajtenberg1990penny,Nuvolari2011,Billington2021}. The main independent variable,  $\text{Innov meas}_{it}$, denotes the innovation measure from equation~\ref{eq:innovation} for patent $i$ at year $t$. For calculating the innovation index from equation~\ref{eq:innovation}, we use the embedding space from three BERT text similarity models for comparison: a) \textit{SteamBERTh} from section~\ref{sec:hist-bert-model} b)  \textit{all-MiniLM-L6-v2}, the most downloaded sentence similarity model on HuggingFace, and c) \textit{paraphrase-mpnet-base-v2}  \citep{reimers-2019-sentence-bert}, a popular larger $109M$ parameter model. Finally, in d) we calculate the tfidf-based breakthrough index from \cite{Kelly2021a}. Each patent is a vector of term counts weighted by a backward-looking inverse document frequency, and the index is the sum of cosine similarities to the patents of the following twenty years divided by the sum over all earlier patents. The model further includes the level and quadratic value of patent word counts $i$, captured by $X'_{it}$, as well as year fixed effects $\alpha_t$.

		\begin{table}[!htbp]
		\centering
		\begin{adjustbox}{max width=1\columnwidth}
			\centering
			\begin{threeparttable}\fontsize{10}{13}\selectfont
				\caption{Model performance, innovation index and historical patent citations}
				\label{tab:valid_steamberth}
				{
\def\sym#1{\ifmmode^{#1}\else\(^{#1}\)\fi}
\begin{tabular}{l*{7}{c}}
\hline\hline
                    &\multicolumn{7}{c}{Log Woodcroft patent citations} \\\cline{2-8}
                    &\multicolumn{1}{c}{(1)}   &\multicolumn{1}{c}{(2)}   &\multicolumn{1}{c}{(3)}   &\multicolumn{1}{c}{(4)}   &\multicolumn{1}{c}{(5)}   &\multicolumn{1}{c}{(6)}   &\multicolumn{1}{c}{(7)}   \\
\hline
\multicolumn{8}{l}{\textbf{Panel A: Patents, 1720--1799}} \\
\textit{Historical model:}&               &               &               &               &               &               &               \\
Log SteamBERTh-based innovation index&       0.217***&               &               &       0.211** &       0.229** &               &       0.225** \\
                    &    (0.0779)   &               &               &    (0.0912)   &    (0.0938)   &               &    (0.0890)   \\
[0.5em]
\textit{Comparison --- contemporary models:}&               &               &               &               &               &               &               \\
Log all-MiniLM-L6-v2-based innovation index&               &       0.167*  &               &      0.0157   &               &               &               \\
                    &               &    (0.0987)   &               &     (0.116)   &               &               &               \\
[0.5em]
Log paraphrase-mpnet-base-v2-based innovation index&               &               &       0.193   &               &     -0.0394   &               &               \\
                    &               &               &     (0.132)   &               &     (0.159)   &               &               \\
[0.5em]
\textit{Comparison --- term-frequency index:}&               &               &               &               &               &               &               \\
Log Kelly et al. (2021) TF-BIDF innovation index&               &               &               &               &               &      0.0446   &     0.00585   \\
                    &               &               &               &               &               &    (0.0304)   &    (0.0331)   \\
[0.5em]
Word count controls &         Yes   &         Yes   &         Yes   &         Yes   &         Yes   &         Yes   &         Yes   \\
[0.5em]
Year fixed effects  &         Yes   &         Yes   &         Yes   &         Yes   &         Yes   &         Yes   &         Yes   \\
\hline
Observations        &        1838   &        1838   &        1838   &        1838   &        1838   &        1783   &        1783   \\
R-squared           &       0.109   &       0.106   &       0.106   &       0.109   &       0.109   &       0.105   &       0.109   \\
\hline
\multicolumn{8}{l}{\textbf{Panel B: Patents, 1800--1841}} \\
\textit{Historical model:}&               &               &               &               &               &               &               \\
Log SteamBERTh-based innovation index&       0.171***&               &               &      0.0771   &      0.0775   &               &      0.0743   \\
                    &    (0.0417)   &               &               &    (0.0507)   &    (0.0500)   &               &    (0.0645)   \\
[0.5em]
\textit{Comparison --- contemporary models:}&               &               &               &               &               &               &               \\
Log all-MiniLM-L6-v2-based innovation index&               &       0.322***&               &       0.246***&               &               &               \\
                    &               &    (0.0693)   &               &    (0.0846)   &               &               &               \\
[0.5em]
Log paraphrase-mpnet-base-v2-based innovation index&               &               &       0.462***&               &       0.361***&               &               \\
                    &               &               &    (0.0973)   &               &     (0.117)   &               &               \\
[0.5em]
\textit{Comparison --- term-frequency index:}&               &               &               &               &               &               &               \\
Log Kelly et al. (2021) TF-BIDF innovation index&               &               &               &               &               &      0.0105   &     0.00157   \\
                    &               &               &               &               &               &    (0.0165)   &    (0.0183)   \\
[0.5em]
Word count controls &         Yes   &         Yes   &         Yes   &         Yes   &         Yes   &         Yes   &         Yes   \\
[0.5em]
Year fixed effects  &         Yes   &         Yes   &         Yes   &         Yes   &         Yes   &         Yes   &         Yes   \\
\hline
Observations        &        6684   &        6684   &        6684   &        6684   &        6684   &        3453   &        3453   \\
R-squared           &       0.177   &       0.178   &       0.178   &       0.178   &       0.179   &       0.218   &       0.218   \\
\hline\hline
\end{tabular}
}

				\begin{tablenotes}
					\item {\footnotesize \emph{Notes:} The table shows coefficients from estimating equation~\ref{eq:validation_ols} via OLS for the sample periods 1720--1799 (panel A) and 1800--1841 (panel B; columns (6) and (7): 1800--1831, since the term-frequency index needs a full twenty-year forward window within the catalogue, which ends in 1851). The dependent variable is patent citations from \cite{Nuvolari2011}. The main explanatory variables is the innovation measure from equation~\ref{eq:innovation}, with text similarity calculated with three different BERT models, \textit{SteamBERTh}, \texttt{all-MiniLM-L6-v2} and \texttt{paraphrase-mpnet-base-v2} \citep{reimers-2019-sentence-bert}, and, in columns (6) and (7), the term-frequency (TF-BIDF) innovation index of \cite{Kelly2021a} computed on the same corpus with the same pools and windows (ratio trimmed at the 0.5th and 99.5th percentiles). The innovation measure is calculated for the full sample of patents, 1700--1851. Because the innovation measure mechanically needs a 20 year pre- and post-period for comparison, it is calculated for the period 1720--1841. The model contains year fixed effects and controls for the level and quadratic value of patent word counts. Robust standard errors in parenthesis. *** denotes statistical significance at the 1\% level, ** at the 5\% level, and * at the 10\% level.}
				\end{tablenotes}
			\end{threeparttable}
		\end{adjustbox}
		\end{table}

		Panel A of table~\ref{tab:valid_steamberth} reports estimates for the time period of 1720--1800, which corresponds to the sample period of the combined ESTC and patent data.\footnote{Given that the patent data starts in 1700 and the innovation index is by definition not defined for the first twenty years of the sample, we get 1720 as a starting year.} Panel B further presents results for nineteenth century patents, 1800--1841. We find that for the period of 1720--1800, the \textit{SteamBERTh}-based innovation measure has a higher $R^2$ than the two models trained on contemporary data. Additionally, in a horse race specification (columns~4–-5), including the \textit{SteamBERTh} measure drives the coefficients on the other innovation measures close to zero, leaving \textit{SteamBERTh} as the only statistically significant predictor. This suggests that \textit{SteamBERTh} not only reduces historical bias but also outperforms modern models when applied to eighteenth century data. Finally, column~6 shows that the \cite{Kelly2021a} index performs significantly worse on historical data, with shorter text length, more abstract vocabulary and varying spelling conventions. As seen in the horse in column~7, all the variation explained through the \cite{Kelly2021a} index is also captured by the \textit{SteamBERTh} innovation measure.

		Next, in contrast to the previous results, panel B of table~\ref{tab:valid_steamberth} shows that for the nineteenth century, the two models trained on modern data outperform \textit{SteamBERTh}. Since both the training data for \textit{MacBERTh} as well as the fine-tuning training data for \textit{SteamBERTh} place a heavy weight on the early modern period, this is not surprising. The results suggest that historically fine-tuned models can outperform contemporary LLMs for specific time periods, but perform less well outside their historical training period. Again, the \cite{Kelly2021a} index performs worse than the embedding-based models on early nineteenth century patent data.

		\begin{table}[!htbp]
			\centering
			\begin{adjustbox}{max width=1\columnwidth}
				\begin{threeparttable}\fontsize{9}{11}\selectfont
					\caption{Top-20 innovative titles in applied physics}
					\label{tab:top-20-innov-phys}
					\begin{tabular}{llp{0.6\textwidth}}
\toprule
Author & Publication year & Title \\
\midrule
Marshall, William, Engraver. & 1640 & [The four elements. 4 engraved plates.] Earth. (Aire. Fire. Water.) \\
Powell, Thomas, 1608-1660. & 1651 & Elements of optics: a new, easy, and convenient method developed. (...) \\
Burgersdijk, Franco Petri, 1590-1635. & 1650 & Physical college, debates 32 absolute; the whole natural philosophy problem has to be put forward. (...) \\
Boyle, Robert, 1627-1691. & 1661 & Some physiological essays at different times and occasions were written by Robert Boyle... translated from English to Latin. \\
Bacon, Francis, 1561-1626. & 1620 & Francis de Verulamius, Chancellor of England, Great Instauration \\
Bacon, Francis, 1561-1626. & 1620 & The great establishment of Francis de Verulam, chancellor of England supreme. \\
--- & 1660 & An excerpt out of a book, shewing, that fluids rise not in the pump, in the syphon, and in the barometer, by the pressure of the air, but propter Fugam vacui. (...) \\
Boyle, Robert, 1627-1691. & 1661 & Certain physiological essays, written at distant times, and on several occasions (...) \\
Bradshaw, Ellis. & 1649 & A nevv and cleer discovery, of the true, and proper, natural cause, of the ebbing and flowing of the main sea (...) \\
Hobbes, Thomas, 1588-1679. & 1662 & Physical Problems (...) \\
Kepler, Johannes, 1571-1630. & 1653 & John Kepler ... Dioptrice (...) \\
Boyle, Robert, 1627-1691. & 1660 & New experiments physico-mechanicall, touching the spring of the air, and its effects, (made, for the most part, in a new pneumatical engine) (...) \\
Descartes, René, 1596-1650. & 1649 & A discourse of a method for the wel-guiding of reason, and the discovery of truth in the sciences. (...) \\
Line, Francis, 1595-1675. & 1661 & A treatise on the inseparability of bodies (...) \\
Gassendi, Pierre, 1592-1655. & 1653 & The astronomical institution of Peter Gassendi, according to the hypotheses of both ancients and moderns. (...) \\
Hobbes, Thomas, 1588-1679. & 1661 & A natural dialogue, or a conjecture about the nature of air (...) \\
Hooke, Robert, 1635-1703. & 1661 & An attempt for the explication of the phænomena observable in an experiment published by the Honourable Robert Boyle (...) \\
Ross, Alexander, 1591-1654. & 1645 & The philosophicall touch-stone: or Observations upon Sir Kenelm Digbie's Discourses of the nature of bodies, and of the reasonable soule. (...) \\
Charleton, Walter, 1619-1707. & 1654 & Physiologia Epicuro-Gassendo-Charltoniana: or A fabrick of science natural, upon the hypothesis of atoms (...) \\
--- & 1663 & A short compendium of the new and much enlarged sea-book, or, Pilots sea-mirror (...) \\
\bottomrule
\end{tabular}

					\begin{tablenotes}
						\item {\footnotesize \emph{Notes:} The 20 ESTC titles in \textit{applied physics} with the highest innovation index from equation~\ref{eq:innovation}. Titles shortened to the main title; omitted text is marked (...).}
					\end{tablenotes}
				\end{threeparttable}
			\end{adjustbox}
		\end{table}

		\section{Validation: Received spillover index}
		
		\label{sec:validation_of_received_spillover_appendix}
		This section formally estimates the relationship between the received spillover measure between $\Omega$ and $\lambda$ from equation~\ref{eq:received_spillover} and authors' share of publications between $\Omega$ and $\lambda$. This exercise is built on the intuition that the received spillover measure should correlate with authors' share of publications between fields as authors publishing in multiple fields would have had more domain knowledge about the other field and therefore would be more likely to receive knowledge spillovers from the other field.
		
		We start with two samples. First, for authors publishing in $\Omega$, we define their publication share in $\lambda$ as:
		
		\begin{equation}
			s_{j\lambda} = \frac{n_{j\lambda}}{n_{j,\Omega\cup \lambda}},
			\label{eq:author_proxy_app1}
		\end{equation}
		
		where $n_{j\lambda}$ denotes the number of publications by author $j$ in field $\lambda$ and $n_{j,\Omega\cup \lambda}\equiv n_{j\Omega}+n_{\lambda}$ denotes total publications in $\Omega \cup \lambda$. Next, for authors publishing in $\lambda$, we define their publication share in $\Omega$ as:
		
		\begin{equation}
			s_{j\Omega} = \frac{n_{j\Omega}}{n_{j,\lambda\cup \Omega}},
			\label{eq:author_proxy_app2}
		\end{equation}

		Then we estimate the association between authors' publication share~\ref{eq:author_proxy_app1} and~\ref{eq:author_proxy_app2} and received knowledge spillovers from equation~\ref{eq:received_spillover}:
		
		\begin{equation}
			\text{Pub. Share $(B)$}_jt = \beta \text{Received spillover meas.}_{ij} + X'_{ijt} \zeta + \alpha_t + \varepsilon_{ijt}
			\label{eq:validation_spillover_ols}
		\end{equation}
		
		where the dependent variable is the combined authors' publication share from equation~\ref{eq:author_proxy_app1} and~\ref{eq:author_proxy_app2} for author $j$ at time $t$ in percent. The main independent variable,  $\text{Received spillover meas.}_{it}$, denotes the received spillover measure from equation~\ref{eq:received_spillover} for publication $i$ and author $j$ at time $t$. The model further includes the level and quadratic value of title word counts $i$, captured by $X'_{it}$, as well as year fixed effects $\alpha_t$.

		Table~\ref{tab:valid_spillover} reports the results. Column~1 reports results for the author proxy measure for all titles in both propositional and prescriptive knowledge. Column~2 and~3 then split the sample into propositional and prescriptive knowledge. For the full sample, a 1\% higher received-spillover index is associated with an approximately 0.023-percentage-point higher author-publication share in the other knowledge category. This corresponds to approximately 0.43\% of the mean share. The results suggest that the received spillover measure is strongly associated with authors with a high publication share in the fields of the received spillover. Moreover, we find that this association is stronger for authors who published in prescriptive knowledge and received spillovers from propositional knowledge than for authors publishing in propositional knowledge and received spillovers from prescriptive knowledge.
		
		Overall, the results offer additional support for the validity of the received spillover index. We expected that authors publishing in multiple fields would also be more likely to implement knowledge from these fields into their works and hence to receive knowledge spillovers. We found that this association is in fact strongly positive and significant.

		\begin{table}[!htbp]
		\centering
		\begin{adjustbox}{max width=1\columnwidth}
			\centering
			\begin{threeparttable}\fontsize{10}{13}\selectfont
				\caption{Validation: Authors' publication share and received spillover index}
				\label{tab:valid_spillover}
				{
\def\sym#1{\ifmmode^{#1}\else\(^{#1}\)\fi}
\begin{tabular}{l*{3}{c}}
\hline\hline
                    &\multicolumn{1}{c}{Prop. \& prescr. knowledge}&\multicolumn{1}{c}{Prop. knowledge}&\multicolumn{1}{c}{Prescr. knowledge}\\\cline{2-2}\cline{3-3}\cline{4-4}
                    &\multicolumn{1}{c}{(1)}   &\multicolumn{1}{c}{(2)}   &\multicolumn{1}{c}{(3)}   \\
                    &Author proxy   &Author proxy   &Author proxy   \\
\hline
Log received spillover index&       2.305***&       2.213***&       2.664** \\
                    &     (0.470)   &     (0.494)   &     (1.208)   \\
[1em]
Word count controls &         Yes   &         Yes   &         Yes   \\
[1em]
Year fixed effects  &         Yes   &         Yes   &         Yes   \\
[1em]
Subject class fixed effects&         Yes   &         Yes   &         Yes   \\
\hline
Observations        &        5826   &        3258   &        2565   \\
R-squared           &        0.11   &        0.09   &        0.20   \\
Mean dep.~var.      &       5.309   &       4.769   &       5.996   \\
\hline\hline
\end{tabular}
}

				\begin{tablenotes}
					\item {\footnotesize \emph{Notes:} The table shows coefficients from estimating equation~\ref{eq:validation_spillover_ols} via OLS for the sample period 1720--1799. The dependent variable is the received spillover index from equation~\ref{eq:received_spillover} for spillovers from both $\Omega$ and $\lambda$. The main explanatory variable is authors' publication share between $\Omega$ and $\lambda$. The share refers to the field they received the spillover from. The model contains year fixed effects and controls for the level and quadratic value of patent word counts.  Robust standard errors in parenthesis. *** denotes statistical significance at the 1\% level, ** at the 5\% level, and * at the 10\% level.}
				\end{tablenotes}
			\end{threeparttable}
		\end{adjustbox}
		\end{table}

		\FloatBarrier

				\section{Modern data: Corpus, measures and validation}
		\label{sec:app:modern}

		This appendix documents the validation on modern data from section~\ref{sec:validation_modern}. It first introduces the corpus and cleaning procedures and documents the construction of the innovation and received spillover index on modern text. Then, it documents results for reverse flows from propositional to prescriptive in the form of research articles citing patents. Lastly, it compares the performance of the embedding-space based innovation measure from equation~\ref{eq:innovation} to the bag-of-words based index from \cite{Kelly2021a}.

		\subsection{Corpus and cleaning}
		The modern corpus is constructed to mirror the two historical corpora. As a counterpart to the propositional knowledge of the eighteenth century, it uses the titles and abstracts of journal articles from OpenAlex within the fields of \textit{physics and astronomy, chemistry}, and \textit{mathematics}. As a counterpart to the prescriptive knowledge of the eighteenth century, it uses \textit{engineering, computer science}, and \textit{materials science}. OpenAlex assigns each work a primary topic nested in a subfield, field and domain. The section uses the field level, restricted to English-language journal articles. For each field and year a fixed random sample is drawn (for engineering 5,000 articles per year, every other field 2,000). For prescriptive knowledge, it uses the abstracts of all US utility patents assigned to publicly listed firms in the data from \cite{kogan2017technological}: 2.9 million patents filed 1985--2024, 2.8 million of which have an abstract in PatentsView. The patent data provide filing and issue dates, forward citations, the market value of each patent and its CPC classification, of which the first code defines the patent's section, class and subclass. Patents are dated by filing year. Altogether, the patent corpus describes the corporate part of the US patent system, which makes up ca. forty per cent of grants.

		Three cleaning steps remove records that are not abstracts. First, records whose abstract or title begins with the boilerplate of abstracting services or scraped web pages (for example ``ChemInform abstract'', ``Advertisement'', ``No abstract available'', ``Get PDF'') and mis-assigned non-scientific documents such as legislative bills are dropped. Second, articles whose first twelve words are shared by four or more other records, the signature of template text, are dropped. Patents are exempt from this step because patent abstracts legitimately share formulaic openings. Third, within each field and year, documents whose embedding has a cosine similarity above 0.95 with at least two other documents are dropped as near duplicates. Documents with fewer than 30 or more than 600 words are excluded. Together these steps remove about 6 per cent of the raw records. Table~\ref{tab:modern_sample_flow} gives the sample sizes at each stage.

		\begin{table}[!htbp]
			\centering
			\begin{threeparttable}\footnotesize
				\caption{Open Alex arcticles and US utility patents corpus}
				\label{tab:modern_sample_flow}
				\begin{tabular}{lr}
\hline\hline
Dataset & $N$ \\
\hline
Patents of listed firms with an abstract, filed 1985--2024 (KPSS $\cap$ PatentsView) & 2,778,447 \\
\quad with a defined innovation index, filed 1990--2019 & 2,349,647 \\
\quad filed 1990--2012 (estimation sample of the modern test) & 1,624,402 \\
Articles in the source pools: prop. / prescr.  & 210,196 / 335,309  \\
\quad basic-science articles with a defined innovation index (validation) & 157,085 \\
\quad engineering articles with a defined innovation index (reverse direction, $\le$ 2012) & 107,900 \\
\hline\hline
\end{tabular}

				\begin{tablenotes}
					\item {\footnotesize \emph{Notes:} Counts after the cleaning steps described in the text. The innovation index of a patent is defined when its CPC class has at least 50 patents in each of the five preceding and the five following years. The spillover validation uses filings up to 2012, so that every patent has at least ten years of citation exposure.}
				\end{tablenotes}
			\end{threeparttable}
		\end{table}

		\subsection{Validation of the received spillover index for scientific articles}
		\label{sec:validation_modern_reverse_flow}

		Section~\ref{sec:validation_modern} documents validation results for the flow from research into patents. This section documents the reverse flow, from patents into research. For this, we draw on references articles make to patents. As this is a highly rare occurrence in most fields, we limit the analysis to engineering articles, published 1990--2012. We recover the references to patents from the unstructured reference strings deposited with Crossref. Of the 40,309 articles with a Crossref reference list, 2.0 per cent cite at least one patent, with 1.9 patents cited on average among the citing articles. Within OpenAlex topic and publication year, a one per cent higher received spillover index from patents is associated with a 2.8 per cent higher probability that an article cites a patent. A second, backward-looking measure uses the OpenAlex reference lists of 49,684 articles. An article builds on science that patents use if at least one of its references is itself cited by a patent in the \cite{marx2020reliance} data, which holds for 79 per cent of the articles; a one per cent higher index is associated with a 0.2 per cent higher probability. A forward-looking measure, whether the article is itself cited by a patent (7.3 per cent of the 107,900 engineering articles with an index), is 1.9 per cent more likely for a one per cent higher index. The spillover index thus tracks the direction of actual knowledge flows on both sides of the boundary. Table~\ref{tab:modern_article_side} reports the three regressions as linear probability models, in percentage points per log point of the index; the elasticities above are these coefficients divided by the mean of the outcome.

		\begin{table}[!htbp]
		\centering
		\begin{adjustbox}{max width=1\columnwidth}
			\centering
			\begin{threeparttable}\fontsize{10}{13}\selectfont
				\caption{Validation of the received spillover index from patents, engineering articles}
				\label{tab:modern_article_side}
				{
\def\sym#1{\ifmmode^{#1}\else\(^{#1}\)\fi}
\begin{tabular}{l*{3}{c}}
\hline\hline
                    &\multicolumn{3}{c}{Engineering articles, 1990--2012: share of articles (per cent) that \ldots}                \\\cline{2-4}
                    &\multicolumn{1}{c}{(1)}   &\multicolumn{1}{c}{(2)}   &\multicolumn{1}{c}{(3)}   \\
                    & Cites a patent&Reference cited by a patent&Cited by a patent\\
\hline
Log received spillover index from patents&       5.553***&      19.345***&      13.568***\\
                    &        (1.189)&        (2.628)&        (1.585)\\
[1em]
Mean of the outcome (per cent)&           1.99&          78.82&           7.30\\
[1em]
Elasticity at the mean (coefficient / mean)&           2.79&           0.25&           1.86\\
[1em]
Word count controls &            Yes&            Yes&            Yes\\
[1em]
Log number of references&            Yes&            Yes&            Yes\\
[1em]
Topic $\times$ year fixed effects&            Yes&            Yes&            Yes\\
\hline
Observations        &       40{,}309&       49{,}684&      107{,}900\\
R-squared           &          0.297&          0.479&          0.239\\
\hline\hline
\end{tabular}
}

				\begin{tablenotes}
					\item {\footnotesize \emph{Notes:} The table shows coefficients from linear probability models, estimated via OLS, of an indicator (in per cent) on the log received spillover index from patents (equation~\ref{eq:received_spillover}, \textit{all-mpnet-base-v2} embeddings, $\tau=5$, $k=20$, pool of engineering-class patents filed in the five preceding years) for engineering articles published 1990--2012. Column (1): the article cites at least one patent, recovered from the unstructured reference strings deposited with Crossref, among articles with a Crossref reference list. Column (2): at least one of the article's references is itself cited by a patent in the data of \cite{marx2020reliance}, among articles with an OpenAlex reference list. Column (3): the article is itself cited by a patent in the data of \cite{marx2020reliance}, among all engineering articles with an index. All models contain OpenAlex topic $\times$ publication year fixed effects and controls for the level and quadratic value of the abstract word count and the log number of references. The elasticity at the mean is the coefficient divided by the mean of the outcome: the per cent change in the probability associated with a one per cent higher index. Standard errors clustered by publication year in parentheses. *** denotes statistical significance at the 1\% level, ** at the 5\% level, and * at the 10\% level.}
				\end{tablenotes}
			\end{threeparttable}
		\end{adjustbox}
		\end{table}

		\subsection{Comparison with the term-frequency index of Kelly et al. (2021)}
		Appendix~\ref{sec:validation_of_innovation} compares the \textit{SteamBERTh}-based innovation index with the term-frequency index of \cite{Kelly2021a} on the historical patents. Table~\ref{tab:modern_kelly} repeats the comparison on the modern data with the validation specification of Section~\ref{sec:validation_modern}. The term-frequency index is built on the same corpus with the same pools and windows as the modern innovation index: each abstract is a vector of unigram counts weighted by a backward-looking inverse document frequency, and the index is the sum of cosine similarities to the documents of the own CPC class (for articles, the own field) in the following five years divided by the sum over the preceding five years. Both indices enter in logs without trimming; within year and class they are correlated at 0.40 for patents and 0.36 for articles.

		\begin{table}[!htbp]
		\centering
		\begin{adjustbox}{max width=1\columnwidth}
			\centering
			\begin{threeparttable}\fontsize{9}{11}\selectfont
				\caption{Model performance, innovation index and modern patent citations, patent value, and article citations}
				\label{tab:modern_kelly}
				{
\def\sym#1{\ifmmode^{#1}\else\(^{#1}\)\fi}
\begin{tabular}{l*{3}{c}}
\hline\hline
                    &\multicolumn{1}{c}{(1)}   &\multicolumn{1}{c}{(2)}   &\multicolumn{1}{c}{(3)}   \\
\hline
\multicolumn{4}{l}{\textbf{Panel A: Log forward citations, patents (1990--2019)}} \\
\textit{Embedding-based index:} &&&\\
Log innovation index (\textit{all-mpnet-base-v2})&       3.744***&               &       2.926***\\
                    &       (0.1227)&               &       (0.0833)\\
[0.5em]
\textit{Comparison --- term-frequency index:} &&&\\
Log Kelly et al. (2021) TF-BIDF innovation index&               &       0.686***&       0.433***\\
                    &               &       (0.0558)&       (0.0579)\\
[0.5em]
Word count controls &            Yes&            Yes&            Yes\\
[0.5em]
CPC section $\times$ year fixed effects &            Yes&            Yes&            Yes\\
\hline
Observations        &      2{,}349{,}647&      2{,}349{,}647&      2{,}349{,}647\\
R-squared           &          0.356&          0.350&          0.361\\
\hline
\multicolumn{4}{l}{\textbf{Panel B: Log patent market value, patents (1990--2019)}} \\
\textit{Embedding-based index:} &&&\\
Log innovation index (\textit{all-mpnet-base-v2})&       2.491***&               &       0.600***\\
                    &       (0.1656)&               &       (0.1447)\\
[0.5em]
\textit{Comparison --- term-frequency index:} &&&\\
Log Kelly et al. (2021) TF-BIDF innovation index&               &       1.054***&       1.002***\\
                    &               &       (0.0834)&       (0.0883)\\
[0.5em]
Word count controls &            Yes&            Yes&            Yes\\
[0.5em]
CPC section $\times$ year fixed effects &            Yes&            Yes&            Yes\\
\hline
Observations        &      2{,}349{,}647&      2{,}349{,}647&      2{,}349{,}647\\
R-squared           &          0.057&          0.066&          0.066\\
\hline
\multicolumn{4}{l}{\textbf{Panel C: Log OpenAlex citations, articles (1990--2019)}} \\
\textit{Embedding-based index:} &&&\\
Log innovation index (\textit{all-mpnet-base-v2})&       2.568***&               &       2.244***\\
                    &       (0.4469)&               &       (0.3351)\\
[0.5em]
\textit{Comparison --- term-frequency index:} &&&\\
Log Kelly et al. (2021) TF-BIDF innovation index&               &       1.031***&         0.551*\\
                    &               &       (0.3564)&       (0.3339)\\
[0.5em]
Word count controls &            Yes&            Yes&            Yes\\
[0.5em]
Field $\times$ year fixed effects &            Yes&            Yes&            Yes\\
\hline
Observations        &        157{,}085&        157{,}085&        157{,}085\\
R-squared           &          0.072&          0.068&          0.072\\
\hline\hline
\end{tabular}
}

				\begin{tablenotes}
					\item {\footnotesize \emph{Notes:} The table shows coefficients from regressing the log outcome on the log innovation index via OLS. Panel A: log forward citations and panel B: log market value of US utility patents of listed firms filed 1990--2019, both from \cite{kogan2017technological}, with CPC section $\times$ filing year fixed effects. Panel C: log OpenAlex citations of articles in \textit{physics and astronomy, chemistry} and \textit{mathematics} published 1990--2019, with field $\times$ publication year fixed effects. All models control for the level and quadratic value of the abstract word count. Column (1): the innovation measure from equation~\ref{eq:innovation} with \textit{all-mpnet-base-v2} embeddings ($\tau=5$, $k=20$, within CPC class for patents and within field for articles); column (2): the term-frequency (TF-BIDF) innovation index of \cite{Kelly2021a} with the same pools and windows; column (3): both. Neither index is trimmed. Standard errors clustered by filing year (patents) or publication year (articles) in parentheses. *** denotes statistical significance at the 1\% level, ** at the 5\% level, and * at the 10\% level.}
				\end{tablenotes}
			\end{threeparttable}
		\end{adjustbox}
		\end{table}

		For patent forward citations and for article citations the embedding-based index has the stronger association, and both indices keep most of their coefficients when entered jointly, so they carry partly different information. For patent market value the term-frequency index dominates and absorbs most of the embedding-based coefficient. Because the log term-frequency index is almost five times as dispersed as the log embedding index within fixed effects for patents (1.7 times for articles), the elasticities of the two are not comparable; per standard deviation of the log index, the effects on forward citations are 0.21 for the embedding-based and 0.18 for the term-frequency index, on market value 0.14 and 0.27, and on article citations 0.12 and 0.08. A plausible reading is that the sum-based term-frequency index rewards patents whose vocabulary is taken up by many later patents, the scale of a technology that the market capitalises, whereas the top-$k$ embedding index rewards closeness to the nearest later documents, which is what a citation link records.

		\subsection{Summary of the validation regressions}
		Table~\ref{tab:validation_summary} collects the validation regressions for the historical and the modern data.

	\begin{table}[!htbp]
			\centering
			\begin{adjustbox}{max width=\columnwidth}
				\begin{threeparttable}\footnotesize
				\caption{Validation of the two indices: historical and modern data}
				\label{tab:validation_summary}
				\begin{tabular}{lcccll}
					\hline\hline
					& Coeff. & (s.e.) & $N$ & Fixed effects & Sample \\
					\hline
					\multicolumn{6}{l}{\textit{Panel A: Historical data (SteamBERTh embeddings, $k=\rho=\tau=20$)}} \\
					Innovation index $\rightarrow$ patent citations (Woodcroft) & 0.18 & (0.04) & 1,922 & Year & Patents, 1720--1799 \\
					Innovation index $\rightarrow$ patent citations (Woodcroft) & 0.17 & (0.02) & 6,684 & Year & Patents, 1800--1841 \\
					Innovation index $\rightarrow$ number of editions & 0.17 & (0.04) & 5,496 & Year & $\Omega$ and $\lambda$ titles, 1620--1779 \\
					Received spillover index $\rightarrow$ author's share of titles in the other type (pp) & 2.82 & (0.49) & 5,827 & Year, subject & $\Omega$ and $\lambda$ titles, 1620--1779 \\
					\multicolumn{6}{l}{\textit{Panel B: Modern data (all-mpnet-base-v2 embeddings, $k=\rho=20$, $\tau=5$)}} \\
					Innovation index $\rightarrow$ article citations (OpenAlex) & 2.57 & (0.45) & 157,085 & Field $\times$ year & Articles, 1990--2019 \\
					Innovation index $\rightarrow$ patent forward citations (KPSS) & 3.74 & (0.12) & 2,348,639 & CPC section $\times$ year & Patents, 1990--2019 \\
					Innovation index $\rightarrow$ patent value (KPSS) & 2.49 & (0.17) & 2,348,639 & CPC section $\times$ year & Patents, 1990--2019 \\
					Received spillover index from basic science $\rightarrow$ share citing a basic science article (pp) & 1.68 & (0.13) & 1,624,402 & CPC class $\times$ year & Patents, 1990--2012 \\
					Received spillover index from engineering $\rightarrow$ share citing a engineering article (pp) & 1.54 & (0.11) & 1,624,402 & CPC class $\times$ year & Patents, 1990--2012 \\
					Received spillover index from patents $\rightarrow$ share citing a patent (pp) & 5.55 & (1.19) & 40,309 & Topic $\times$ year & Engineering articles, 1990--2012 \\
					Received spillover index from patents $\rightarrow$ share with a reference cited by patents (pp) & 19.34 & (2.63) & 49,684 & Topic $\times$ year & Engineering articles, 1990--2012 \\
					\hline\hline
				\end{tabular}
				\begin{tablenotes}
					\item {\footnotesize \emph{Notes:} Each row is a separate OLS regression of the outcome on the log index with the fixed effects indicated and word-count controls (modern rows). Rows 1--3 and the modern innovation rows are elasticities (log outcome on log index). The spillover rows are linear probability models in percentage points per log point of the index: historically, the share of an author's titles that fall in the other type of knowledge (Appendix~\ref{sec:validation_of_received_spillover_appendix}); in modern data, an indicator for citing at least one article of the source group among the paper's OpenAlex pools (mean 0.39 per cent for basic science, 0.87 per cent for engineering). Standard errors clustered by year (historical rows: robust; modern rows: 30 and 23 year clusters).}
				\end{tablenotes}
			\end{threeparttable}
			\end{adjustbox}
		\end{table}
		
		\FloatBarrier
		
		\section{\label{sec:app:comparing-sim}Comparing similarity functions}
		This section validates the following two similarity functions for the calculation of the innovation index from section~\ref{sec:defining-similarity}.
		
		Average similarity index:
		\begin{equation*}
			f(v_{it},Y)= \frac{1}{n} \sum_{y' \in Y} sim(v_{it},y)
		\end{equation*}
		
		K-top similarity index:
		\begin{equation*}
			f(v_{it},\,Y;\,k\bigr)
			\;=\;
			\frac{1}{k}
			\sum_{\,y\in\operatorname{Top}_{k}
				\{\!sim(v_{it},y') : y'\in Y\}}
			\!sim(v_{it},y),
			\qquad
			k:=\min\{k,|Y|\}
		\end{equation*}
		
\begin{figure}[!htbp]
			\centering
			\begin{subfigure}[t]{0.95\textwidth}
				\includegraphics[width=\linewidth]{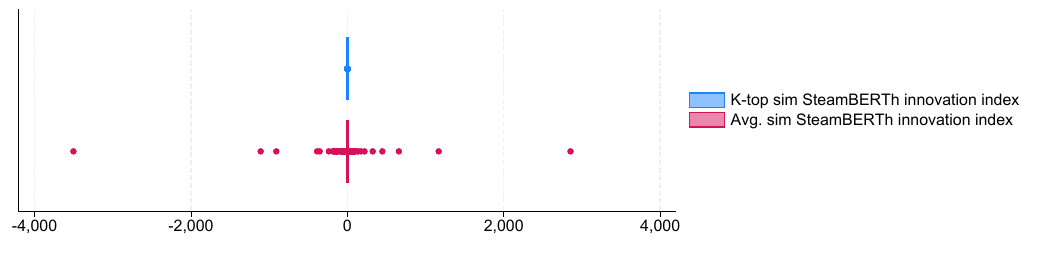}
				\subcaption{Boxplot, k-top and average similarity function}
				\label{fig:boxplot}
			\end{subfigure}

			\medskip
			\begin{subfigure}[t]{0.32\textwidth}
				\includegraphics[width=\linewidth]{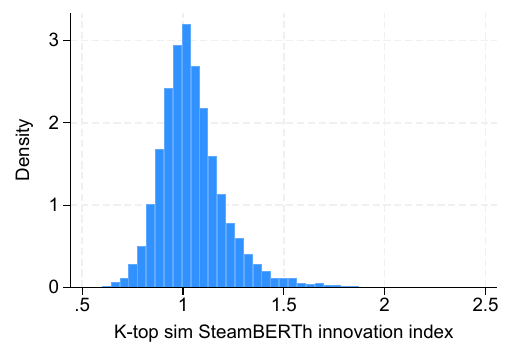}
				\subcaption{Level, k-top}
				\label{fig:histogram_ktop}
			\end{subfigure}
			\begin{subfigure}[t]{0.32\textwidth}
				\includegraphics[width=\linewidth]{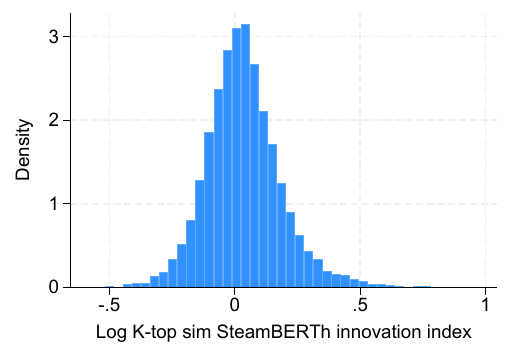}
				\subcaption{Log, k-top}
			\end{subfigure}
			\begin{subfigure}[t]{0.32\textwidth}
				\includegraphics[width=\linewidth]{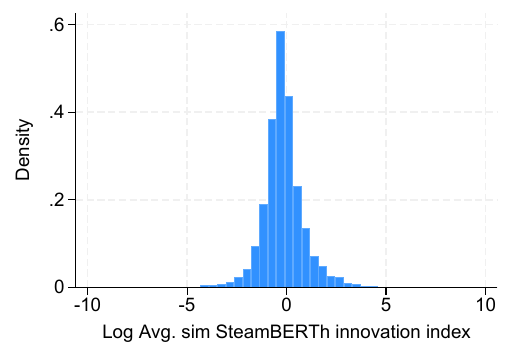}
				\subcaption{Log, average}
			\end{subfigure}
			\caption{Distribution of the innovation index using (1) the k-top and (2) the average similarity function \vspace{4pt} \newline \emph{\footnotesize Notes: Panel a) shows boxplots of both innovation indices, panel b) the histogram of the k-top innovation index, and panels c) and d) histograms of the natural logarithm of both indices. The k-top similarity function is defined in equation~\ref{eq:similarity-ktop}. Throughout the paper, we use the k-top similarity function for the calculation of the innovation index.}}
			\label{fig:similarity_functions}
		\end{figure}

		Using both measures, we create two innovation indices and plot their distribution in figure \ref{fig:boxplot}. We see that the average similarity index is significantly more likely to produce extreme values, while the k-top index is well-centered around 1 (see figure \ref{fig:histogram_ktop}). 
		
		The problem of extreme values can be mitigated by using the natural logarithm, see panels c) and d) of figure \ref{fig:similarity_functions}. However, note that for the average similarity index values equal to or below zero had to be excluded.
		
		Next, we test the performance of the innovation indices as predictors of patent citations using the model from equation~\ref{eq:validation_ols} from section~\ref{sec:validation}. We see that the innovation index based on the k-top similarity function outperforms the innovation index based on the average similarity function. The innovation index based on the k-top similarity function has a larger R-squared value. Additionally, its coefficient is a significant predictor of patent citations, while the innovation index based on the average similarity function is insignificant within this setting.
		
		Overall, the results of this section indicate that the k-top similarity function yields an innovation index with a tighter distribution and more predictive power for the test case of patent citations. Therefore, for the other validation exercises as well as for the main empirical framework in section~\ref{sec:empirical-framework}, the paper adopts the k-top innovation index for the calculation of innovation and knowledge spillovers.

		\FloatBarrier

		\begin{table}[!htbp]
		\centering
		\begin{adjustbox}{max width=.95\columnwidth}
			\centering
			\begin{threeparttable}\fontsize{10}{13}\selectfont
				\caption{Predicting patent citations: Comparison of innovation index using (1) the k-top and (2) the average similarity function}
				\label{tab:sim_innov_comparison}
				{
\def\sym#1{\ifmmode^{#1}\else\(^{#1}\)\fi}
\begin{tabular}{l*{2}{c}}
\hline\hline
                    &\multicolumn{2}{c}{Log Woodcroft patent citations (1720--1799)}\\\cline{2-3}
                    &\multicolumn{1}{c}{(1)}   &\multicolumn{1}{c}{(2)}   \\
                    &   Pat. cit.   &   Pat. cit.   \\
\hline
Log K-top sim SteamBERTh innovation index&       0.176** &               \\
                    &    (0.0780)   &               \\
[1em]
Log Avg. sim SteamBERTh innovation index&               &      0.0304   \\
                    &               &    (0.0212)   \\
[1em]
Word count controls &         Yes   &         Yes   \\
[1em]
Year fixed effects  &         Yes   &         Yes   \\
\hline
Observations        &        1787   &        1787   \\
R-squared           &       0.111   &       0.110   \\
\hline\hline
\end{tabular}
}

				\begin{tablenotes}
					\item {\footnotesize \emph{Notes:} The table shows coefficients for estimating equation~\ref{eq:validation_ols} via OLS. The dependent variable is patent citations from \cite{Nuvolari2011}. The main explanatory variables is the innovation measure from equation \ref{eq:innovation} using (1) the k-top and (2) the average similarity function. The model contains year fixed effects and controls for the level and quadratic value of patent word counts. The model is estimated for the sample period of 1720--1799. Applying the natural logarithm to values below zero for the innovation index based on the average similarity function leads to excluded observations. To make the models comparable, we apply the same sample restriction to the model in column~1 and~2. Robust standard errors in parenthesis. *** denotes statistical significance at the 1\% level, ** at the 5\% level, and * at the 10\% level.}
				\end{tablenotes}
			\end{threeparttable}
		\end{adjustbox}
		\end{table}

		\FloatBarrier
		\section{Spillovers from propositional to prescriptive knowledge: From negative to positive coefficients in a hundred years}
		\label{sec:app:negative_coefficient}
		
		This section discusses one of the stylized findings of section~\ref{sec:empirical_results}, the transition from negative to positive coefficents for the association between spillovers from propositional knowledge and innovation in prescriptive knowledge. 				
		To interpret these findings, this section decomposes the influence of spillovers and innovation for individual titles and interprets the previous findings by discussing individual titles and the way in which they incorporated spillovers from propositional knowledge in different periods.
		
		To identify the influence of individual titles, we decompose the pooled coefficient into the exact contributions of individual titles, $\beta = \sum_i \tilde{x}_i \tilde{y}_i / \sum_i \tilde{x}_i^2$, where $\tilde{x}_i$ and $\tilde{y}_i$ are the log received spillover and the log innovation index of title $i$, residualized on all other regressors and fixed effects (Frisch--Waugh--Lovell).

		\begin{table}[!htbp]
			\centering
			\begin{adjustbox}{max width=1\columnwidth}
				\begin{threeparttable}\scriptsize
					\caption{High spillovers, low innovation: the ten titles of type (a) with the largest contributions to the negative coefficient of 1660--1719}
					\label{tab:negcoef_top10}
					\begin{tabular}{rrlp{0.72\textwidth}r}
\toprule
Rank & Year & Field & Author and title & Contribution \\
\midrule
2 & 1668 & Instruments & Hues: \textit{A treatise on the heavenly and earthly globes and their use} (...) & $-0.013$ \\
7 & 1704 & Agriculture & \textit{Geoponika. Geoponicum, sive de re rustica, libri XX} (...) & $-0.006$ \\
9 & 1686 & Agriculture & Markham: \textit{Markham's faithful farrier} (...) & $-0.005$ \\
12 & 1671 & Trades & \textit{The Queens closet opened. Incomparable secrets in physick, chirurgery, preserving and candying, \&c.} (...) & $-0.005$ \\
13 & 1718 & Instruments & Villette: \textit{A description of the great burning-glass made by Mr. Villette and his two sons, born at Lyons.} (...) & $-0.005$ \\
15 & 1702 & Trades & Neve: \textit{Apopiroscopy: or, a compleat and faithful history of experiments and observations} (...) & $-0.004$ \\
17 & 1665 & Instruments & Moxon: \textit{A tutor to astronomy \& geography, or, The use of the Copernican spheres} (...) & $-0.004$ \\
20 & 1691 & Agriculture & E. R.: \textit{The experienc'd farrier: or, A compleat treatise of horsemanship.} (...) & $-0.004$ \\
21 & 1682 & Agriculture & Butler: \textit{The monarchy of the women, or the history of the Bees} (...) & $-0.004$ \\
22 & 1661 & Trades & Rabisha: \textit{The whole body of cookery dissected} (...) & $-0.004$ \\
\midrule
\multicolumn{4}{l}{Sum of the ten titles (coefficient without them: $-0.089$)} & $-0.054$ \\
\bottomrule
\end{tabular}

					\begin{tablenotes}
						\item {\footnotesize \emph{Notes:} Contributions $\tilde{x}_i \tilde{y}_i / \sum_j \tilde{x}_j^2$ of individual titles to the pooled 1660--1719 coefficient of equation~\ref{eq:ols_spillover} for spillovers from $\Omega$ to $\lambda$ ($-0.146$), for titles of type (a), which receive more spillovers and are less innovative than predicted. Rank: rank among the contributions of all 717 titles of 1660--1719. Titles as translated in the data, shortened to the main title; omitted text is marked (...).}
					\end{tablenotes}
				\end{threeparttable}
			\end{adjustbox}
		\end{table}
		
		To understand what was driving the spillover coefficients in 1660--1719, we first explore the content of titles with high received spillovers but low innovation. For these, table~\ref{tab:negcoef_top10} lists the top-10 titles that contributed most to the $\beta$ coefficient. Most of the titles present practical knowledge in the register of secrets and physick, or in practical instructions of use. Examples are \textit{The Queens closet opened}, describing the Royal household's recipes and medicines, or Joseph Moxon's \textit{Tutor to astronomy and geography} (1665 and 1698), or Vilette's concave mirror that could burn metals. Overall, these texts draw on many traditional forms of propositional knowledge, such as medicine or, in the case of the \textit{Geoponika}, ancient Byzantine practices of agriculture. Other cases demonstrate the application of recent scientific discoveries, yet without them being economically viable. We can illustrate this with Vilette's \textit{great burning glass}, an invention that could melt metals through a large concave glass lens. While successfully demonstrating the application of geometrical optics, it was still useless for practical purposes, depending on strong sunlight and yielding relatively little heat.\footnote{Note that the instrument had been originally demonstrated in 1669 --- the English printing of the French original corresponds to the new interest in the method coinciding with John Harris' and John Theophilus Desaguliers experiments with Vilette's apparatus that were published in the Philosophical Transactions of the Royal Society, volume 30 (1719), pp. 976–977.} Overall, it appears that the propositional knowledge of the seventeenth century was still insufficient for technological breakthrough inventions.
		
		We further explore the the difference between the knowledge economy of the pre 1720 and post 1740 world, by reporting top-10 titles with high-spillovers and high innovation for 1660--1719 and 1740--1779 in table~\ref{tab:negcoef_hh_early} and~\ref{tab:negcoef_hh_late}.		
		We find that in 1660--1719, titles clustered around a few problems of the restoration period: Finding the longitude at sea, Robert Fitzgerald's patented method of making sea water fresh, and William Oughtred's mathematical instruments. In contrast, for 1740--1779, we find plenty of applications of experimental natural philosophy: Henry Baker's \textit{Microscope made easy}, William Henley's electrical machine, John Dollond's patent for the achromatic telescope lens (1758), Francis Home's \textit{Principles of agriculture and vegetation}, and Michael Combrune's \textit{Essay on brewing} ``with a view of establishing the principles of the art''. It appears that, by 1740, practical applications of propositional knowledge had become plentiful.

		\begin{table}[!htbp]
			\centering
			\begin{adjustbox}{max width=1\columnwidth}
				\begin{threeparttable}\scriptsize
					\caption{High spillovers, high innovation: the ten titles with the largest contributions, 1660--1719}
					\label{tab:negcoef_hh_early}
					\begin{tabular}{rrlp{0.72\textwidth}r}
\toprule
 & Year & Field & Author and title & Contribution \\
\midrule
1 & 1707 & Agriculture & Mortimer: \textit{The whole art of husbandry, or, the way of managing and improving of land.} (...) & $0.0047$ \\
2 & 1678 & Navigation & Blackborow: \textit{The longitude not found: or, An answer to a treatise, written by Henry Bond Senior, shewing a way to find the longitude by the magnetical inclinatory needle.} (...) & $0.0040$ \\
3 & 1715 & Trades & Panciroli: \textit{The history of many memorable things lost, which were in use among the ancients: and an account of many excellent things found, now in use among the moderns, both natural and artificial.} (...) & $0.0029$ \\
4 & 1683 & Trades & Grew: \textit{New experiments, and useful observations concerning sea-water made fresh, according to the patentees invention} (...) & $0.0028$ \\
5 & 1717 & Agriculture & Solleysel: \textit{The compleat horseman: discovering the surest marks of the beauty, goodness, faults, and imperfections of horses} (...) & $0.0027$ \\
6 & 1680 & Navigation & Holden: \textit{The improvement of navigation a great cause of the increase of knowledge.} (...) & $0.0026$ \\
7 & 1718 & Trades & \textit{The Practical distiller: Or, A brief treatise of practical distillation.} (...) & $0.0024$ \\
8 & 1714 & Navigation & R. B.: \textit{Longitude to be found out with a new invented instrument, both by sea and land.} (...) & $0.0023$ \\
9 & 1660 & Instruments & Oughtred: \textit{The circles of proportion and the horizontall instrument \&c.} (...) & $0.0022$ \\
10 & 1703 & Navigation & \textit{Some remarks and observations on the report of the five commissioners to the Honourable House of Commons, dated the 25th of November, 1702.} (...) & $0.0022$ \\
\midrule
\multicolumn{4}{l}{All 186 titles with high spillover and high innovation (26 per cent of 717)} & $0.110$ \\
\bottomrule
\end{tabular}

					\begin{tablenotes}
						\item {\footnotesize \emph{Notes:} Titles with more received spillovers and more innovation than predicted, ranked by their contribution $\tilde{x}_i \tilde{y}_i / \sum_j \tilde{x}_j^2$ to the 1660--1719 coefficient of equation~\ref{eq:ols_spillover} for spillovers from $\Omega$ to $\lambda$, estimated with the periods 1660--1719, 1720--1739, and 1740--1779.}
					\end{tablenotes}
				\end{threeparttable}
			\end{adjustbox}
		\end{table}
		
		\begin{table}[!htbp]
			\centering
			\begin{adjustbox}{max width=1\columnwidth}
				\begin{threeparttable}\scriptsize
					\caption{High spillovers, high innovation: the ten titles with the largest contributions, 1740--1779}
					\label{tab:negcoef_hh_late}
					\begin{tabular}{rrlp{0.72\textwidth}r}
\toprule
 & Year & Field & Author and title & Contribution \\
\midrule
1 & 1742 & Instruments & Baker: \textit{The microscope made easy} (...) & $0.0024$ \\
2 & 1776 & Trades & Henley: \textit{Experiments and observations on a new apparatus, called, a machine for exhibiting perpetual electricity.} (...) & $0.0017$ \\
3 & 1773 & Patents & Barber (patent): \textit{machine, with the apparatus thereunto belonging, which, by fire, water, air, and steam, will purify fossil coal, extract metals from their ores} (...) & $0.0016$ \\
4 & 1758 & Trades & Combrune: \textit{An essay on brewing. With a view of establishing the principles of the art.} & $0.0013$ \\
5 & 1770 & Patents & Vallotton (patent): \textit{manufacture of stocking or hose gloves, night caps, waist- coat or breeches pieces, \&c, to be wove on a common stocking or other frame, or knit with needles} (...) & $0.0013$ \\
6 & 1756 & Agriculture & Home: \textit{The principles of agriculture and vegetation.} (...) & $0.0013$ \\
7 & 1758 & Patents & Dollond (patent): \textit{method of making the object glasses of refracting telescopes, by compounding mediums of different refractive qualitys} (...) & $0.0012$ \\
8 & 1768 & Patents & Whateley (patent): \textit{plating gold upon silver plated wire, \& drawing the same into wires of very fine sizes} (...) & $0.0012$ \\
9 & 1777 & Trades & Walker: \textit{A philosophical estimate of the causes, effects, and cure, of unwholsome air in large cities.} (...) & $0.0012$ \\
10 & 1747 & Navigation & Langwith: \textit{Observations on Doctor Arbuthnot's dissertations on coins, weights, and measures.} (...) & $0.0012$ \\
\midrule
\multicolumn{4}{l}{All 545 titles with high spillover and high innovation (28 per cent of 1,954)} & $0.140$ \\
\bottomrule
\end{tabular}

					\begin{tablenotes}
						\item {\footnotesize \emph{Notes:} As table~\ref{tab:negcoef_hh_early}, for the 1740--1779 coefficient. Patents are listed with the patentee's name and the description of the invention. }
					\end{tablenotes}
				\end{threeparttable}
			\end{adjustbox}
		\end{table}
		
		To capture the fields whose spillovers are driving the results, table~\ref{tab:negcoef_sources} estimates the baseline model from equation~\ref{eq:ols_spillover} with each field as the only source. For spillovers into $\lambda$ (panel A), the negative coefficients before 1720 come from \textit{astronomy}, which then included almanacs and astrology, from \textit{mathematics}, and from \textit{encyclopedias}. Spillovers from the experimental sciences, \textit{applied physics} and \textit{chemistry}, are never significantly negative, and by 1760--1779 the positive coefficient comes mainly from \textit{applied physics} and \textit{astronomy}. For spillovers into $\Omega$ (panel B), single $\lambda$ fields give small coefficients before 1720. After 1740, the positive coefficient comes from \textit{navigation} and \textit{scientific instruments}. Altogether, we find that the transition from negative to positive $\beta$ coefficients reflected structural changes in the knowledge economy, with a reversal of fields' impact, and crucially the emergence of applied physics as a source for innovation-inducing spillovers. 
		
		\begin{table}[!htbp]
			\centering
			\begin{adjustbox}{max width=1\columnwidth}
				\begin{threeparttable}\footnotesize
					\caption{Coefficients with a single field as the source of spillovers}
					\label{tab:negcoef_sources}
					\begin{tabular}{lcccccc}
\toprule
 & 1660--1679 & 1680--1699 & 1700--1719 & 1720--1739 & 1740--1759 & 1760--1779 \\
\midrule
\multicolumn{7}{l}{\textit{Panel A: Spillovers from a single $\Omega$ field $\rightarrow$ innovation in $\lambda$}} \\
\midrule
Applied physics & $-0.052$ (0.055) & $0.017$ (0.061) & $-0.027$ (0.035) & $-0.014$ (0.041) & $0.103^{***}$ (0.032) & $0.148^{***}$ (0.018) \\
Astronomy & $-0.067$ (0.085) & $-0.163^{***}$ (0.058) & $-0.139^{**}$ (0.062) & $0.016$ (0.037) & $0.082^{**}$ (0.034) & $0.106^{***}$ (0.017) \\
Mathematics & $0.029$ (0.085) & $-0.223^{***}$ (0.070) & $0.064$ (0.088) & $-0.006$ (0.030) & $-0.013$ (0.027) & $0.025$ (0.025) \\
Chemistry & $-0.036$ (0.027) & $0.047$ (0.035) & $-0.032$ (0.024) & $-0.010$ (0.011) & $0.046^{***}$ (0.011) & $0.030^{***}$ (0.010) \\
Encyclopedias and dictionaries & $-0.091$ (0.059) & $-0.219^{***}$ (0.051) & $-0.129^{*}$ (0.070) & $-0.037$ (0.040) & $-0.017$ (0.033) & $0.003$ (0.016) \\
All $\Omega$ fields (baseline) & $-0.153$ (0.103) & $-0.171^{**}$ (0.076) & $-0.122^{**}$ (0.058) & $-0.007$ (0.044) & $0.122^{***}$ (0.035) & $0.146^{***}$ (0.027) \\
\midrule
\multicolumn{7}{l}{\textit{Panel B: Spillovers from a single $\lambda$ field $\rightarrow$ innovation in $\Omega$}} \\
\midrule
Technical instructions: trades & $-0.021$ (0.049) & $-0.053$ (0.047) & $-0.021$ (0.042) & $-0.005$ (0.042) & $0.065$ (0.057) & $-0.027$ (0.044) \\
Technical instructions: agriculture & $-0.005$ (0.028) & $-0.033$ (0.024) & $-0.050^{***}$ (0.019) & $-0.012$ (0.030) & $0.043$ (0.039) & $-0.007$ (0.027) \\
Navigation & $-0.006$ (0.031) & $0.033$ (0.029) & $-0.003$ (0.027) & $-0.059^{*}$ (0.035) & $0.033$ (0.039) & $0.123^{***}$ (0.024) \\
Scientific instruments & $-0.053^{**}$ (0.024) & $-0.003$ (0.014) & $0.007$ (0.010) & $0.021^{*}$ (0.012) & $0.044^{***}$ (0.016) & $0.056^{**}$ (0.023) \\
All $\lambda$ fields (baseline) & $-0.014$ (0.036) & $0.014$ (0.027) & $-0.019$ (0.028) & $0.021$ (0.037) & $0.131^{***}$ (0.049) & $0.114^{**}$ (0.048) \\
\bottomrule
\end{tabular}

					\begin{tablenotes}
						\item {\footnotesize \emph{Notes:} Coefficients $\beta_p$ of equation~\ref{eq:ols_spillover} with the received spillover index computed from a single source field. Standard errors clustered by year in parentheses. Panel A: $N = 3{,}119$ (\textit{chemistry}: 2,672). Panel B: $N = 2{,}940$ (\textit{agriculture}: 2,930; \textit{scientific instruments}: 2,792). *** $p<0.01$, ** $p<0.05$, * $p<0.1$.}
					\end{tablenotes}
				\end{threeparttable}
			\end{adjustbox}
		\end{table}

		\FloatBarrier

		\section{Robustness: Main results}

		\subsection{Mechanical effects of changing pool size}
		\label{sec:mechanical}
		
		As argued in section~\ref{sec:empirical-framework}, for the interpretation of the estimated $\beta$ coefficients in equation~\ref{eq:ols_spillover}, we have to assume that the $\beta$ coefficients are not influenced by mechanical features of the innovation and spillover measure. As both, the innovation and received spillover measures rest on the comparison of forward and backward pool, it is important to ensure that changes in the pool size do not produce spurious results. 
		
		As shown in section~\ref{sec:timetrends}, the number of published titles in Britain grew over time, thereby mechanically leading to larger forward than backward pools for most periods. Assuming that titles are evenly embedded distributed across the embedding space, this would mechanically increase a title's likelihood of having similar top-20 forward pairs and thereby mechanically raise the innovation and spillover index.
				
		The paper directly addresses this feature of the innovation and spillover measure by adding subject and year fixed effects. As pool growing pool size should affect all titles within a subject-year simultaneously, such year-level effects are fully absorbed.		
		However, we might worry about heterogeneous effects of increasing pool size. For example, if titles are not evenly distributed across the embedding space, pool size effects might be larger for smaller sub-groups of titles. If these sub-groups are the ones that are more likely to receive spillovers, the relationship could mechanically drive the $\beta$ coefficients. 
		
		To test whether such patterns exist in the data, this section implements a twofold approach. First, both indices are recomputed on pools of fixed size. We ensure fixed size by replacing every pool that enters an index (backward, forward, source, and peer pool) by a random subsample of $n^{*}$ titles, drawn repeatedly, so that pool size cannot vary over time or across fields and only the geometry of the neighbourhood does.
		In the absence of spurious effects from increasing pool size, the estimated coefficients from this exercise should remain similar to the baseline results. Second, we directly test whether the current data structure could give rise to mechanical artifacts. For this, we permute publication years within fields, while drawing from the existing year distribution. This preserves pool size and growth patterns but removes the temporal structure of content, so that any coefficient it produces is mechanical. Any significant results would point towards spurious results.

\begin{figure}[!htbp]
			\centering
			\begin{subfigure}[t]{0.49\textwidth}
				\includegraphics[width=1\linewidth]{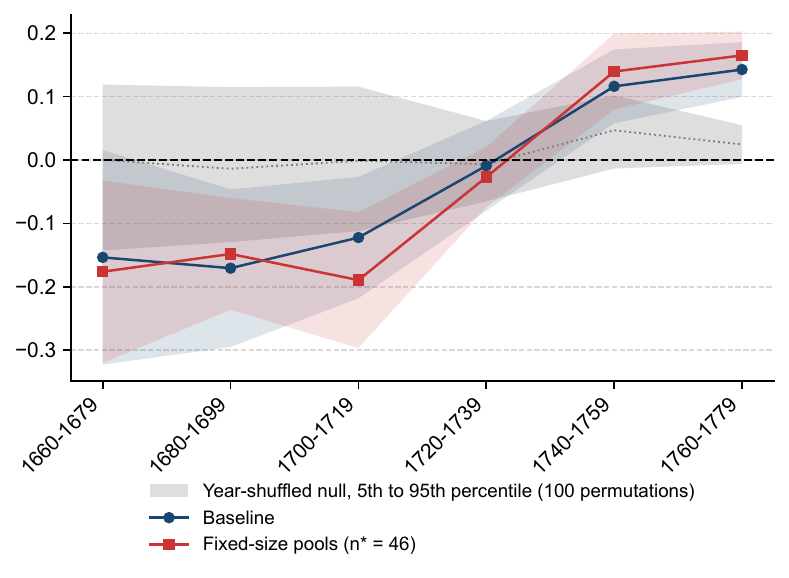}
				\subcaption{Spillovers from $\Omega$ $\rightarrow$ innovation in $\lambda$}
			\end{subfigure}
			\begin{subfigure}[t]{0.49\textwidth}
				\includegraphics[width=1\linewidth]{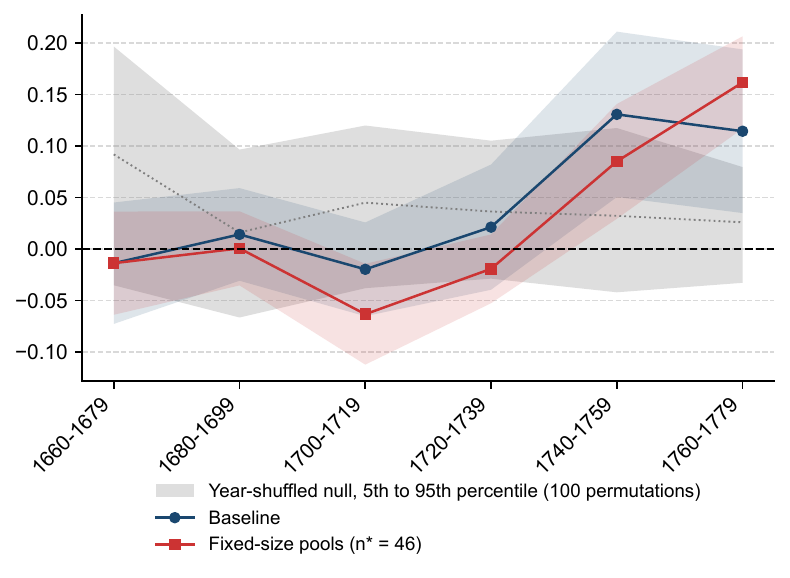}
				\subcaption{Spillovers from $\lambda$ $\rightarrow$ innovation in $\Omega$}
			\end{subfigure}
			\caption{Baseline coefficients, fixed-size pools, and the year-shuffled null \vspace{4pt} \newline \footnotesize \emph{Notes:} Coefficients $\beta_p$ from equation~\ref{eq:ols_spillover}, estimated on the baseline indices (navy) and on indices computed from pools of fixed size $n^{*} = $ 46 (red: each pool replaced by a random subsample of $n^{*}$ titles, 50 draws). The grey band is the 5th to 95th percentile of the coefficients under a null in which publication years are permuted within field. $N = $ 3,119 (a) and 2,940 (b). 90\% confidence intervals estimated using standard errors clustered at the publication year.}
			\label{fig:pool_size_check}
		\end{figure}

		Figure~\ref{fig:pool_size_check} presents the results from these two exercises. We find no evidence of mechanical artifacts in either approach. The figure plots the baseline coefficients from section~\ref{sec:empirical_results} together with estimated results when mechanically constructing fixed pool sizes. The results are almost identical, indicating that the baseline results were not driven by pool size artifacts. As even stronger evidence, we permute the years within our sample by drawing from the same year-distribution, preserving the temporal structure, but removing the structure of semantic information. As expected, this exercise yield precisely estimated coefficients (in gray) that are close to zero. Hence, these exercises show that we do not find empirical evidence that growing pool size is mechanically affecting our estimates from the main section.

		\subsection{Title length}
		\label{sec:app:title_length}

		This section addresses the concern that similarities in the embedding space might be driven by the length of titles rather than by their content. Like most sentence transformers, \textit{SteamBERTh} represents a title as the mean of its token embeddings. Averaging over more tokens removes idiosyncratic variation, so that the embeddings of long titles lie closer to the common direction of the corpus and closer to each other. Similarity then rises with length, independently of content. For the results of section~\ref{sec:results}, this matters if length changes the relationship between the innovation and the received spillover index over time. The word count controls in equation~\ref{eq:ols_spillover} address this only in part. They absorb the effect of a title's own length, but not the effect of the length of the titles it is compared to.

		To remove the length component from the embedding space, the paper constructs length-neutral embeddings. Each of the 768 dimensions of the \textit{SteamBERTh} embeddings is regressed on the log word count of the title and its square. We then subtract the fitted component and re-normalise each embedding to unit length. This removes the part of the embedding that is linearly predictable from length, while keeping the full text of every title. Both indices from equations~\ref{eq:innovation} and~\ref{eq:received_spillover} are then recomputed from the length-neutral embeddings with the baseline parameters.

\begin{figure}[!htbp]
			\centering
			\includegraphics[width=0.85\linewidth]{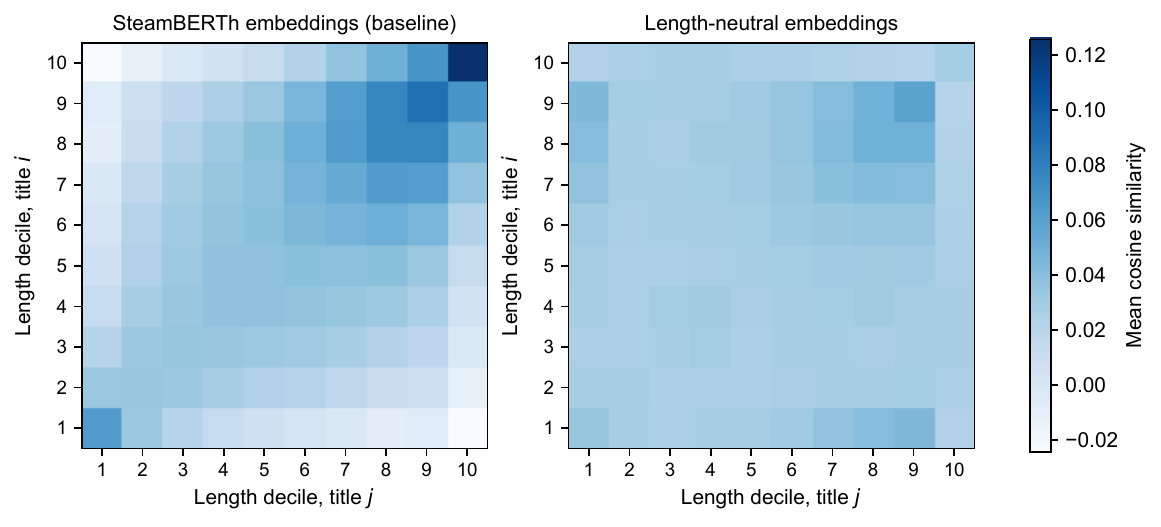}
			\caption{Mean pairwise cosine similarity by title length \vspace{4pt} \newline \footnotesize \emph{Notes:} One million random pairs of $\Omega$ and $\lambda$ titles and patents, 1600--1800. Each cell shows the mean cosine similarity of all pairs whose two titles fall into the given deciles of word count. Left panel: baseline \textit{SteamBERTh} embeddings. Right panel: length-neutral embeddings.}
			\label{fig:title_length_heatmap}
		\end{figure}

		Figure~\ref{fig:title_length_heatmap} shows that this correction works. For one million random pairs of titles, the figure plots the mean cosine similarity by the length deciles of the two titles. With the baseline embeddings (left panel), similarity rises steadily with the length of both titles. Holding the pair of subject fields and the distance between publication years fixed, one log point more in the mean word count of a pair raises its cosine similarity by 0.014 (s.e. 0.0002), and length raises the $R^2$ from 0.057 to 0.102. With length-neutral embeddings (right panel), this gradient disappears. The coefficient falls to $-0.0005$, and length no longer adds to the $R^2$.\footnote{Within the same field and twenty-year period, a smaller length gradient remains (0.021 against 0.036 with the baseline embeddings), since the correction removes the length component common to the whole corpus.} At the same time, the correction leaves the content of the embedding space intact. The subject field of 76 per cent of titles is correctly predicted by a majority vote of their 20 nearest neighbours with both embeddings, and the elasticity of ESTC editions with respect to the innovation index barely changes (0.174 against 0.186).

\begin{figure}[!htbp]
			\centering
			\begin{subfigure}[t]{0.49\textwidth}
				\includegraphics[width=1\linewidth]{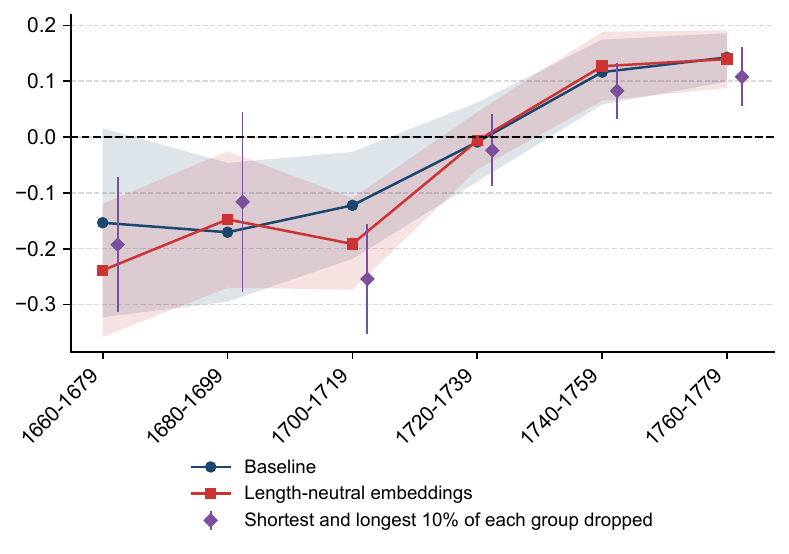}
				\subcaption{Spillovers from $\Omega$ $\rightarrow$ innovation in $\lambda$}
			\end{subfigure}
			\begin{subfigure}[t]{0.49\textwidth}
				\includegraphics[width=1\linewidth]{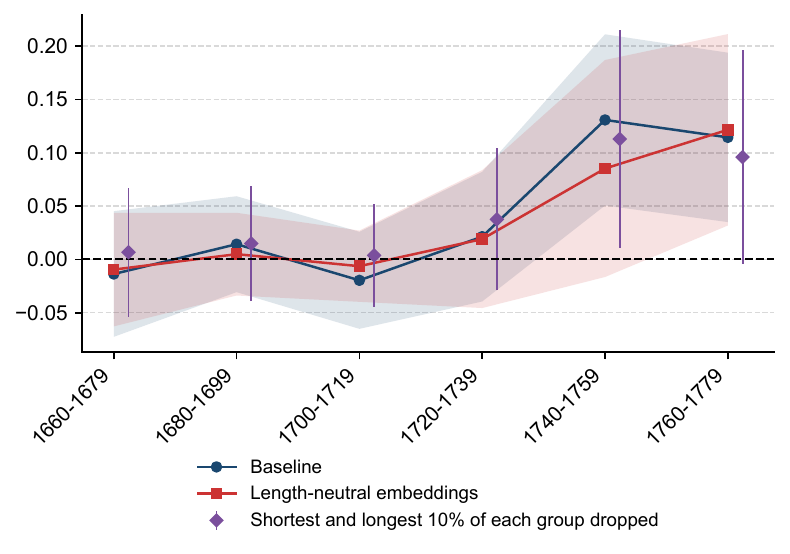}
				\subcaption{Spillovers from $\lambda$ $\rightarrow$ innovation in $\Omega$}
			\end{subfigure}
			\caption{Main results with length-neutral embeddings and without titles of extreme length \vspace{4pt} \newline \footnotesize \emph{Notes:} Coefficients $\beta_p$ from equation~\ref{eq:ols_spillover}. Navy: Baseline. Red: Both indices recomputed from length-neutral embeddings, from which the component of each embedding that is linearly predictable from log word count and its square has been removed. Purple: baseline indices, with the shortest and longest 10 per cent of titles of each group ($\Omega$, $\lambda$ titles from the ESTC, and patents) dropped.   $N = $ 3,119 (a) and 2,940 (b) in the baseline, 2,461 (a) and 2,346 (b) without titles of extreme length. 90\% confidence intervals estimated using standard errors clustered at the publication year level.}
			\label{fig:title_length_robust}
		\end{figure}

		Figure~\ref{fig:title_length_robust} reports the coefficients of equation~\ref{eq:ols_spillover} with both indices recomputed from the length-neutral embeddings (red), together with the baseline (navy). The results are practically unchanged. In 1760--1779, the coefficient for spillovers from $\Omega$ to $\lambda$ is 0.140 (baseline: 0.143) and the coefficient for spillovers from $\lambda$ to $\Omega$ is 0.122 (baseline: 0.114). The negative seventeenth-century coefficients for spillovers from $\Omega$ to $\lambda$ remain. 
		
		As a second test, we drop the titles at the extremes of the length distribution, where the effect of length on similarity is strongest (figure~\ref{fig:title_length_heatmap}). For this, we exclude the shortest and longest 10 per cent of titles within each group ($\Omega$, $\lambda$ titles from the ESTC, and patents) and re-estimate equation~\ref{eq:ols_spillover} with the baseline indices. This reduces the sample by about a fifth. Figure~\ref{fig:title_length_robust} reports the results in purple. The coefficients follow the baseline pattern, with positive spillovers emerging after 1740. In 1760--1779, the coefficients are 0.108 for spillovers from $\Omega$ to $\lambda$ and 0.096 for spillovers from $\lambda$ to $\Omega$. Both lie within the confidence bands of the baseline, although the smaller sample makes the latter coefficient, as well as its counterpart for 1740--1759, less precise.

		Altogether, we find no evidence that title length drives the results. The emergence of positive spillovers in both directions at the end of the eighteenth century holds when length is removed from the embedding space. Dropping titles of extreme length preserves the pattern, at some cost in precision.

		\subsection{LLM-based rewriting of original titles in a simplified style}
		\label{appendix:section:llm_rewrite}
		
		This section addresses the concern that changes in cosine similarities used for the innovation measure from equation~\ref{eq:innovation} and the spillover measure from equation~\ref{eq:spillover} might also be driven by changes in linguistic style. As a main concern, we can imagine that more sophisticated writing styles could have been associated with a higher author capability and hence with higher innovative capability. Likewise changes in writing style might have had field specific spillovers, especially from high prestige fields such as science. The previous placebo results can only address this issue indirectly.
		
		Therefore, the paper introduces an additional robustness test where we directly simplify the writing style of titles by directing an LLM to rewrite the titles \textit{in a simpler style}. For this, a \texttt{gpt-4o-2024-08-06} model was prompted to rewrite the titles using a simpler style while preserving the semantic content. The full prompt is documented in figure~\ref{fig:system_prompt}. The following example illustrates the output:
		
		\begin{quote}
			(Original title:)
			Hydrostatical tryals and conclusions. That fluids gravitate in proprio loco, the upper parts continually pressing upon the lower; that this pressure is not only propagated downwards, but even upwards and side-ways, according to all possible directions
		\end{quote}

		\begin{quote}
			(Rewritten title:)
			Experiments and findings in hydrostatics. Fluids have weight in their own place, with the top layers constantly pressing on the ones below. This pressure spreads not just downward, but also upward and sideways, in every possible direction
		\end{quote}
		
		As can be seen, the rewriting exercise leads both to a simplified style as well as a significant permutation of writing style. We first document this using conjunction based measures of textual complexity in table~\ref{tab:syntax_complexity_rewrite}. Here we find that rewritten titles have less conjunctional complexity, i.e. use simpler sentence structures. Next, we validate the extent to which the LLM permutated titles employing Jaccard similarity as a simple measure of word overlap.\footnote{Jaccard similarity between title A and B is defined as: $J(A,B) = \frac{|A \cap B|}{|A \cup B|}$.} We document this in table~\ref{tab:jaccard_distance_rewrite}. We find an average Jaccard similarity of  $0.485$ indicating that the LLM-based rewriting approach exchanged more than half of all words in the original titles.
		
		Hence, this rewriting exercise makes it possible to re-estimate the model from section~\ref{sec:empirical_results} on the rewritten titles. Following the rewriting exercise, any results that were purely driven by linguistic style should disappear. 
		The approach therefore serves as a useful test to rule out confounders from non-semantic content. However, as a cautionary note, we add that it is not clear whether this approach is fully successful in preserving the original semantic information. Alterations might include modern interpretations imposed by the LLM processing the text. Moreover, any rewriting exercise comes at the cost of some informational loss. Hence, this exercise should be mainly interpreted as a robustness test.
		
		\begin{table}[!htbp]
			\centering
			\begin{adjustbox}{max width=8.5\columnwidth}
				\begin{threeparttable}\fontsize{10}{13}\selectfont
					\caption{Comparison: Syntactic complexity between original and LLM rewritten text}
					\label{tab:syntax_complexity_rewrite}
					\begin{tabular}{lrrr}
\toprule
Metric & Translated & Simplified & $\Delta$ (S-T) \\
\midrule
Preposition density & 0.170900 & 0.148500 & -0.022500 \\
Conjunction density & 0.088700 & 0.088100 & -0.000600 \\
Comma density & 0.088600 & 0.084200 & -0.004400 \\
Colon presence rate & 0.141100 & 0.120100 & -0.021000 \\
\midrule
\bottomrule
\end{tabular}

					\begin{tablenotes}
						\item {\footnotesize \emph{Notes:} The table presents measures of tokenized synctactic complexity. Titles are tokenized into lowercase words, and each measure is express relative to the number of tokens. Preposition density counts occurrences of relational prepositions, \textit{ of, for, with, in, on, by, from, upon, into, over, under, between, among, through, during, before, after}, while conjunction density counts coordinating conjunctions \textit{and, or, with} and comma density counts \textit{commas} per token. Colon presence indicates whether a colon separator (``:'') appears. }
					\end{tablenotes}
				\end{threeparttable}
			\end{adjustbox}
		\end{table}

		\begin{table}[!htbp]
			\centering
			\begin{adjustbox}{max width=8.5\columnwidth}
				\begin{threeparttable}\fontsize{10}{13}\selectfont
					\caption{Jaccard distance between original and LLM rewritten text}
					\label{tab:jaccard_distance_rewrite}
					\begin{tabular}{lrrr}
\toprule
Subject class & Mean & Median & Std. dev. \\
\midrule
Applied physics & 0.535 & 0.525 & 0.149 \\
Astronomy & 0.509 & 0.500 & 0.142 \\
Chemistry & 0.541 & 0.543 & 0.165 \\
Encyclopedias and dictionaries & 0.527 & 0.514 & 0.139 \\
Mathematics & 0.496 & 0.485 & 0.144 \\
Navigation & 0.532 & 0.527 & 0.139 \\
Scientific instruments & 0.549 & 0.523 & 0.153 \\
Technical instructions Agriculture & 0.476 & 0.463 & 0.154 \\
Technical instructions Trades & 0.497 & 0.485 & 0.140 \\
Patents & 0.466 & 0.455 & 0.169 \\
\midrule
All titles & 0.485 & 0.476 & 0.161 \\
\bottomrule
\end{tabular}

					\begin{tablenotes}
						\item {\footnotesize \emph{Notes:} The table reports Jaccard distances between the original ESTC titles and titles with simpler style. Jaccard similarity between title A and B is defined as: $J(A,B) = \frac{|A \cap B|}{|A \cup B|}$. Titles with simpler style are generated by an LLM with instruction to write in simpler style. The prompt is documented in figure~\ref{fig:system_prompt}. }
					\end{tablenotes}
				\end{threeparttable}
			\end{adjustbox}
		\end{table}

\begin{figure}[!htbp]
			\centering
			\tiny
			\begin{subfigure}[t]{0.49\textwidth}
				\includegraphics[width=1\linewidth]{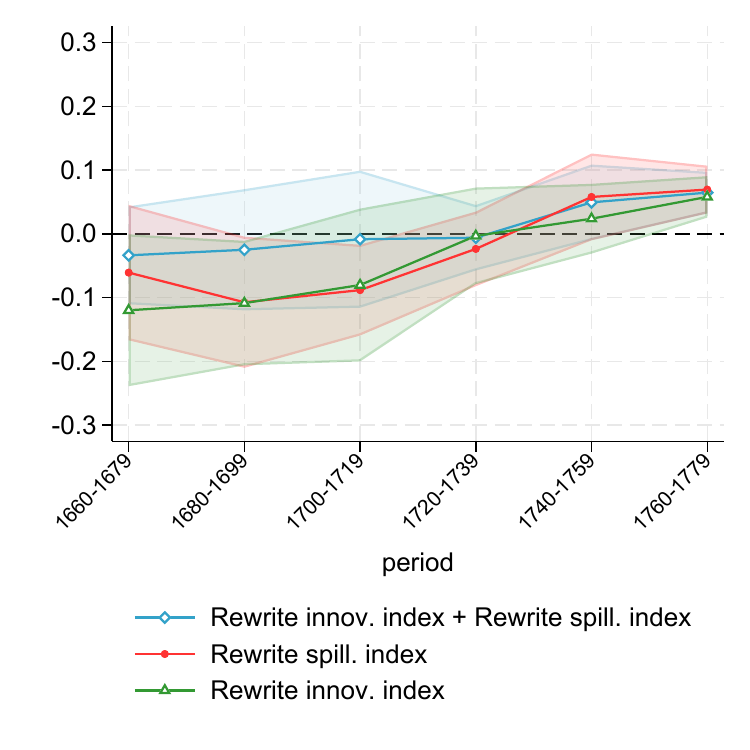}
				\subcaption{Spillovers from $\Omega$ $\rightarrow$ innovation in $\lambda$}
			\end{subfigure}
			\begin{subfigure}[t]{0.49\textwidth}
				\includegraphics[width=1\linewidth]{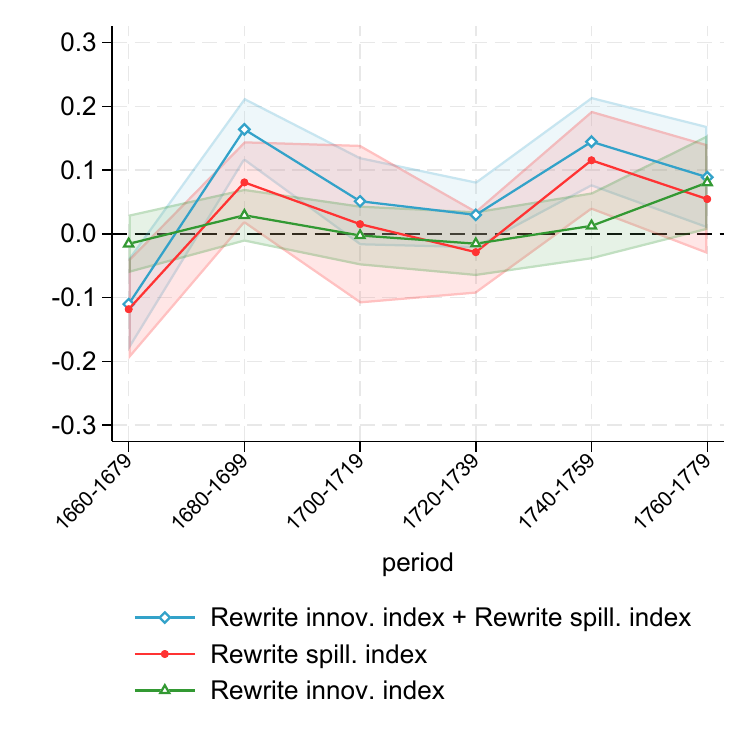}
				\subcaption{Spillovers from $\lambda$ $\rightarrow$ innovation in $\Omega$ }
			\end{subfigure}
			\caption{Results for text data with LLM-rewritten simplified style \vspace{4pt} \newline \footnotesize \emph{Notes:}  The figure reports results baseline results from equation~\ref{eq:ols_spillover} with alternative innovation and received spillover measures based on LLM-rewritten text. Each subfigure presents three specifications where either the innovation, spillover, or both measures are constructed from the rewritten text. 90\% confidence intervals estimated using standard errors clustered at the publication year level.}
			\label{fig:rewriting_exercise}	
		\end{figure}

		Figure~\ref{fig:rewriting_exercise} presents the estimation results using the innovation and spillover indices constructed from the rewritten texts. We report three specifications. First, we re-estimate the baseline model in equation~\ref{eq:ols_spillover}, using the innovation measure derived from the rewritten text as the dependent variable. Second, we present results in which the received spillover measure derived from the rewritten text is the sole explanatory variable. Finally, we report a specification in which both the innovation and received spillover measures are constructed from the rewritten text. Note that in itself, it should be sufficient to change the style in the construction of only the dependent or independent variable to rule out that similar style produced a spurious relationship between the dependent and independent variable. At the same time, using rewritten text for both the independent and dependent variable would introduce most noise from semantic changes during the rewriting process. We find it reassuring that our main results holds throughout all of these specifications.

		Results are presented in Figure~\ref{fig:rewriting_exercise}. They confirm the main patterns found in our baseline analysis. We find a clear positive association between knowledge spillovers and innovation at the end of our period. We can also note some second order differences. First, for spillovers from propositional ($\Omega$) to prescriptive ($\lambda$) knowledge, in one of the three specification, the negative coefficient found for 1660-1669 disappears. Second, for spillovers from prescriptive ($\lambda$) to propositional ($\Omega$) knowledge, we find evidence of a one-time positive spillover in the period 1680--1699 in two of the three specifications.\footnote{Note that this is the period of the great scientific breakthroughs in English science, including the publication of Newton's \textit{Principia} in 1687.} Generally, deviations from the baseline results seem to occur in the earlier periods where we have less observations. Yet overall, the results from rewriting exercise show little deviations from the earlier baseline results from section~\ref{sec:empirical_results}. Most of all, the main pattern, positive knowledge spillovers at the end of the period, remains robust under all different specifications.  We take this as additional evidence that the empirical framework captures informational flows beyond purely stylistic changes.

		\begin{figure}[!htbp]
			\noindent\textbf{System prompt:} 
			
			\begin{quote}
				\ttfamily
				\small
				You are given a 17th/18th century text. Rewrite the text using a simpler
				style while containing all original details. Make sure to contain all the
				original information. Change only the style but make sure the content stays
				the same. Don't unnecessarily shorten or lengthen the text or exclude
				technical details. Do not add details that were not in the original. Also
				stick closely to the original 17th/18th century language you find in the
				source.
			\end{quote}
			
			\noindent\textbf{Model:} \texttt{gpt-4o-2024-08-06}
			
			\caption{Documentation of system prompt}
			\label{fig:system_prompt}
		\end{figure}

		\FloatBarrier

		\subsection{Quantile regression for 1760--1779 coefficients}

		\begin{figure}[H]
			\centering
			\tiny
			\begin{subfigure}[t]{0.4\textwidth}
				\includegraphics[width=1\linewidth]{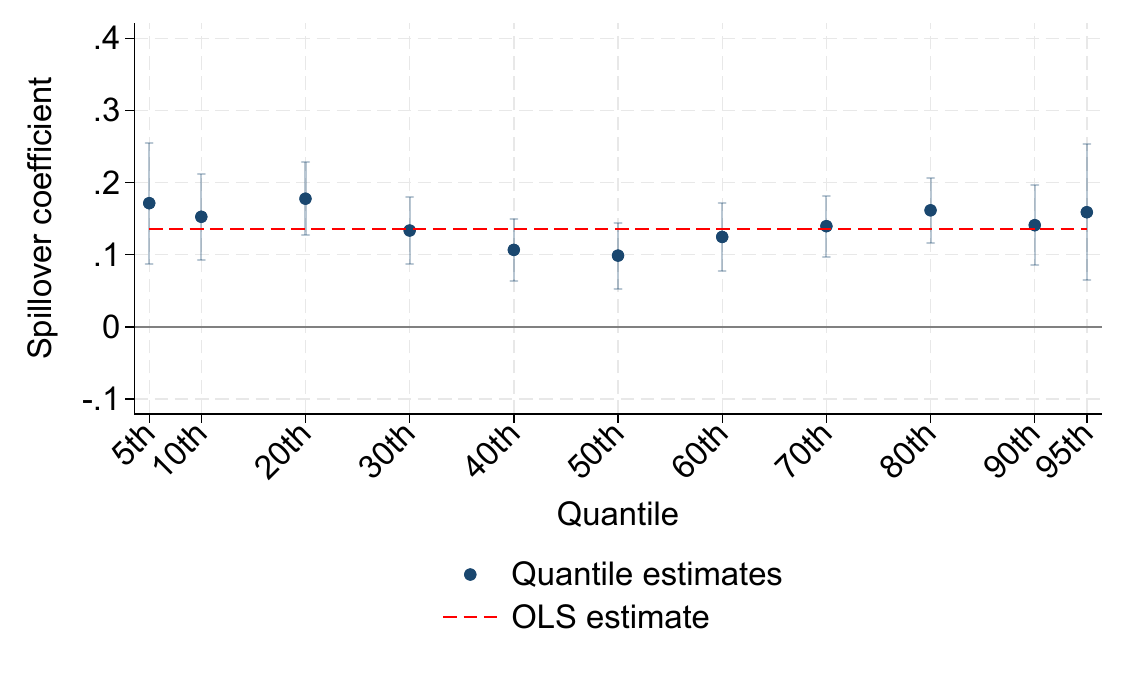}
				\subcaption{Spillovers from $\Omega$}
			\end{subfigure}
			\begin{subfigure}[t]{0.4\textwidth}
				\includegraphics[width=1\linewidth]{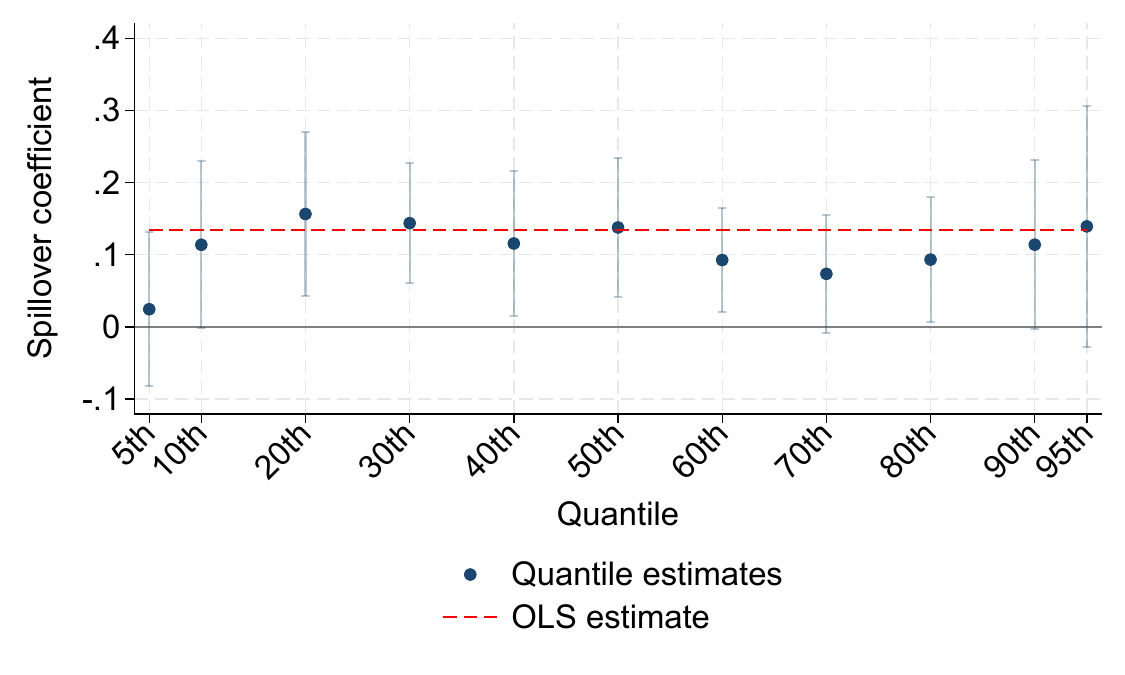}
				\subcaption{Spillovers from $\lambda$}
			\end{subfigure}
			\caption{Quantile regression for 1760--1779 coefficients \vspace{4pt} \newline \footnotesize \emph{Notes:}  The figure reports estimated coefficients for the  model from equation~\ref{eq:ols_spillover} using a quantile regressions. First, the figure reports OLS coefficients corresponding to figure~\ref{fig:results_prop_prescr}. Bootstrapped standard error reported at the 90\% level.}
			\label{fig:results_quantile}	
		\end{figure}
		
		\FloatBarrier

		\subsection{Different \textit{k} and \textit{p} in innovation index formula}
		\label{app:sec:robustness_k_rho}

\begin{figure}[!htbp]
			\centering
			\tiny
			{\footnotesize \textit{Spillovers from $\Omega$ $\rightarrow$ innovation in $\lambda$: different $k$ in the innovation index}}\par\smallskip
			\begin{subfigure}[t]{0.25\textwidth}
				\includegraphics[width=1\linewidth]{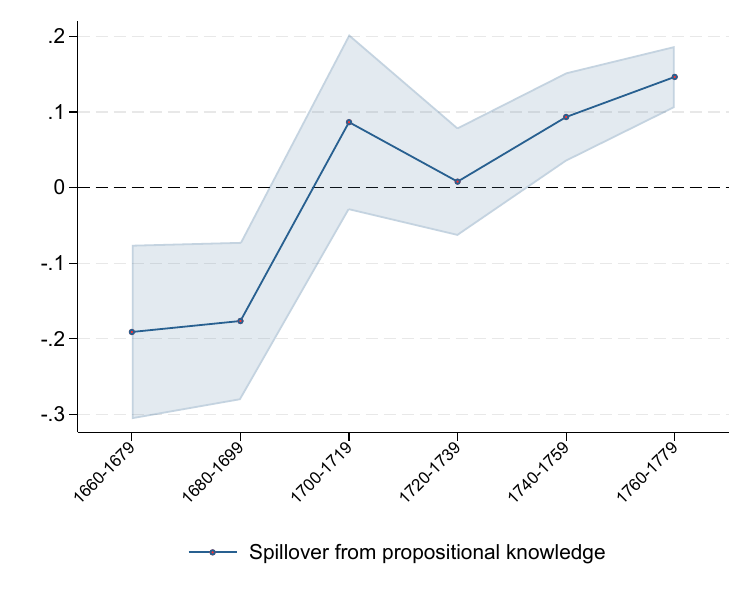}
				\subcaption{$k=10$}
			\end{subfigure}
			\begin{subfigure}[t]{0.25\textwidth}
				\includegraphics[width=1\linewidth]{graphs/spillover_prop_to_prescr_subset_20yrs.pdf}
				\subcaption{$k=20$}
			\end{subfigure}
			\begin{subfigure}[t]{0.25\textwidth}
				\includegraphics[width=1\linewidth]{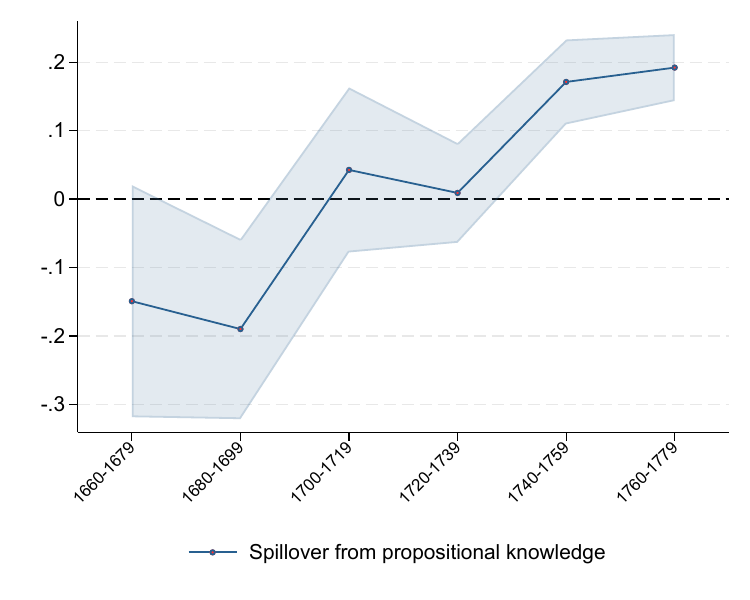}
				\subcaption{$k=30$}
			\end{subfigure}

			\medskip
			{\footnotesize \textit{Spillovers from $\Omega$ $\rightarrow$ innovation in $\lambda$: different $k$ and $p$}}\par\smallskip
			\begin{subfigure}[t]{0.25\textwidth}
				\includegraphics[width=1\linewidth]{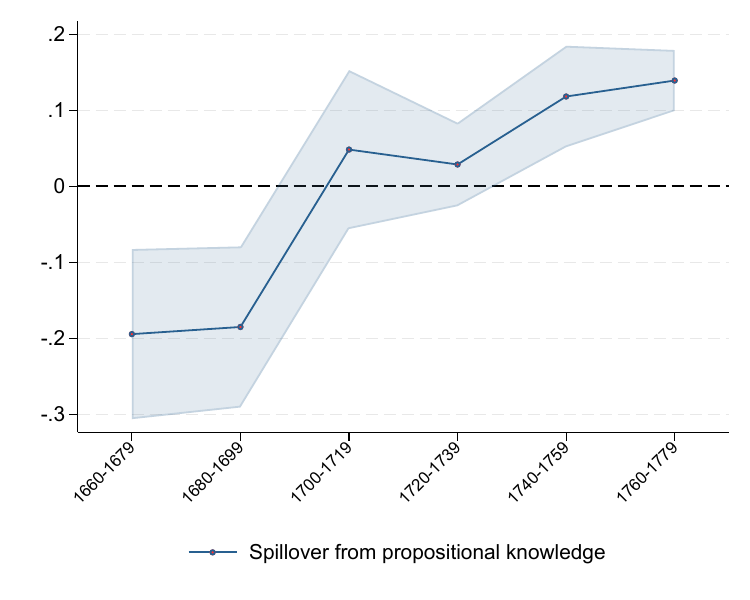}
				\subcaption{$k,p=10$}
			\end{subfigure}
			\begin{subfigure}[t]{0.25\textwidth}
				\includegraphics[width=1\linewidth]{graphs/spillover_prop_to_prescr_subset_20yrs.pdf}
				\subcaption{$k,p=20$}
			\end{subfigure}
			\begin{subfigure}[t]{0.25\textwidth}
				\includegraphics[width=1\linewidth]{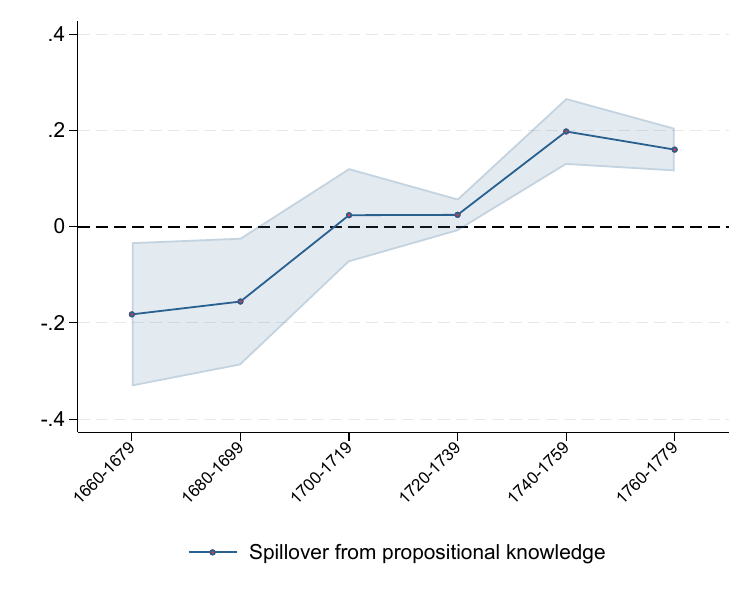}
				\subcaption{$k,p=30$}
			\end{subfigure}

			\medskip
			{\footnotesize \textit{Spillovers from $\lambda$ $\rightarrow$ innovation in $\Omega$: different $k$ in the innovation index}}\par\smallskip
			\begin{subfigure}[t]{0.25\textwidth}
				\includegraphics[width=1\linewidth]{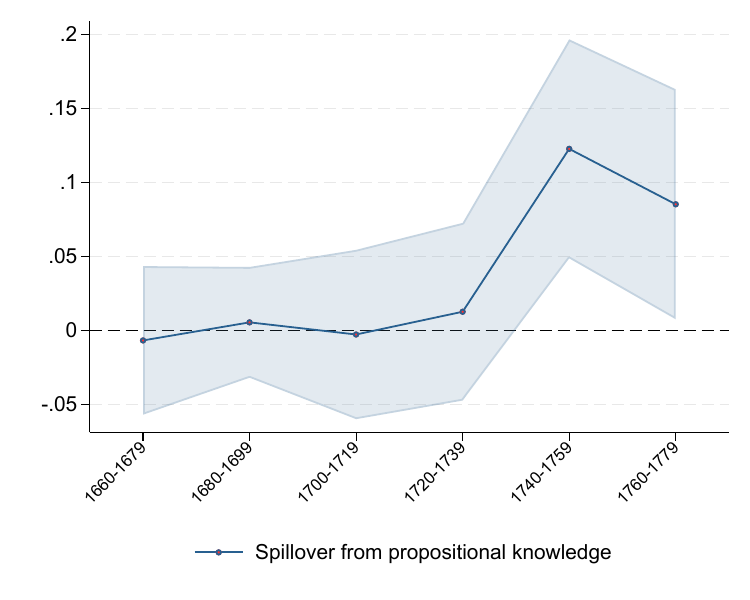}
				\subcaption{$k=10$}
			\end{subfigure}
			\begin{subfigure}[t]{0.25\textwidth}
				\includegraphics[width=1\linewidth]{graphs/spillover_presc_to_prop_subset_20yrs.pdf}
				\subcaption{$k=20$}
			\end{subfigure}
			\begin{subfigure}[t]{0.25\textwidth}
				\includegraphics[width=1\linewidth]{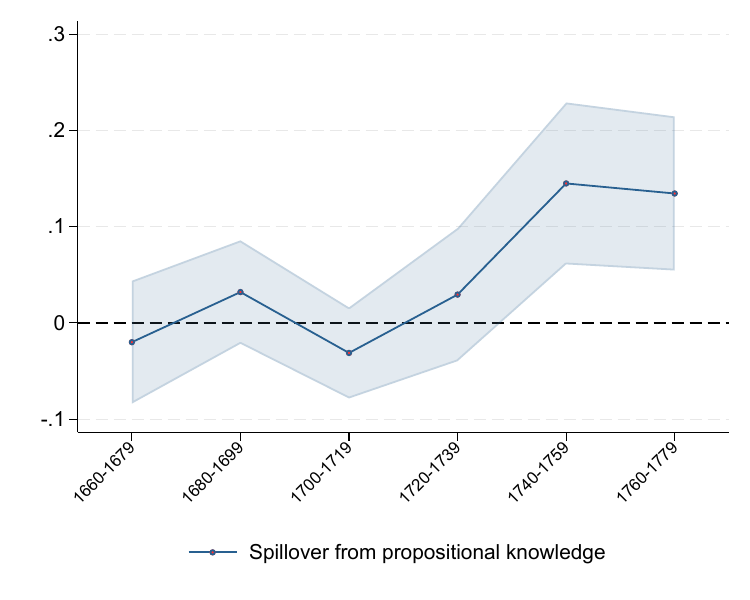}
				\subcaption{$k=30$}
			\end{subfigure}

			\medskip
			{\footnotesize \textit{Spillovers from $\lambda$ $\rightarrow$ innovation in $\Omega$: different $k$ and $p$}}\par\smallskip
			\begin{subfigure}[t]{0.25\textwidth}
				\includegraphics[width=1\linewidth]{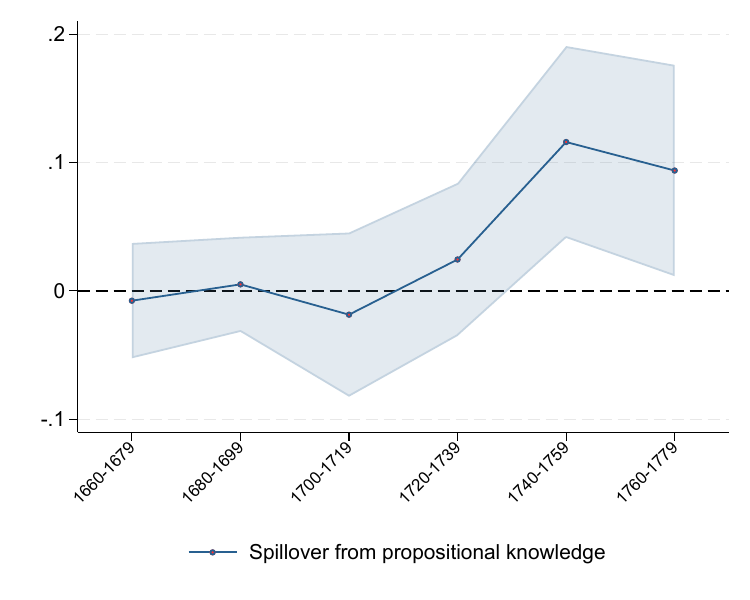}
				\subcaption{$k,p=10$}
			\end{subfigure}
			\begin{subfigure}[t]{0.25\textwidth}
				\includegraphics[width=1\linewidth]{graphs/spillover_presc_to_prop_subset_20yrs.pdf}
				\subcaption{$k,p=20$}
			\end{subfigure}
			\begin{subfigure}[t]{0.25\textwidth}
				\includegraphics[width=1\linewidth]{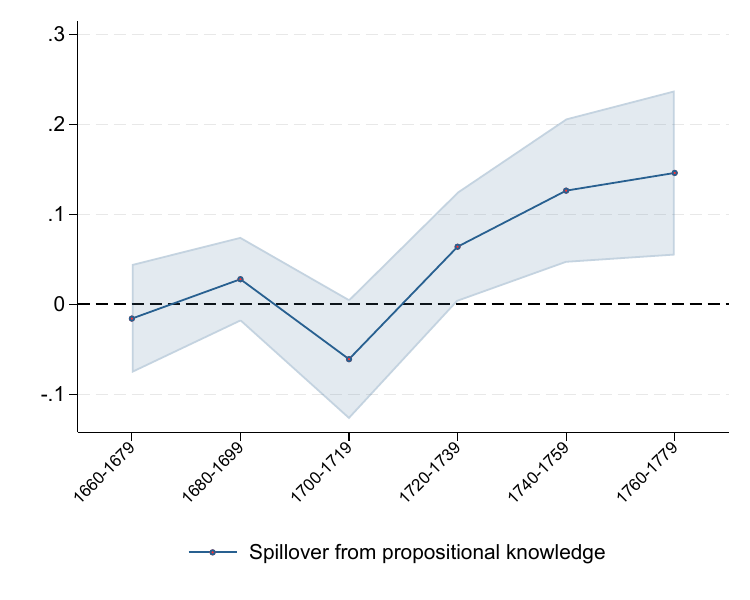}
				\subcaption{$k,p=30$}
			\end{subfigure}
			\caption{Robustness for different values of $k$ and $p$ in the calculation of the innovation and received spillover index \vspace{4pt} \newline \footnotesize \emph{Notes:} The figure reports the estimated coefficients from equation~\ref{eq:ols_spillover}. The dependent variable is the innovation index from equation~\ref{eq:innovation}. The plotted independent variable is an interaction term between the received spillover index from equation~\ref{eq:received_spillover} and twenty year time periods. The model further controls for the word count of titles and includes year fixed effects. Panels a)--f) report spillovers from $\Omega$ to $\lambda$, panels g)--l) spillovers from $\lambda$ to $\Omega$. Panels a)--c) and g)--i) change the $k$ parameter controlling the top-$k$ comparisons for calculating the innovation index from equation~\ref{eq:innovation}. Panels d)--f) and j)--l) in parallel also change the $p$ parameter controlling the size of the within-subject backwards counterfactual group from equation~\ref{eq:received_spillover}. The middle column reports the baseline with $k,p=20$. 90\% confidence intervals estimated using standard errors clustered at the publication year.}
			\label{fig:robustness_k_p}
		\end{figure}

		Figure~\ref{fig:robustness_k_p} reports the baseline results for different values of the top-$k$ parameter of the innovation index and of the peer parameter $p$ of the received spillover index.

		\subsection{Longer $\tau$ time windows for calculation of innovation and spillover index}
		\label{sec:different_time_windows}
		
\begin{figure}[!htbp]
			\centering
			\tiny
			{\footnotesize \textit{Spillovers from $\Omega$ $\rightarrow$ innovation in $\lambda$}}\par\smallskip
			\begin{subfigure}[t]{0.3\textwidth}
				\includegraphics[width=1\linewidth]{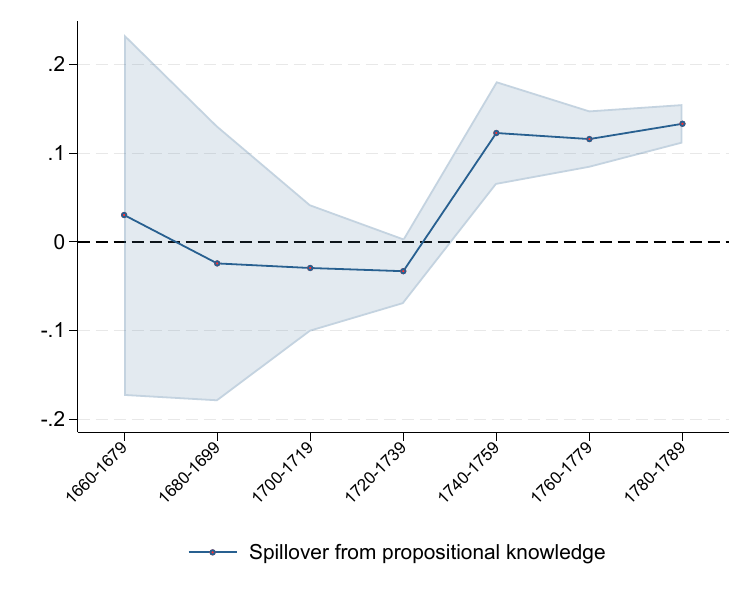}
				\subcaption{$\tau=10$}
			\end{subfigure}
			\begin{subfigure}[t]{0.3\textwidth}
				\includegraphics[width=1\linewidth]{graphs/spillover_prop_to_prescr_subset_20yrs.pdf}
				\subcaption{$\tau=20$}
			\end{subfigure}
			\begin{subfigure}[t]{0.3\textwidth}
				\includegraphics[width=1\linewidth]{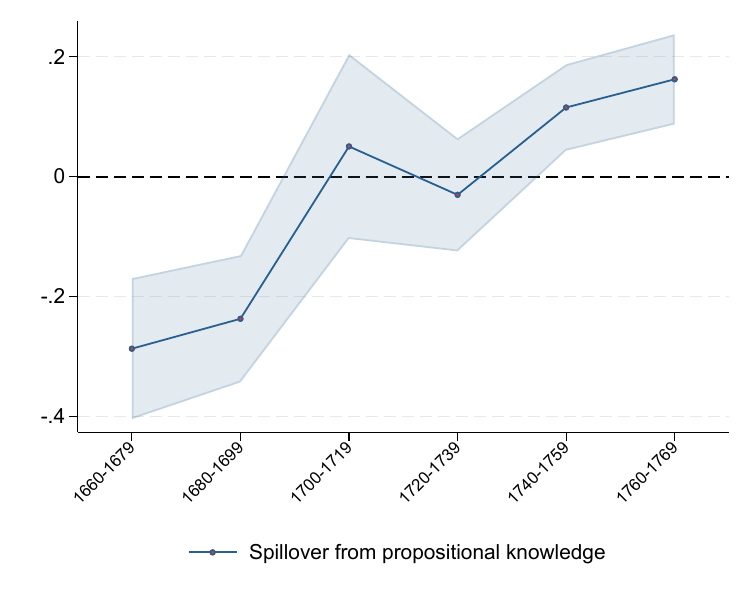}
				\subcaption{$\tau=30$}
			\end{subfigure}

			\medskip
			{\footnotesize \textit{Spillovers from $\lambda$ $\rightarrow$ innovation in $\Omega$}}\par\smallskip
			\begin{subfigure}[t]{0.3\textwidth}
				\includegraphics[width=1\linewidth]{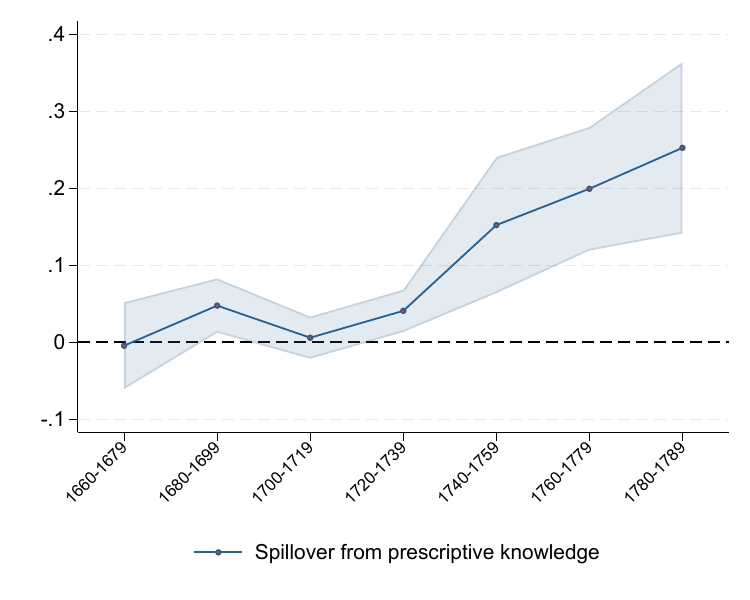}
				\subcaption{$\tau=10$}
			\end{subfigure}
			\begin{subfigure}[t]{0.3\textwidth}
				\includegraphics[width=1\linewidth]{graphs/spillover_presc_to_prop_subset_20yrs.pdf}
				\subcaption{$\tau=20$}
			\end{subfigure}
			\begin{subfigure}[t]{0.3\textwidth}
				\includegraphics[width=1\linewidth]{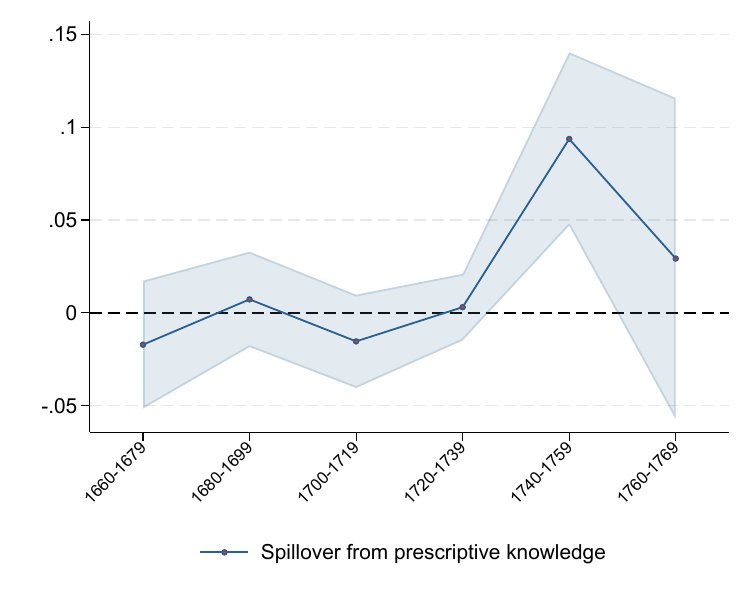}
				\subcaption{$\tau=30$}
			\end{subfigure}
			\caption{Robustness for different values of the time windows $\tau$ for the calculation of the innovation and spillover index \vspace{4pt} \newline \footnotesize \emph{Notes:} The figure reports the estimated coefficients from equation~\ref{eq:ols_spillover}. The dependent variable is the innovation index from equation~\ref{eq:innovation}. The plotted independent variable is an interaction term between the received spillover index from equation~\ref{eq:received_spillover} and twenty year time periods. The model further controls for the word count of titles and includes year fixed effects. Panels a)--c) report spillovers from $\Omega$ to $\lambda$, panels d)--f) spillovers from $\lambda$ to $\Omega$, when changing the $\tau$ parameter for the backward and forward comparison windows in equation~\ref{eq:innovation} and~\ref{eq:received_spillover}. Panels b) and e) report the baseline results with $\tau=20$. 90\% confidence intervals estimated using standard errors clustered at the publication year.}
			\label{fig:robustness_time_windows}
		\end{figure}

		Figure~\ref{fig:robustness_time_windows} reports results for different lengths of the time windows $\tau$.

		\subsection{Spillovers from individual fields}
		
\begin{figure}[!htbp]
			\centering
			\tiny
			{\footnotesize \textit{Spillovers from individual fields in $\Omega$ $\rightarrow$ innovation in $\lambda$}}\par\smallskip
			\begin{subfigure}[t]{0.19\textwidth}
				\includegraphics[width=1\linewidth]{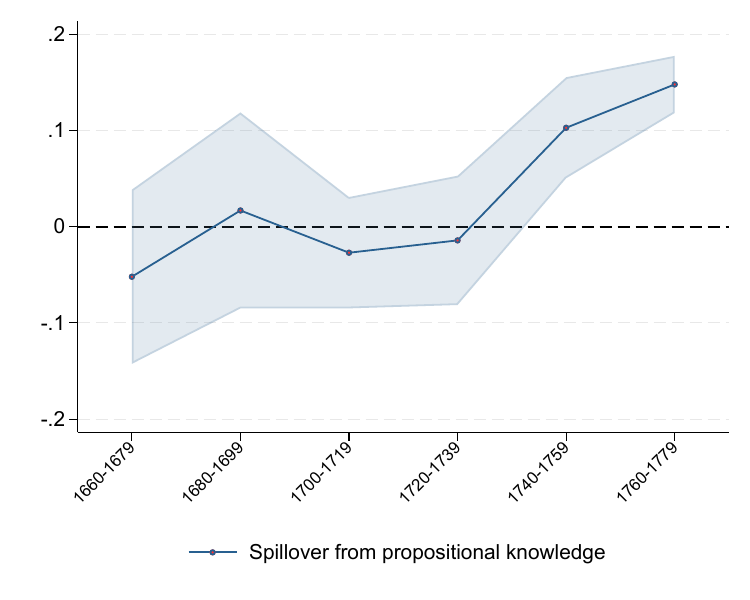}
				\subcaption{Applied physics}
			\end{subfigure}
			\begin{subfigure}[t]{0.19\textwidth}
				\includegraphics[width=1\linewidth]{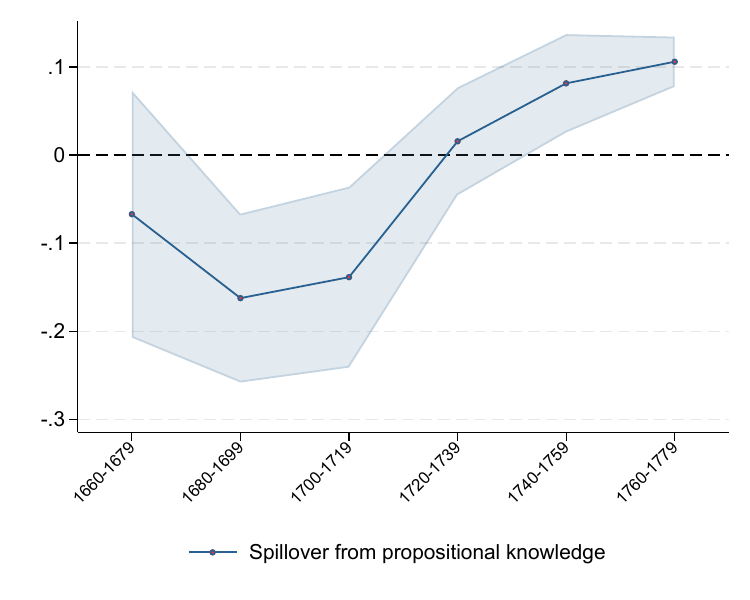}
				\subcaption{Astronomy}
			\end{subfigure}
			\begin{subfigure}[t]{0.19\textwidth}
				\includegraphics[width=1\linewidth]{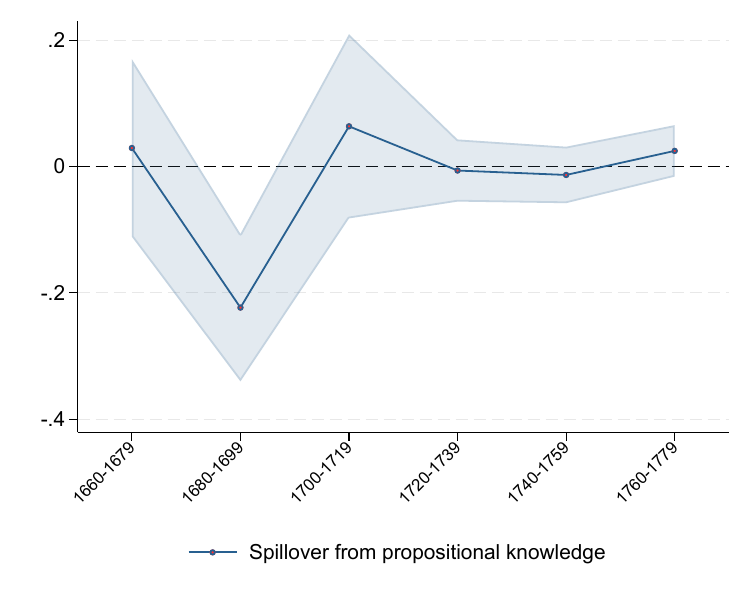}
				\subcaption{Mathematics}
			\end{subfigure}
			\begin{subfigure}[t]{0.19\textwidth}
				\includegraphics[width=1\linewidth]{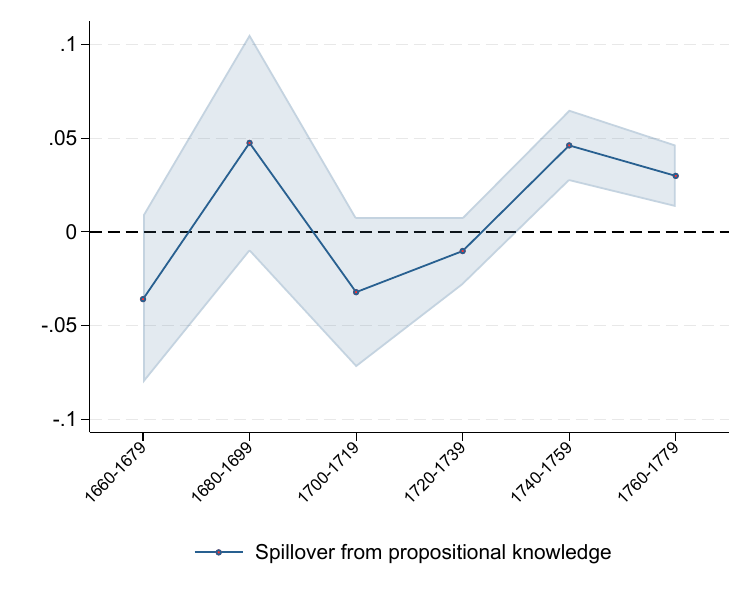}
				\subcaption{Chemistry}
			\end{subfigure}
			\begin{subfigure}[t]{0.19\textwidth}
				\includegraphics[width=1\linewidth]{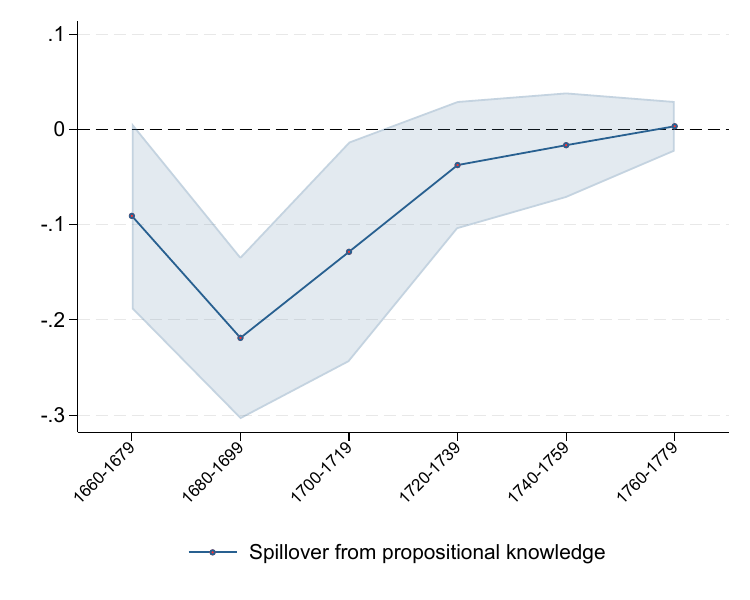}
				\subcaption{Encyclopedias}
			\end{subfigure}

			\medskip
			{\footnotesize \textit{Spillovers from individual fields in $\lambda$ $\rightarrow$ innovation in $\Omega$}}\par\smallskip
			\begin{subfigure}[t]{0.24\textwidth}
				\includegraphics[width=1\linewidth]{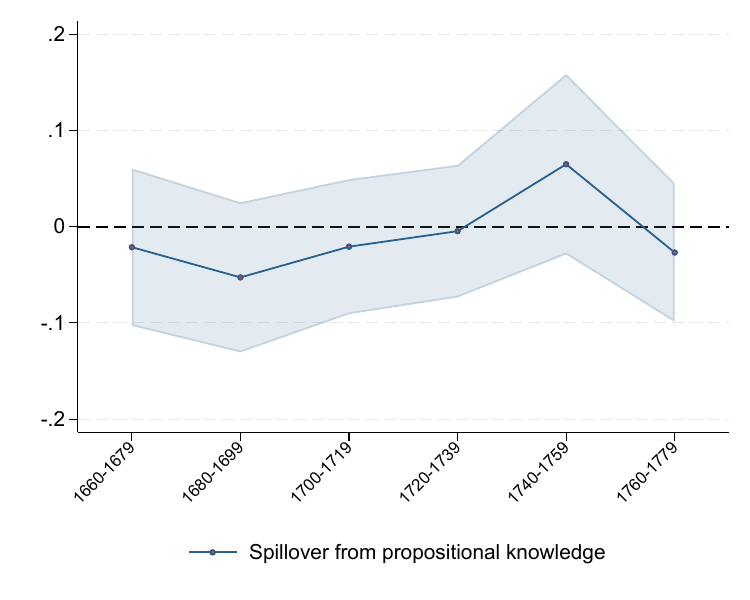}
				\subcaption{Technology in trades}
			\end{subfigure}
			\begin{subfigure}[t]{0.24\textwidth}
				\includegraphics[width=1\linewidth]{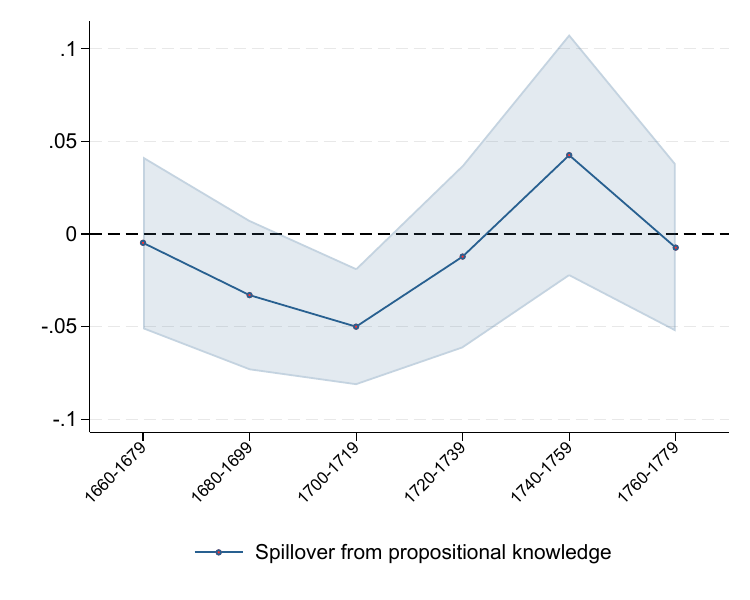}
				\subcaption{Technology in agriculture}
			\end{subfigure}
			\begin{subfigure}[t]{0.24\textwidth}
				\includegraphics[width=1\linewidth]{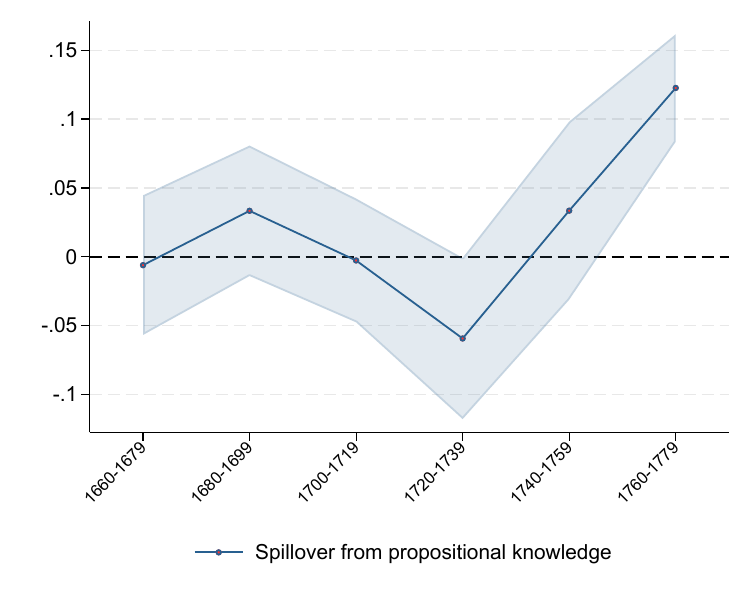}
				\subcaption{Navigation}
			\end{subfigure}
			\begin{subfigure}[t]{0.24\textwidth}
				\includegraphics[width=1\linewidth]{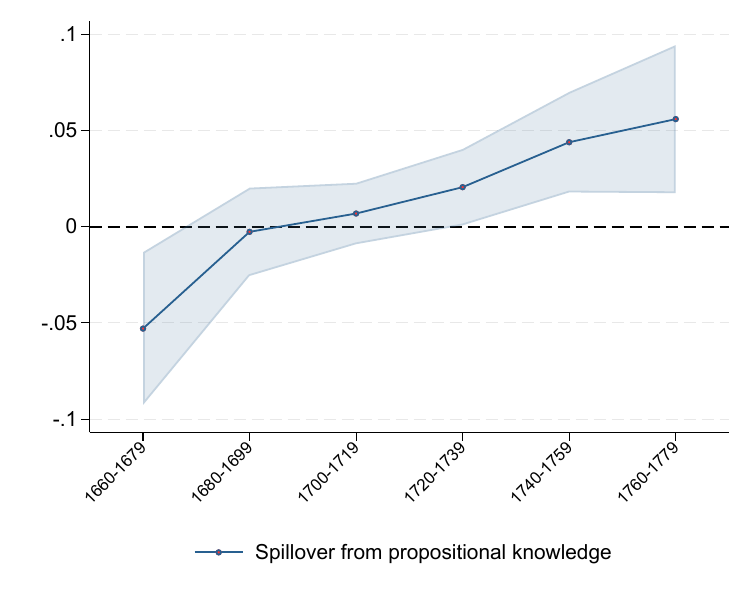}
				\subcaption{Scientific instruments}
			\end{subfigure}
			\caption{Results for individual spillovers from $\Omega$ and $\lambda$ \vspace{4pt} \newline \footnotesize \emph{Notes:} The figure reports the estimated coefficients from equation~\ref{eq:ols_spillover}. Instead of estimating the association between spillovers from \textit{all} fields in $\Omega$ ($\lambda$) and innovation in $\lambda$ ($\Omega$), this figure reports individual results for spillovers from each field. Panels a)--e) report spillovers from applied physics, astronomy, mathematics, chemistry, and encyclopedias in $\Omega$. Panels f)--i) report spillovers from technology in trades, technology in agriculture, navigation, and scientific instruments in $\lambda$. 90\% confidence intervals estimated using standard errors clustered at the publication year level.}
			\label{fig:results_individual_fields}
		\end{figure}

		Figure~\ref{fig:results_individual_fields} reports results for spillovers from individual fields in $\Omega$ and $\lambda$.

		\subsection{10 year intervals, compositional bias, and long-run results}
		\label{sec:10-year-interval}
		\label{sec:long-run}
		
\begin{figure}[!htbp]
			\centering
			\tiny
			{\footnotesize \textit{Spillovers from $\Omega$ $\rightarrow$ innovation in $\lambda$}}\par\smallskip
			\begin{subfigure}[t]{0.3\textwidth}
				\includegraphics[width=1\linewidth]{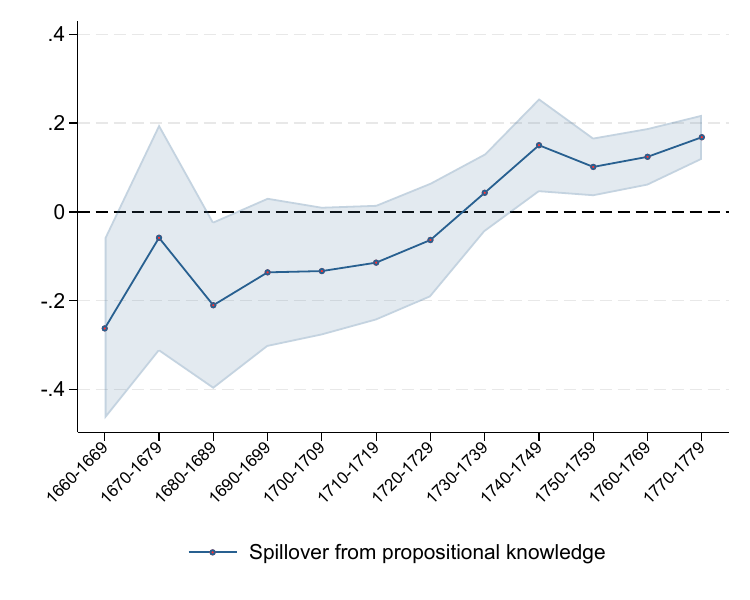}
				\subcaption{10 year intervals}
			\end{subfigure}
			\begin{subfigure}[t]{0.3\textwidth}
				\includegraphics[width=1\linewidth]{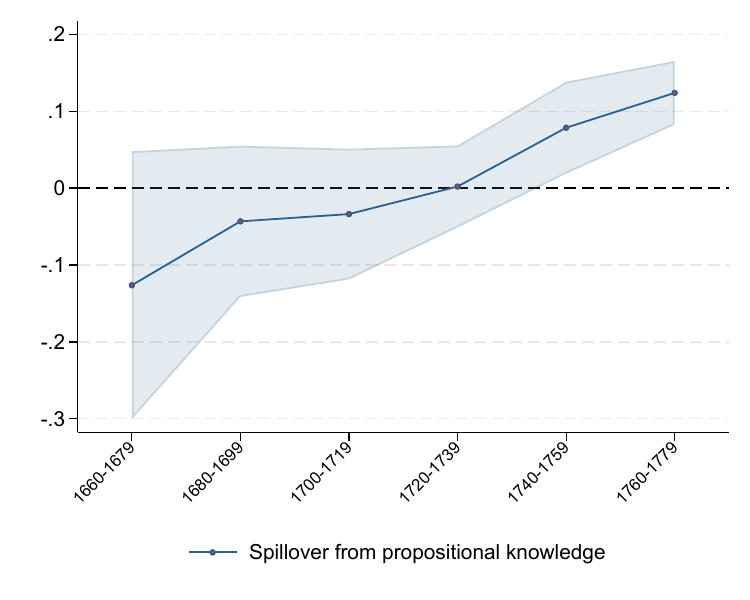}
				\subcaption{Subject $\times$ year fixed effects}
			\end{subfigure}
			\begin{subfigure}[t]{0.3\textwidth}
				\includegraphics[width=1\linewidth]{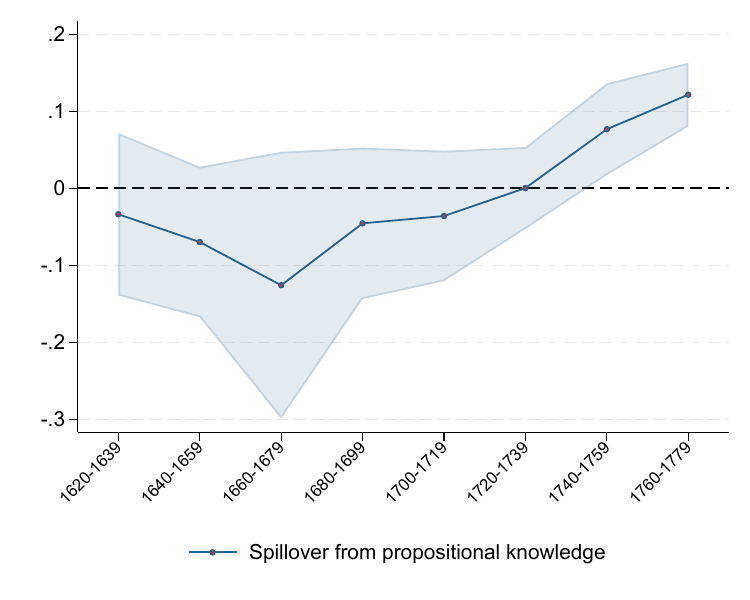}
				\subcaption{Full period, 1620--1779}
			\end{subfigure}

			\medskip
			{\footnotesize \textit{Spillovers from $\lambda$ $\rightarrow$ innovation in $\Omega$}}\par\smallskip
			\begin{subfigure}[t]{0.3\textwidth}
				\includegraphics[width=1\linewidth]{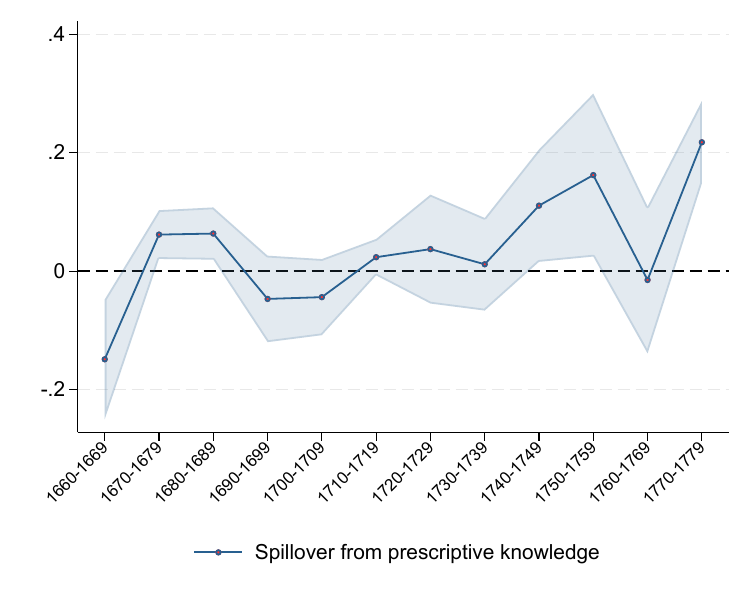}
				\subcaption{10 year intervals}
			\end{subfigure}
			\begin{subfigure}[t]{0.3\textwidth}
				\includegraphics[width=1\linewidth]{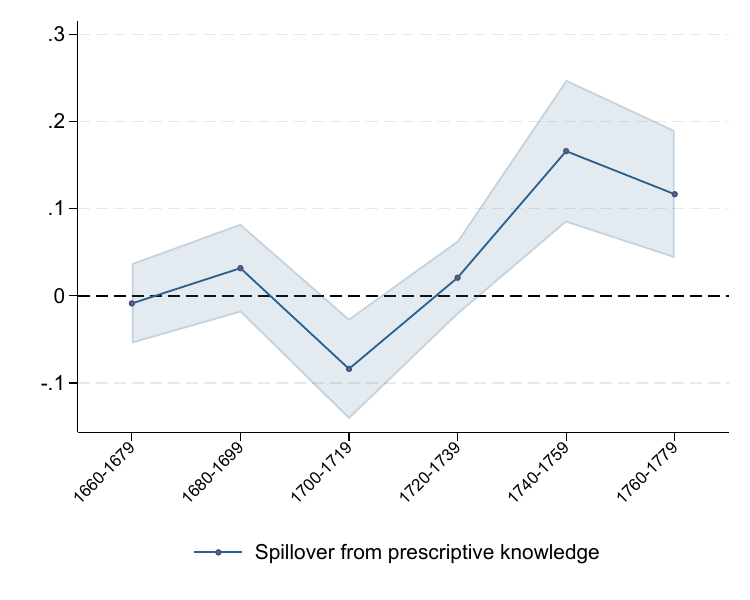}
				\subcaption{Subject $\times$ year fixed effects}
			\end{subfigure}
			\begin{subfigure}[t]{0.3\textwidth}
				\includegraphics[width=1\linewidth]{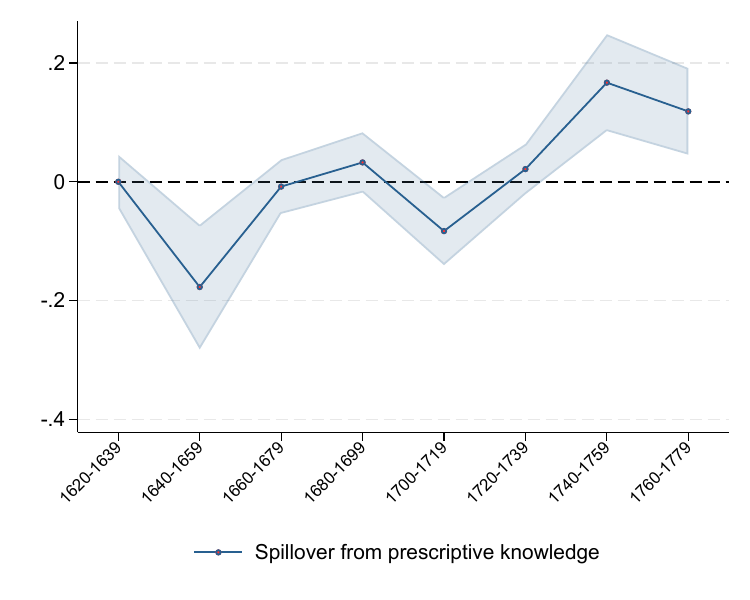}
				\subcaption{Full period, 1620--1779}
			\end{subfigure}
			\caption{10 year intervals, subject $\times$ year fixed effects, and full sample period --- Spillovers from propositional ($\Omega$) and prescriptive knowledge ($\lambda$) $\rightarrow$ innovation \vspace{4pt} \newline \footnotesize \emph{Notes:} The figure reports the estimated coefficients from equation~\ref{eq:ols_spillover}. The dependent variable is the innovation index from equation~\ref{eq:innovation}. The model further controls for the level and quadratic count of words in titles and patents and includes year fixed effects. Panels a)--c) report spillovers from $\Omega$ to $\lambda$, panels d)--f) spillovers from $\lambda$ to $\Omega$. Panels a) and d) plot the coefficients from interacting the received spillover index from equation~\ref{eq:received_spillover} with ten-year time periods. Panels b) and e) interact it with twenty-year time periods and, to account for compositional effects, further include subject class $\times$ year fixed effects. Panels c) and f) additionally extend the sample to the full period 1620--1779, with twenty-year time periods and subject class $\times$ year fixed effects. 90\% confidence intervals estimated using standard errors clustered at the publication year level.}
			\label{fig:robustness_periods_fe}
		\end{figure}

		Figure~\ref{fig:robustness_periods_fe} reports results for ten-year intervals, with subject $\times$ year fixed effects, and for the full period 1620--1779.

		\subsection{Alternative definitions of propositional and prescriptive knowledge}
		\label{sec:excluding-ency}
		\label{sec:robustness_patent_discont}
		\label{sec:robustness_patent_excl}
		
\begin{figure}[!htbp]
			\centering
			\tiny
			\begin{subfigure}[t]{0.3\textwidth}
				\includegraphics[width=1\linewidth]{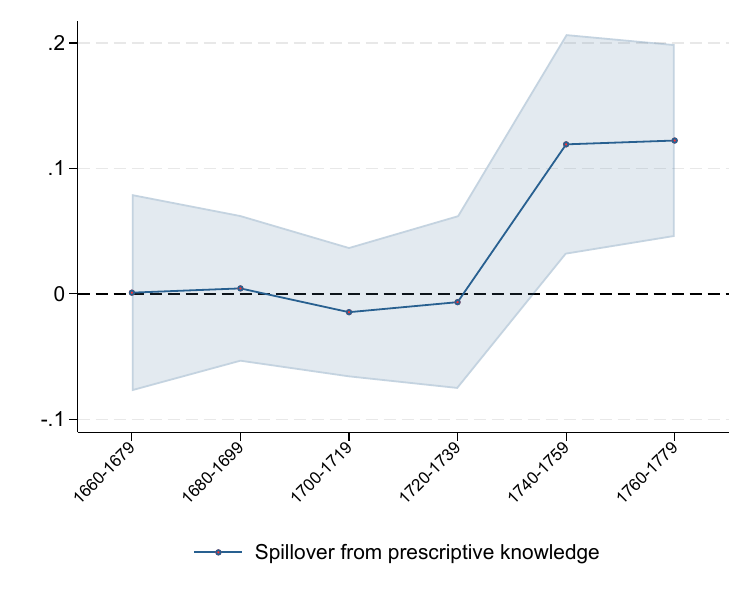}
				\subcaption{$\lambda$ $\rightarrow$ innovation in $\Omega$, excl. \textit{encyclopedias and dictionaries}}
				\label{fig:results_prop_prescr_excluding_ency}
			\end{subfigure}
			\begin{subfigure}[t]{0.3\textwidth}
				\includegraphics[width=1\linewidth]{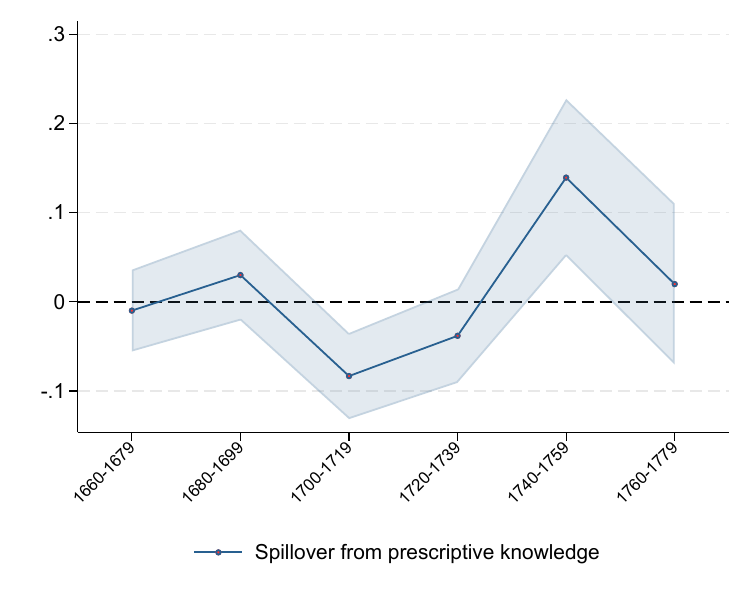}
				\subcaption{$\lambda$ $\rightarrow$ innovation in $\Omega$, incl. patents in the spillover index}
				\label{fig:results_prescr_prop_no_pat}
			\end{subfigure}
			\begin{subfigure}[t]{0.3\textwidth}
				\includegraphics[width=1\linewidth]{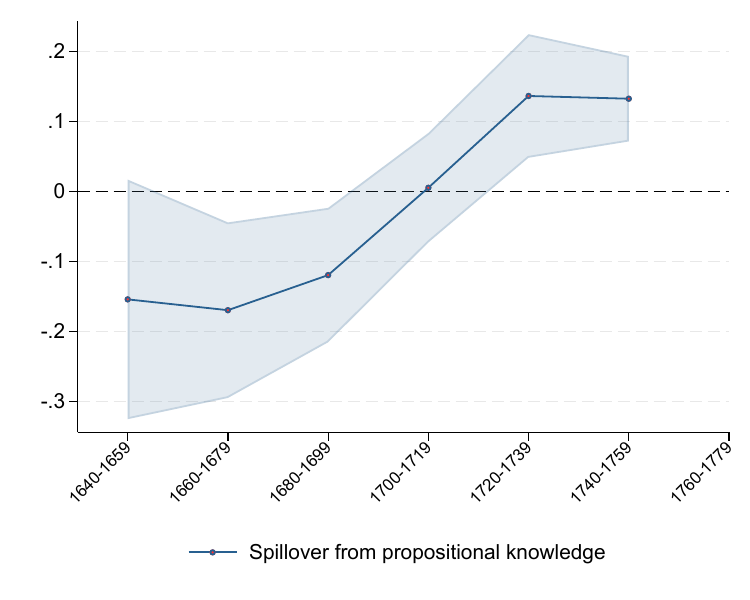}
				\subcaption{$\Omega$ $\rightarrow$ innovation in $\lambda$, excl. patents from $\lambda$}
				\label{fig:results_prop_prescr_no_pat}
			\end{subfigure}
			\caption{Alternative definitions of propositional ($\Omega$) and prescriptive knowledge ($\lambda$) \vspace{4pt} \newline \footnotesize \emph{Notes:} The figure reports the estimated coefficients from equation~\ref{eq:ols_spillover}. The dependent variable is the innovation index from equation~\ref{eq:innovation}. The graph plots the coefficients from interacting the received spillover index from equation~\ref{eq:received_spillover} with twenty-year time periods. The model further controls for the level and quadratic count of words in titles and patents and includes year fixed effects. Panel a) excludes all observations within \textit{Encyclopedias and dictionaries}, the most ambiguous category of $\Omega$. Panel b) includes patents in the calculation of the spillover index, which the baseline excludes. Because patent data is not available before 1700 (see data section~\ref{sec:data_patents}), this creates an undesirable discontinuity; to mitigate it, the specification in panel b) further adds subject class $\times$ year fixed effects. Panel c) excludes patents from the set of prescriptive knowledge ($\lambda$). 90\% confidence intervals estimated using standard errors clustered at the publication year level.}
			\label{fig:robustness_definitions}
		\end{figure}

		Figure~\ref{fig:robustness_definitions} reports results when excluding \textit{encyclopedias and dictionaries} from $\Omega$, when including patents in the spillover index, and when excluding patents from $\lambda$.

		\subsection{Patents: 10-year periods and patent citations as alternative outcome}
		
\begin{figure}[!htbp]
			\centering
			\tiny
			\begin{subfigure}[t]{0.24\textwidth}
				\includegraphics[width=1\linewidth]{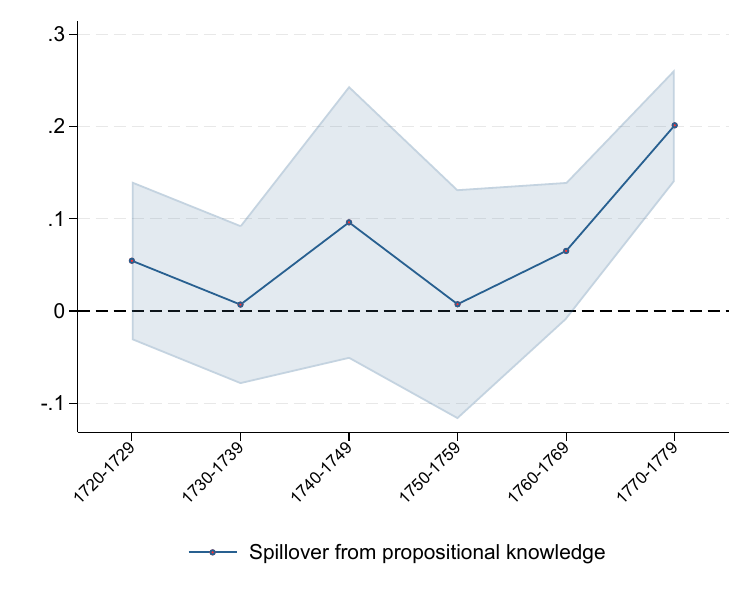}
				\subcaption{$\Omega$ $\rightarrow$ patent innovation}
			\end{subfigure}
			\begin{subfigure}[t]{0.24\textwidth}
				\includegraphics[width=1\linewidth]{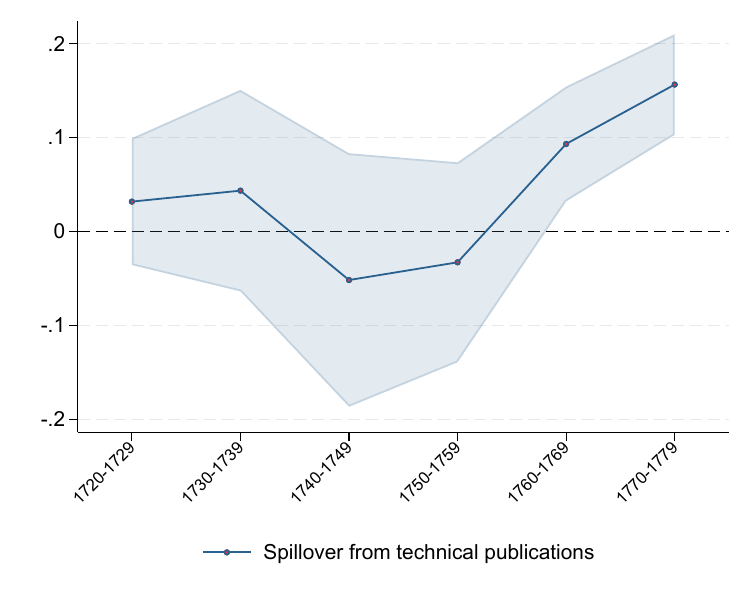}
				\subcaption{Applied physics $\rightarrow$ patent innovation}
			\end{subfigure}
			\begin{subfigure}[t]{0.24\textwidth}
				\includegraphics[width=1\linewidth]{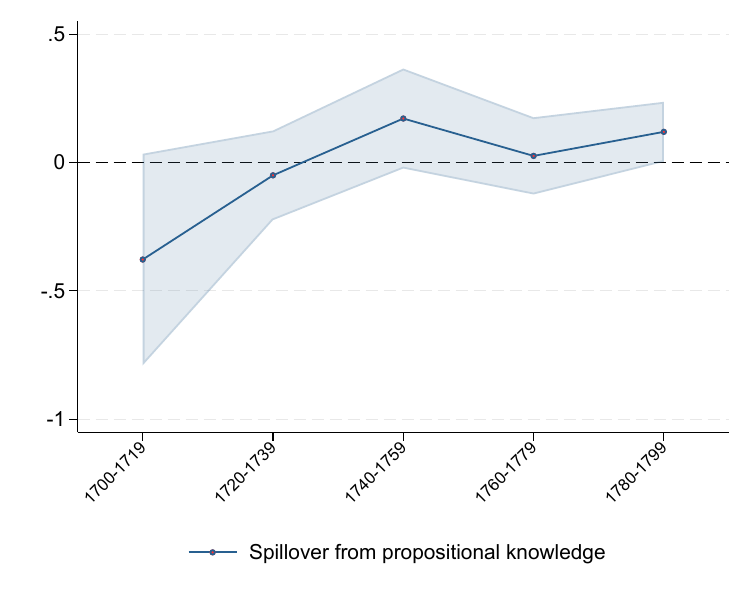}
				\subcaption{$\Omega$ $\rightarrow$ patent citations}
			\end{subfigure}
			\begin{subfigure}[t]{0.24\textwidth}
				\includegraphics[width=1\linewidth]{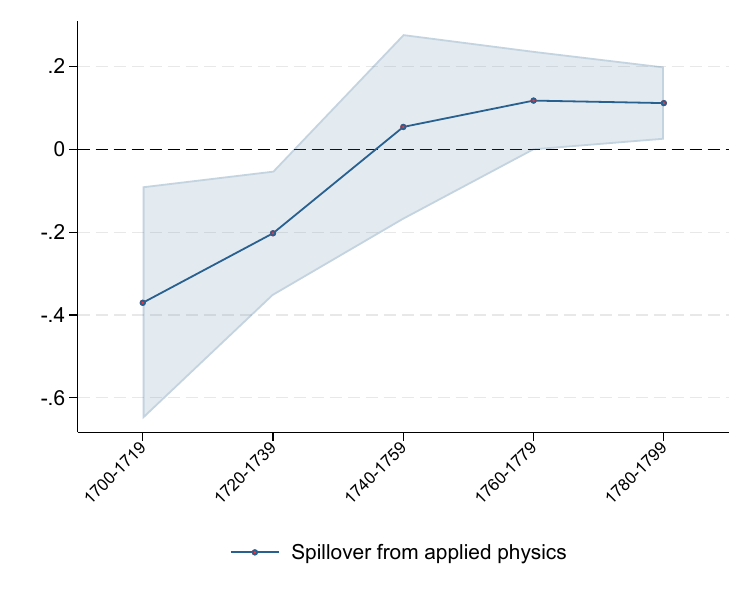}
				\subcaption{Applied physics $\rightarrow$ patent citations}
			\end{subfigure}
			\caption{Spillovers from propositional knowledge ($\Omega$) and applied physics $\rightarrow$ patent innovation (10 year periods) and patent citations \vspace{4pt} \newline \footnotesize \emph{Notes:} The figure reports the estimated coefficients from equation \ref{eq:ols_spillover}. Panel a) and b) show patent results from figure~\ref{fig:results_patents} for 10-year periods. Panel c) and d) show results from figure~\ref{fig:results_patents} with patent citations from \cite{Nuvolari2011}  as an alternative outcome. 90\% confidence intervals estimated using standard errors clustered at the publication year level.}
			\label{fig:results_patents_robustness}
		\end{figure}

		Figure~\ref{fig:results_patents_robustness} reports results for patents with 10-year periods and with patent citations as the outcome.

		\subsection{Patents: By-industry results}
		\label{sec:by-industry}
		
\begin{figure}[!htbp]
			\centering
			\tiny
			\begin{subfigure}[t]{0.4\textwidth}
				\includegraphics[width=1\linewidth]{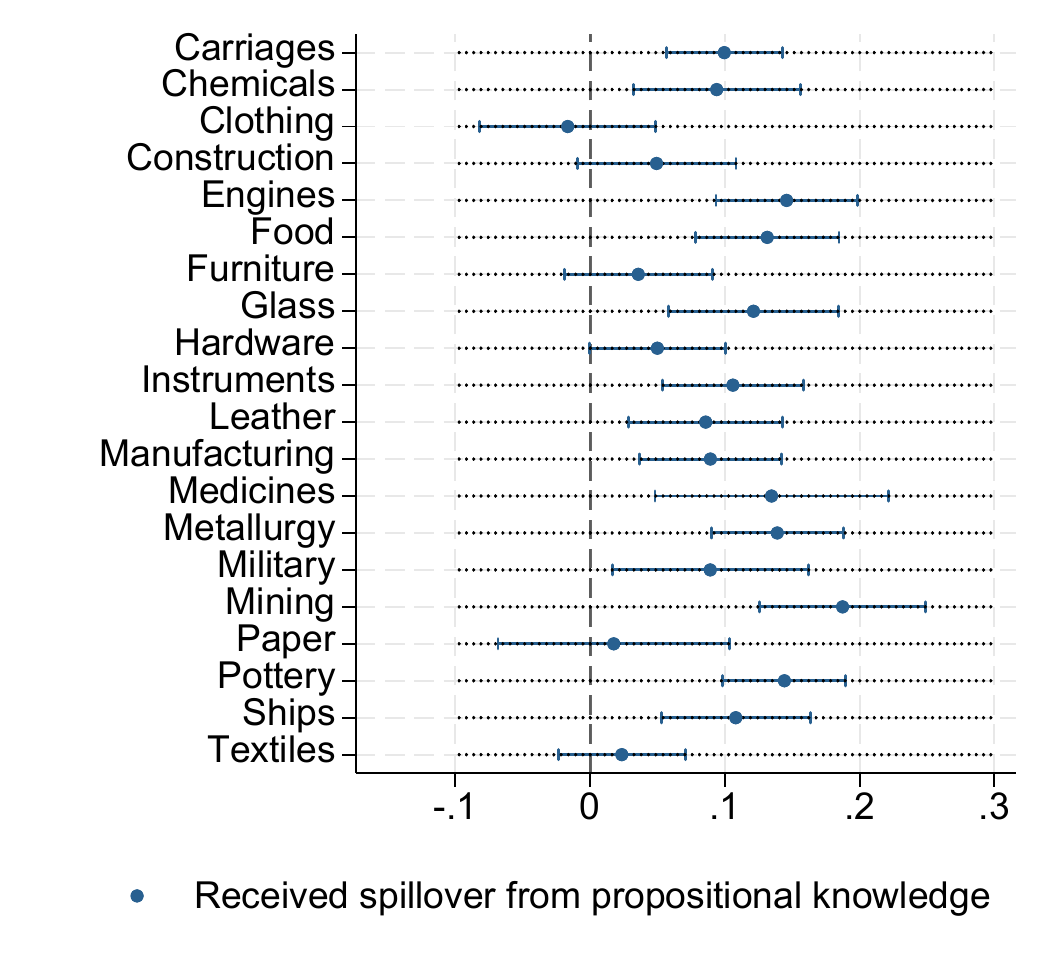}
				\subcaption{Spillovers in $\Omega$ $\rightarrow$ patent innovation}
			\end{subfigure}
			\begin{subfigure}[t]{0.4\textwidth}
				\includegraphics[width=1\linewidth]{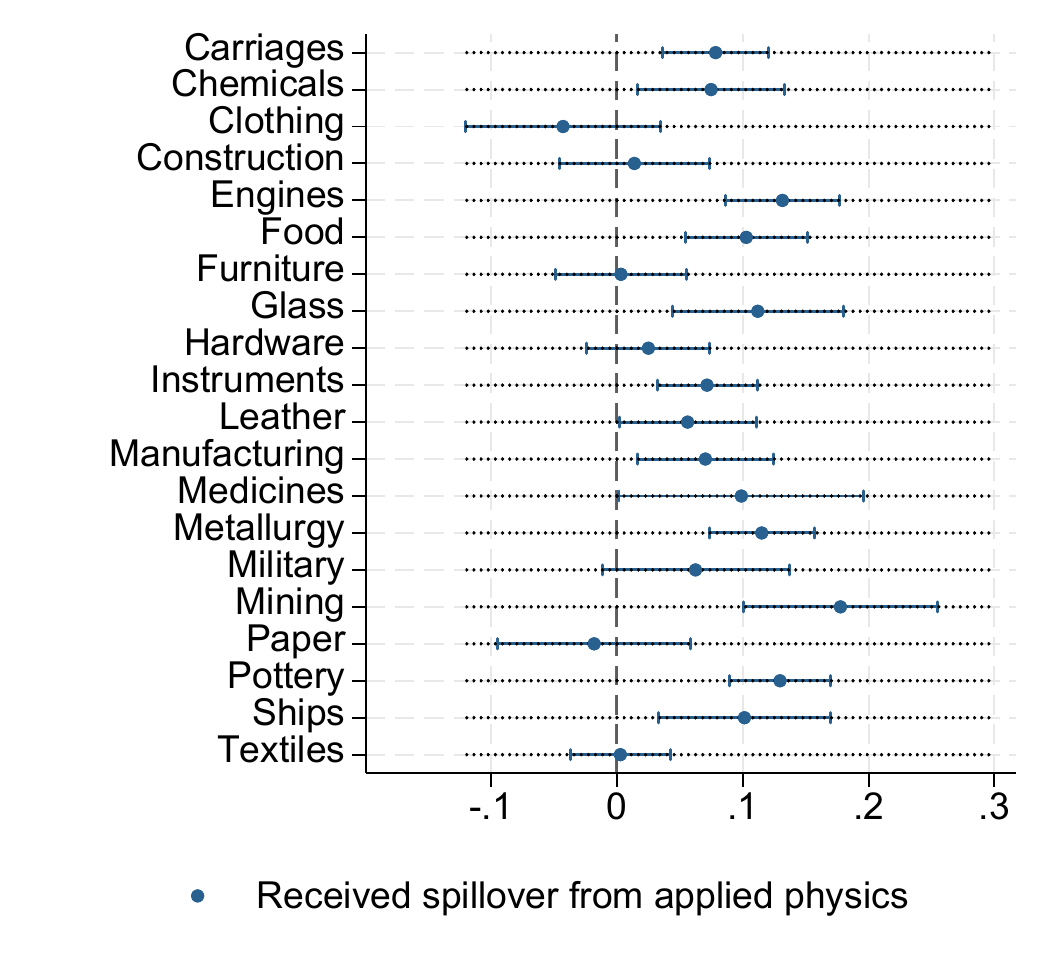}
				\subcaption{Spillovers in applied physics $\rightarrow$ patent innovation}
			\end{subfigure}
			\caption{By-industry results: The association between innovation and spillovers form $\Omega$ and applied physics\vspace{4pt} \newline \footnotesize \emph{Notes:} The figure reports the estimated coefficients from equation \ref{eq:ols_spillover}. The dependent variable is the innovation index from equation~\ref{eq:innovation}. The graph plots the coefficients from interacting the received spillover index from equation~\ref{eq:spillover} with twenty-year time periods. The model further controls for the level and quadratic count of words and includes year fixed effects. 90\% confidence intervals estimated using standard errors clustered at the publication year level.}
			\label{fig:patents_results_by_industry}	
		\end{figure}

		This section presents by-industry results for the association between patents and received spillovers from $\Omega$ and applied physics. We estimate a model similar to equation~\ref{eq:ols_spillover}, with the adjustment of interacting received spillovers with indicator variables for each industry from \cite{Nuvolari2011}:
		
		\begin{equation}
			\label{eq:ols_per_industry}
			\text{Innovation}^{A}_{ijt} = \sum_{p\in \text{Industry}}^{} ( \beta_p \cdot 	\text{Received spillover}_{B \to A}(v_{ijt}) \times I_p) + \boldsymbol{X'_{ijt}}\zeta	+ \delta_j + \alpha_t + \varepsilon_{ijt}
		\end{equation}
		
		the model specification is identical to equation~\ref{eq:ols_spillover}, with the difference of dropping time interaction terms and adding industry interaction term, $I_p$. The explanatory variable captures received spillovers from either propositional knowledge or applied physics. Results are shown in figure~\ref{fig:patents_results_by_industry}.

		\subsection{Specific types of contents}	
		
\begin{figure}[!htbp]
			\centering
			\begin{subfigure}[t]{0.49\textwidth}
				\includegraphics[width=1\linewidth]{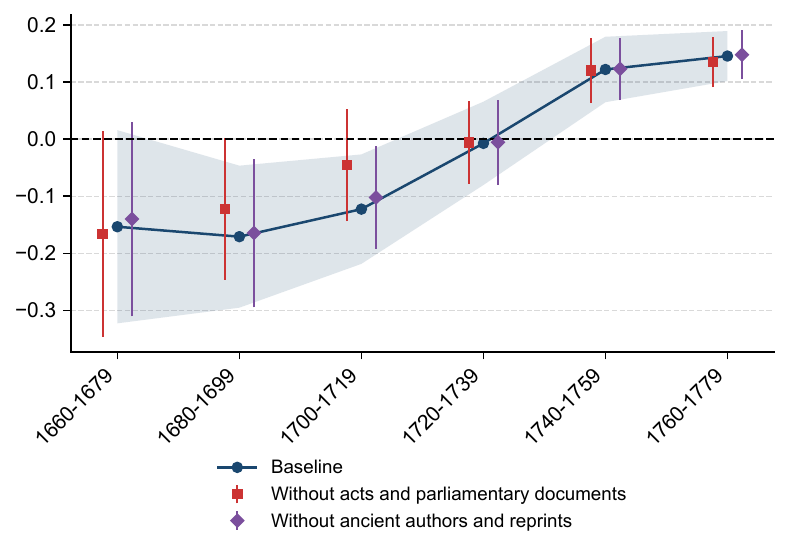}
				\subcaption{Spillovers from $\Omega$ $\rightarrow$ innovation in $\lambda$}
			\end{subfigure}
			\begin{subfigure}[t]{0.49\textwidth}
				\includegraphics[width=1\linewidth]{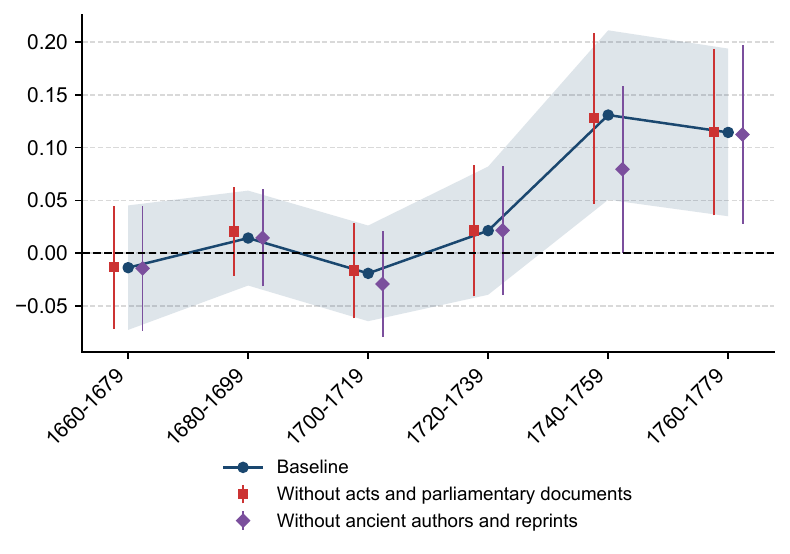}
				\subcaption{Spillovers from $\lambda$ $\rightarrow$ innovation in $\Omega$}
			\end{subfigure}
			\caption{Main results without parliamentary documents and without ancient authors and reprints \vspace{4pt} \newline \footnotesize \emph{Notes:} Coefficients $\beta_p$ from equation~\ref{eq:ols_spillover}. Navy: baseline. Red: acts and parliamentary documents dropped. Purple: titles by ancient authors (e.g.\ Euclid, Archimedes, Xenophon) and reprints (cosine similarity above 0.9 to an earlier title of the same field) dropped.  $N = $ 3,119 (a) and 2,940 (b) in the baseline, 2,989 (a) and 2,931 (b) without acts and parliamentary documents, 3,066 (a) and 2,797 (b) without ancient authors and reprints. 90\% confidence intervals estimated using standard errors clustered at the publication year level.}
			\label{fig:negcoef_robustness}
		\end{figure}

		Figure~\ref{fig:negcoef_robustness} reports robustness checks for excluding specific types of texts. First, official acts that announced engineering projects, and second reprints of ancient writers --- none of these is driving the results. 
		
		\section{Mechanism}
		\subsection{Upper-tail human capital}
		\label{sec:appendix_mechanism_robustness}
		
		As additional evidence for section~\ref{sec:uthc}, which studies the role of upper-tail human capital in generating knowledge spillovers, this section provides additional evidence on the composition of different groups of upper-tail human capital as well as long-run trends in the share of Oxford and Cambridge educated authors among publications.

		First, table~\ref{tab:mechanism_descriptive} illustrates how much the different occupational categories overlapped. First, for membership in the Society of Arts \citep[see][]{Howes2020}, we find that more than 63\% of the publishing members of the \textit{Society for the Encouragement of Arts, Manufactures and Commerce} were also fellows of the Royal Society.\footnote{Vice versa, only a minority of fellows of the Royal Society were also members of the \textit{Society for the Encouragement of Arts, Manufactures and Commerce}. However this is mainly an artifact of the later foundation of the \textit{Society for the Encouragement of Arts, Manufactures and Commerce} in 1754.} Unsurprisingly, a high number of the Fellows of the Royal Society also held academic or teaching positions.\footnote{Nonetheless, the Society was broader than just the academic community and included fellows from e.g. the nobility, clergy, the navy or other broad walks of life.} Next, we find that publishing engineers were very likely to also be Fellows of the Royal Society (especially given its status as an elite scientific society). Moreover, 18.2\% of engineers also held an academic or teaching position. This finding highlights the importance of engineers for bridging the spheres of scholarly knowledge and applied practical work \citep[see also][]{hanlonengineer}.
		
		\begin{table}[!htbp]
		\centering
			\begin{adjustbox}{max width=0.8\columnwidth}
				\centering
				\begin{threeparttable}\fontsize{10}{13}\selectfont
					\caption{Determinants of spillovers, excl. indicator for university enrollment}
					\label{tab:mechanism_descriptive}
					\begin{tabular}{lcccc}
  \toprule
  & \multicolumn{4}{c}{\textit{Percentage of overlap between \ldots}} \\
  \cmidrule(lr){2-5}
  & Royal Soc. & Soc. Arts. & Engineer & Academic \\
  \midrule
  Fellowship in Royal Society & --- & 2.5\% & 10.4\% & 39.9\% \\
  Society of Arts & 63.2\% & --- & 10.5\% & 0.0\% \\
  Engineer & 24.4\% & 1.0\% & --- & 18.2\% \\
  Academic Career & 26.7\% & 0.0\% & 5.2\% & --- \\
  \bottomrule
\end{tabular}

					\begin{tablenotes}
						\item {\footnotesize \emph{Notes:} The table shows the overlap between different occupational categories. Each cell reports the share of individuals in the row category who also belong to the column category. Sample restricted to authors of publications in propositional and prescriptive knowledge published from 1740 onwards.}
					\end{tablenotes}
				\end{threeparttable}
			\end{adjustbox}
		\end{table}
		
\begin{figure}[!htbp]
			\centering
			\tiny
			\includegraphics[width=0.6\linewidth]{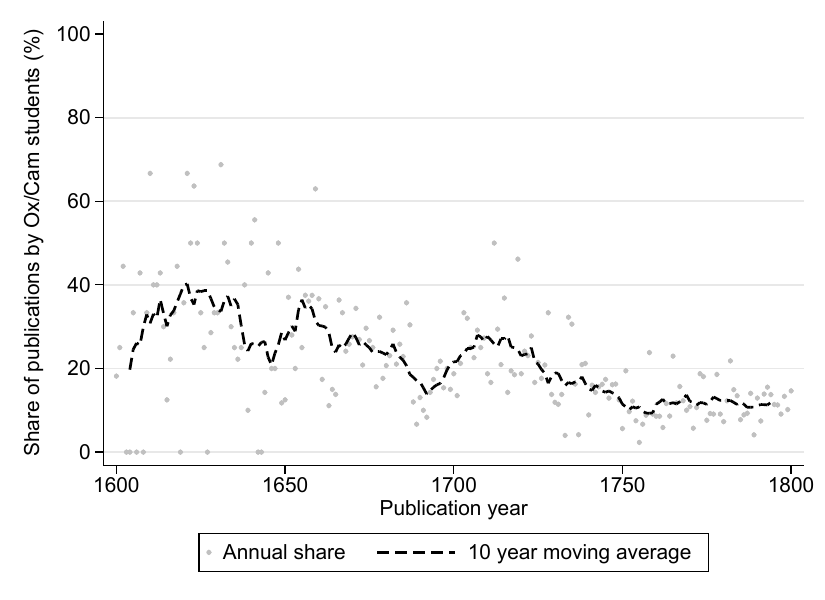}
			\caption{Share of publications in propositional and prescriptive knowledge by Oxford and Cambridge students over time \vspace{4pt} \newline \footnotesize \emph{Notes:}   The figure reports the share of publications by authors who enrolled at either the University of Oxford or Cambridge. It includes both the yearly average share as well as a plot of the 10 year moving average. The sample consists of the joint set of prescriptive and propositional knowledge. Publications without a recorded author name are excluded from the sample. Student data obtained from \cite{Koschnick2025}.}
			\label{fig:share_ox_cam_over_time}	
		\end{figure}

\begin{figure}[!htbp]
			\centering
			\tiny
			\begin{subfigure}[t]{0.49\textwidth}
				\includegraphics[width=1\linewidth]{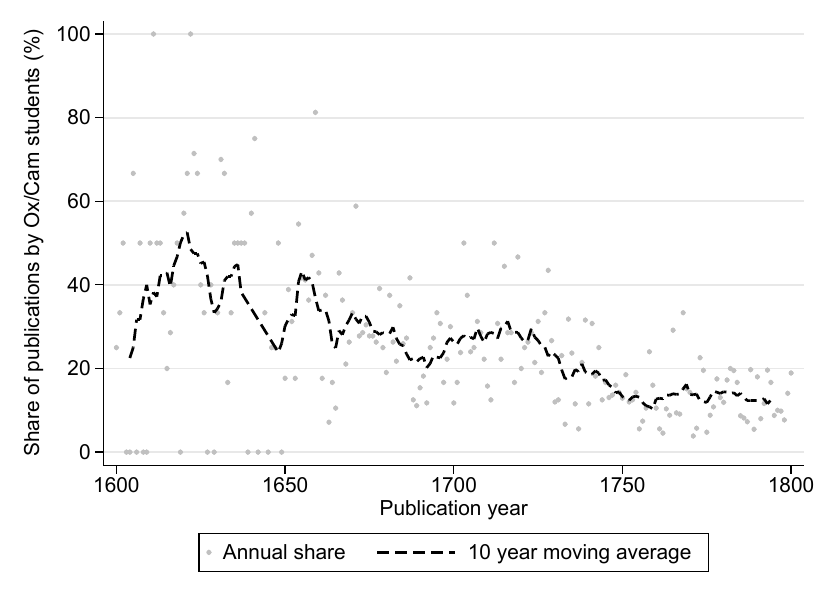}
				\subcaption{Propositional knowledge}
			\end{subfigure}
			\begin{subfigure}[t]{0.49\textwidth}
				\includegraphics[width=1\linewidth]{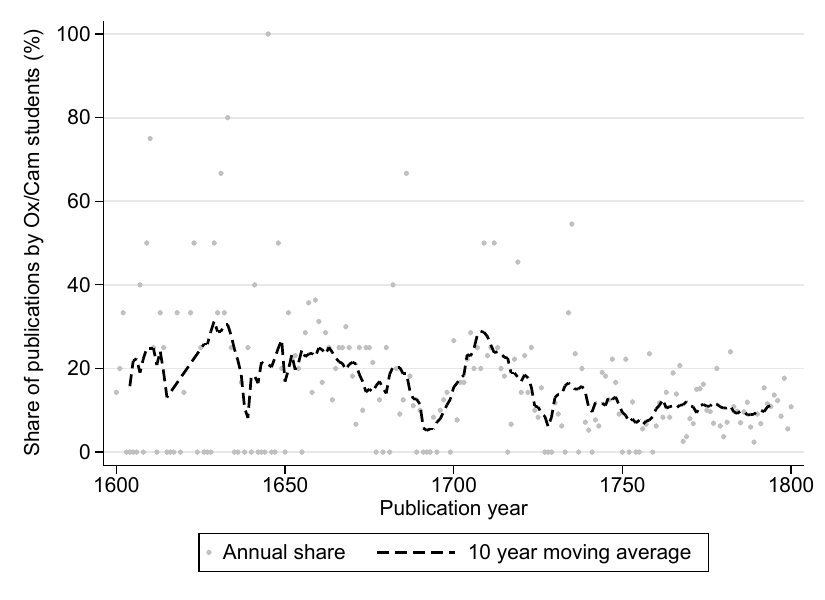}
				\subcaption{Prescriptive knowledge}
			\end{subfigure}
			\caption{Share of publications in propositional and prescriptive knowledge by Oxford and Cambridge students over time by types of knowledge \newline \footnotesize \emph{Notes:} The figure reports the share of publications by authors who enrolled at either the University of Oxford or Cambridge by type of knowledge. Subfigure a) reports the Oxford and Cambridge share propositional knowledge and subfigure b) reports it for prescriptive knowledge. The figures further include both the yearly average share as well as a plot of the 10 year moving average. Publications without a recorded author name are excluded from the sample. Student data obtained from \cite{Koschnick2025}.}
			\label{fig:share_ox_cam_over_time_by-type}	
		\end{figure}

		Next, to help interpret the positive Oxford/Cambridge education coefficient in table~\ref{tab:mechanism}, this the paper presents broad trends on publications in propositional and prescriptive knowledge that were written by Oxford or Cambridge educated authors. Figure~\ref{fig:share_ox_cam_over_time} presents results for the full set of both propositional and prescriptive knowledge, while figure~\ref{fig:share_ox_cam_over_time_by-type} shows trends for each propositional and prescriptive knowledge separately.

		The share of titles written by Oxford- or Cambridge-educated authors is highest in the seventeenth century, reaching up to 50\% for propositional knowledge and 30\% for prescriptive knowledge. This is followed by a decline over the eighteenth century. By 1800, the share in both categories falls to around 15--18\%. While this constitutes a large decline in comparison to the levels seen during the seventeenth century, it is still a meaningful share. Hence, a sizable share of authors in both prescriptive and propositional knowledge therefore had formal education that enabled them to read works in Latin and access scholarly works.

		In explaining the decline of the share of Oxford or Cambridge educated authors, three factors stand out. First, the trends in figure~\ref{fig:share_ox_cam_over_time} and~\ref{fig:share_ox_cam_over_time_by-type} broadly track the overall decline in student numbers at the universities of Oxford and Cambridge throughout the late seventeenth and eighteenth century \citep{Koschnick2025}. Second, it is important to note that enrollment at the universities of Oxford or Cambridge does not capture the universe of university education during this time. During the early eighteenth century, the Scottish universities became a new center of academic learning, especially in applied and experimental sciences. Third, vernacularization and the spread of the Industrial Enlightenment would likely have broadened the social spheres of authors \citep{Mokyr2016}. Also note that the ESTC book catalog includes works by foreign authors who might also have had a university education in their country of origin.\footnote{Likewise, we naturally also exclude English students at foreign universities such as Douai or Leiden.} 
		
		Hence, the numbers reported in figure~\ref{fig:share_ox_cam_over_time} and~\ref{fig:share_ox_cam_over_time_by-type} should be interpreted as an absolute lower bound on the share of publications by university educated authors in Britain. Yet, even this lower bound points to the relative importance of university educated authors for knowledge production in prescriptive and propositional knowledge. It remains clear that a significant share of authors had access to higher learning which, as argued in section~\ref{sec:uthc}, would likely have helped them to incorporate knowledge from other knowledge domains. Hence, these statistics complement the previous findings on the role of upper-tail human capital as a catalyst for knowledge spillovers.

		Finally, this appendix section reports robustness checks on the main mechanism section from table~\ref{tab:mechanism}. First, we report spillovers for the extended period 1740--1779. This becomes possible since the spillover innovation index, in contrast to the innovation index, is defined backwards. While this sample period does not longer corresponds to the association between spillovers and innovation from section~\ref{sec:results_spillovers_innovation}, it is still highly informative since, following \cite{Mokyr2002}, we expect the importance of positive spillovers to have increased over time. Table~\ref{tab:mechanism_1740_1800} reports the results.
		
		We find that results are highly similar to the shorter period 1740--1779 in table~\ref{tab:mechanism}. The largest difference is that for 1740--1799, coefficients for engineers for spillovers from propositional to prescriptive knowledge become significant. This is in line with the literature \citep{Reyonolds1983,Mokyr2022,Maloney2022,hanlonengineer} that have stressed the importance of engineers for incorporating abstract and scientific knowledge into experimental work --- it also fits with the relatively late emergence of the engineer as an independent profession within the First Industrial Revolution.

		\begin{table}[!htbp]
		\centering
			\begin{adjustbox}{max width=0.85\columnwidth}
				\begin{threeparttable}\fontsize{9}{11}\selectfont
					\caption{Determinants of spillovers between propositional ($\Omega$) and prescriptive knowledge ($\lambda$) for 1740--1800}
					\label{tab:mechanism_1740_1800}
					{
\def\sym#1{\ifmmode^{#1}\else\(^{#1}\)\fi}
\begin{tabular}{l*{6}{c}}
\hline\hline
                    &\multicolumn{6}{c}{Time frame: 1740--1799}                                                     \\\cline{2-7}
                    &\multicolumn{1}{c}{(1)}   &\multicolumn{1}{c}{(2)}   &\multicolumn{1}{c}{(3)}   &\multicolumn{1}{c}{(4)}   &\multicolumn{1}{c}{(5)}   &\multicolumn{1}{c}{(6)}   \\
\hline
\multicolumn{7}{l}{\textbf{Panel A: Spillovers from $\Omega$ to $\lambda$}} \\
Fellowship in Royal Society&      0.0879***&               &               &               &               &      0.0788***\\
                    &    (0.0146)   &               &               &               &               &    (0.0144)   \\
[0.5em]
Membership in Society of Arts&               &      0.0908***&               &               &               &      0.0494   \\
                    &               &    (0.0304)   &               &               &               &    (0.0319)   \\
[0.5em]
Engineer            &               &               &      0.0254   &               &               &      0.0246   \\
                    &               &               &    (0.0178)   &               &               &    (0.0189)   \\
[0.5em]
University enrollment&               &               &               &      0.0587***&               &      0.0210   \\
                    &               &               &               &    (0.0152)   &               &    (0.0146)   \\
[0.5em]
Academic career     &               &               &               &               &      0.0526***&      0.0470***\\
                    &               &               &               &               &    (0.0170)   &    (0.0169)   \\
[0.5em]
Word count controls &         Yes   &         Yes   &         Yes   &         Yes   &         Yes   &         Yes   \\
[0.5em]
Year fixed effects  &         Yes   &         Yes   &         Yes   &         Yes   &         Yes   &         Yes   \\
[0.5em]
Subject class fixed effects&         Yes   &         Yes   &         Yes   &         Yes   &         Yes   &         Yes   \\
\hline
Observations        &        2018   &        2018   &        2018   &        2783   &        2018   &        2018   \\
R-squared           &        0.30   &        0.29   &        0.29   &        0.30   &        0.29   &        0.31   \\
\hline
\multicolumn{7}{l}{\textbf{Panel B: Spillovers from $\lambda$ to $\Omega$}} \\
Fellowship in Royal Society&      0.0357***&               &               &               &               &      0.0310***\\
                    &   (0.00922)   &               &               &               &               &   (0.00975)   \\
[0.5em]
Membership in Society of Arts&               &       0.183***&               &               &               &       0.194***\\
                    &               &   (0.00883)   &               &               &               &    (0.0206)   \\
[0.5em]
Engineer            &               &               &      0.0115   &               &               &    -0.00357   \\
                    &               &               &    (0.0180)   &               &               &    (0.0190)   \\
[0.5em]
University enrollment&               &               &               &      0.0228** &               &      0.0153   \\
                    &               &               &               &    (0.0105)   &               &   (0.00995)   \\
[0.5em]
Academic career     &               &               &               &               &      0.0222***&      0.0135   \\
                    &               &               &               &               &   (0.00814)   &   (0.00831)   \\
[0.5em]
Word count controls &         Yes   &         Yes   &         Yes   &         Yes   &         Yes   &         Yes   \\
[0.5em]
Year fixed effects  &         Yes   &         Yes   &         Yes   &         Yes   &         Yes   &         Yes   \\
[0.5em]
Subject class fixed effects&         Yes   &         Yes   &         Yes   &         Yes   &         Yes   &         Yes   \\
\hline
Observations        &        2053   &        2053   &        2053   &        2685   &        2053   &        2053   \\
R-squared           &        0.38   &        0.38   &        0.38   &        0.38   &        0.38   &        0.39   \\
\hline\hline
\end{tabular}
}

					\begin{tablenotes}
						\item {\footnotesize \emph{Notes:} OLS estimates of equation~\ref{eq:ols_spillover_mechanism}. The dependent variable is the log received spillover index from equation~\ref{eq:received_spillover}, from $\Omega$ to $\lambda$ (panel A) and from $\lambda$ to $\Omega$ (panel B). The explanatory variables are indicators for authors' membership, occupations, and education, entered one at a time in columns~1--5 and jointly in column~6. All models control for the level and quadratic word count and include year and subject class fixed effects. Standard errors clustered by publication year. *** $p<0.01$, ** $p<0.05$, * $p<0.1$.}
					\end{tablenotes}
				\end{threeparttable}
			\end{adjustbox}
		\end{table}

		\subsection{List of terms for cosine similarity to methods}
		
		\begin{table}[!htbp]
			\centering
			\begin{adjustbox}{max width=\columnwidth, valign=t}
				\begin{threeparttable}
					\fontsize{10}{13}\selectfont
					\caption{Conceptual term lists for revolutions in methods}
					\label{tab:revolution_method_terms}
					\small
					
					\begin{tabular}{%
							>{\raggedright\arraybackslash}p{0.32\linewidth}
							>{\raggedright\arraybackslash}p{0.32\linewidth}
							>{\raggedright\arraybackslash}p{0.32\linewidth}
						}
						\textbf{Newtonian mechanics} &
						\textbf{Precise measurement \& toolset} &
						\textbf{Scientific method} \\[0.3em]
						\hline
						\begin{itemize}[leftmargin=1em, itemsep=0em, parsep=0em, topsep=0.2em]
							\item Newtonian mechanics
							\item laws of motion
							\item force, acceleration, momentum, mass
							\item gravity, center of gravity
							\item levers, pulleys
							\item projectile motion, uniform motion
							\item resistance of media, friction
							\item impact, pressure
							\item hydrostatics, statics, dynamics
							\item centripetal force, centrifugal force
							\item impulse
							\item fluid resistance, flow of fluids
						\end{itemize}
						&
						\begin{itemize}[leftmargin=1em, itemsep=0em, parsep=0em, topsep=0.2em]
							\item precise measurement
							\item precise instruments
							\item precision instruments
							\item exact measurement
							\item standardized scales
							\item calibration
						\end{itemize}
						&
						\begin{itemize}[leftmargin=1em, itemsep=0em, parsep=0em, topsep=0.2em]
							\item observation
							\item measurement
							\item mathematical formalization
							\item experiment
						\end{itemize}
						\\ 
					\end{tabular}
					
					\begin{tablenotes}
						\footnotesize
						\item \emph{Notes:} The table reports the list of terms used for calculating title-level similarity to the three methods of \textit{Newtonian mechanics}, \textit{precise measurement}, and the \textit{scientific method}. See section~\ref{sec:steamberth}.
					\end{tablenotes}
					
				\end{threeparttable}
			\end{adjustbox}
		\end{table}

	\end{appendices}

\end{document}